\documentclass[fleqn,usenatbib]{mnras}

\pdfoutput=1

\usepackage{mathptmx}

\usepackage[T1]{fontenc}
\usepackage{ae,aecompl}

\DeclareRobustCommand{\VAN}[3]{#2}
\let\VANthebibliography\thebibliography
\def\thebibliography{\DeclareRobustCommand{\VAN}[3]{##3}\VANthebibliography}

\usepackage{graphicx}	
\usepackage{amsmath}	
\usepackage{amssymb}	
\usepackage{subfig}

\newcommand{\angstrom}{\text{\normalfont\AA}}
\newcommand{\myparagraph}[1]{\paragraph{#1}\mbox{}\\}

\title[Assembly history of S0s at large radii]{The SLUGGS survey: combining stars, globular clusters and planetary nebulae to understand the assembly history of early-type galaxies from their large radii kinematics}

\author[Dolfi et al.]{Arianna Dolfi,$^{1}$\thanks{E-mail: adolfi@swin.edu.au}
Duncan A. Forbes,$^{1}$
Warrick J. Couch,$^{1}$
Kenji Bekki,$^{3}$
Anna Ferr\'e-Mateu,$^{2,1}$
\and Aaron J. Romanowsky,$^{4,5}$
Jean P. Brodie,$^{1,5}$
\\
\\
$^{1}$ Centre for Astrophysics \& Supercomputing, Swinburne University of Technology, Hawthorn VIC 3122, Australia\\
$^{2}$ Institut de Ciencies del Cosmos (ICCUB), Universitat de Barcelona (IEEC-UB), E02028 Barcelona, Spain\\
$^{3}$ ICRAR, M468, The University of Western Australia 35 Stirling Highway, Crawley Western Australia, 6009, Australia\\
$^{4}$ Department of Physics $\&$ Astronomy, San Jos\'e State University, One Washington Square, San Jose, CA 95192, USA\\
$^{5}$ University of California Observatories, 1156 High St., Santa Cruz, CA 95064, USA\\
}

\date{Accepted 2021 April 09. Received 2021 March 29; in original form 2021 February 01}

\pubyear{2021}

\begin{document}
\label{firstpage}
\pagerange{\pageref{firstpage}--\pageref{lastpage}}
\maketitle

\begin{abstract}
We investigate the kinematic properties of nine nearby early-type galaxies with evidence of a disk-like component. Three of these galaxies are located in the field, five in the group and only one in the cluster environment. By combining the kinematics of the stars with those of the globular clusters (GCs) and planetary nebulae (PNe), we probe the outer regions of our galaxies out to $\sim4$-$6\, R_{\mathrm{e}}$. Six galaxies have PNe and red GCs that show good kinematic alignment with the stars, whose rotation occurs along the photometric major-axis of the galaxies, suggesting that both the PNe and red GCs are good tracers of the underlying stellar population beyond that traced by the stars. Additionally, the blue GCs also show rotation that is overall consistent with that of the red GCs in these six galaxies. The remaining three galaxies show kinematic twists and misalignment of the PNe and GCs with respect to the underlying stars, suggesting recent galaxy interactions. From the comparison with simulations, we propose that all six \textit{aligned} galaxies that show similar dispersion-dominated kinematics at large radii ($>2$-$3\, R_{\mathrm{e}}$) had similar late ($z \lesssim 1$) assembly histories characterised by mini mergers (mass-ratio $<$ 1:10). The diﬀerent $V_{\mathrm{rot}}/\sigma$ proﬁles are then the result of an early ($z>1$) minor merger (1:10 $<$ mass-ratio $<$ 1:4) for the four galaxies with \textit{peaked and decreasing} $V_{\mathrm{rot}}/\sigma$ proﬁles and of a late minor merger for the two galaxies with \textit{flat} $V_{\mathrm{rot}}/\sigma$ proﬁles. The three \textit{mis-aligned} galaxies likely formed through multiple late minor mergers that enhanced their velocity dispersion at all radii, or a late major merger that spun-up both the GC sub-populations at large radii. Therefore, lenticular galaxies can have complex merger histories that shape their characteristic kinematic proﬁle shapes.
\end{abstract}



\section{Introduction}
Early-type galaxies (ETGs; i.e. elliptical and lenticular) are found to be more numerous towards the denser regions of rich galaxy clusters \citep{Dressler1980}, with the fraction of lenticular (S0) galaxies increasing by a factor of $2$-$3$ in such regions between $z\simeq0.5$ and the present day at the expenses of spirals \citep{Dressler1997,Fasano2000}. 
This suggests that S0 galaxies could be formed from spirals with quenched star formation and faded spiral arms, as a result of cluster-related physical processes, such as ram-pressure stripping, tidal interactions and starvation, that stripped the gas envelope surrounding the spiral galaxy or suppressed its gas supply from the Intergalactic Medium (IGM) (e.g. \citealt{Gunn1972,Larson1980,Bekki2009,Bekki2011,Merluzzi2016}).

However, all these physical processes require a high density environment in order to be most effective. 
The observations of S0s in low-density environments, where the interactions between galaxies are limited or non-existent, require alternative formation pathways. Mergers and accretion events have been shown to be able to produce S0 galaxies (e.g. \citealt{Bekki1998,Bournaud2005,Naab2014,Tapia2017,ElicheMoral2018}). Specifically, \citet{Bournaud2005} showed that minor mergers with mass-ratio 1:4.5-1:10 can produce disky ETGs, resembling the kinematics and morphology of S0s, while major mergers with mass-ratio 1:1-1:3 are likely to produce elliptical (E) galaxies with hotter kinematics. However, \citet{Tapia2017} and \citet{ElicheMoral2018} have shown that a few $\mathrm{Gyrs}$ after a major merger, many galaxy remnants displayed relaxed morphology and properties typical of S0s, suggesting that major mergers are also a possible S0 formation pathway.

Alternatively, a secular evolution scenario has been proposed for the formation of S0 galaxies, where the spiral galaxy evolves mostly passively, consuming its own in-situ gas through star formation, and slowly fades due to internal processes and instabilities. This scenario is expected to produce more rotationally-supported S0s, as they preserve a higher degree of rotation of their progenitor spiral. Internal processes, such as the collapse of a cold disk that becomes gravitationally unstable at high redshifts, feedback events from active galactic nuclei (AGN) and quenching of star formation due to the stabilization of the gas disk (i.e. morphological quenching), have been proposed as possible formation pathways for S0 galaxies \citep{Martig2009,Saha2018,Mishra2018}.
\citet{Bellstedt2017} and \citet{Rizzo2018} have found S0 galaxies displaying high values of stellar angular momentum, which are more consistent with those of disk galaxies rather than with those of merger remnants. 

Finally, an alternative scenario involves the build-up of a disk component around a massive compact quiescent galaxy at high redshift \citep{Damjanov2014} from the accretion of a gas-rich dwarf galaxy \citep{Diaz2018}.

Many physical processes have been thus shown to be able to produce S0 galaxies, therefore a wide range of stellar kinematic and population properties should be expected in S0s formed through these different formation pathways, as well as in the different environments.

Early results from the ATLAS$^{\mathrm{3D}}$ survey \citep{Cappellari2011_ATLAS3D} classified a sample of $260$ nearby ETGs (i.e. S0s and Es) into the two families of fast and slow rotators, based on the stellar kinematics of the central $1\, R_{\mathrm{e}}$ \citep{Emsellem2011}. 
The fast rotators typically showed regular rotation patterns with aligned kinematic and photometric position angles, while the slow rotators showed more complex velocity fields with misaligned kinematic and photometric position angles and were typically found among the most massive galaxies and in higher density environments. These results suggested a wide range of different processes for the formation of the fast/slowly rotating ETGs (e.g. mergers, interactions, gas stripping and secular evolution; \citealt{Krajnovic2011}).     

\citet{Coccato2020} studied the stellar kinematics out to $\sim1.5$-$2\, R_{\mathrm{e}}$ for a sample of $21$ S0 galaxies located in field and cluster environments. Their results showed that cluster S0s were more rotationally supported (i.e. high $V_{\mathrm{rot}}/\sigma$) and consistent with a faded spiral scenario, while field S0s were more pressure supported (i.e. low $V_{\mathrm{rot}}/\sigma$) and consistent with a formation through minor mergers. They also performed a stellar population analysis of such S0s, but they did not find any clear environmental dependence as for the stellar kinematics. However, the subsequent work by \citet{Johnston2020}, that specifically focused on the stellar population analysis of $8$ of the S0s from \citet{Coccato2020}, found that the field S0s had a younger stellar population than the cluster S0s, therefore suggesting a minor merger formation pathway. 
Similar conclusions to \citet{Coccato2020} were reached by \citet{Deeley2020}, who studied the stellar and gas kinematics of a sample of $219$ S0 galaxies from the SAMI survey \citep{Croom2012}. \citet{Deeley2020} found that pressure supported S0s were located in low-density environments (i.e. field and small groups) and had misaligned stellar and gas kinematics, consistent with a minor merger formation scenario. On the other hand, rotationally supported S0s were located in high-density environments and had kinematic properties more similar to spiral galaxies, therefore suggesting a faded spiral formation scenario.
\citet{Tous2020} have identified two sub-populations of S0s, one being characterized by S0s with star formation rates similar to those of late-type spirals and the other being characterized by quiescent S0s. They find that the sub-population of active S0s is slightly less massive with a younger and more metal-poor stellar population than the quiescent S0s, and is predominantly located in low-density regions \citep{Tous2020}, suggesting that the environment has likely played some role in the formation of S0s.  

\citet{McKelvie2018} suggested that the stellar mass of the galaxies, rather than the environment, is the main discriminator in the formation of S0s from the stellar population analysis of a sample of S0 galaxies in the MaNGA survey \citep{Bundy2015}. Specifically, \citet{McKelvie2018} found that more massive S0s had an older and more metal-rich stellar population in the bulge than in the disk, while low-mass S0s had a younger stellar population in the bulge than in the disk, suggesting an inside-out quenching scenario for the more massive S0s and an outside-in quenching scenario for the low-mass S0s.
A more recent stellar population analysis of a sample of S0 galaxies in the MaNGA survey has also revealed a bimodal S0 population, with massive S0s (i.e. $\mathrm{M}_{*} > 3 \times 10^{10}\, \mathrm{M_{\odot}}$) having mostly flat metallicity gradients and low-mass S0s having steeper metallicity gradients, while the age gradients show the opposite trend \citep{Dominguez2020}. These results suggest that gas-rich mergers are more important in the formation of massive S0s. 

The SAGES Legacy Unifying Globulars and GalaxieS (SLUGGS; \citealt{Brodie2014}) survey was able to obtain extended stellar kinematic measurements out to $\sim2$-$4\, R_{\mathrm{e}}$ for a sample of $25$ nearby ETGs (including S0s), as well as radial velocities for a large number of globular clusters (GCs; \citealt{Forbes2017_GC}).
From the stellar kinematics, \citet{Arnold2014} and \citet{Foster2016} found that the kinematic behaviour of the central $1\, R_{\mathrm{e}}$ of ETGs can be quite different from the kinematic behaviour of the outer galaxy regions, with centrally fast rotators often showing decreasing or rapidly increasing rotation at larger radii. 
Specifically, \citet{Arnold2014} found that many of their ETGs showed evidence of a fast-rotating stellar disk embedded in a more slowly rotating spheroidal component, which is consistent with the model of a two-phase formation scenario (e.g. \citealt{Oser2010,Zolotov2015,Rodriguez2016}). According to the two-phase model, massive ETGs formed at early times (i.e. $z\geq2$) with a rapid collapse or wet merger event during which they formed most of their stellar mass, in a compact object know as 'red-nugget' \citep{Damjanov2014,Zolotov2015}, i.e. in-situ formation. At later times (i.e. $z\simeq0$), they mostly grew in size through the accretion of low-mass dwarf galaxies deposited onto their outskirts (ex-situ formation). These low-mass mergers are responsible for producing the decreasing rotation in the outer regions of ETGs. 
However, stellar kinematic studies of different samples of ETGs, extending beyond $2\, R_{\mathrm{e}}$, found different results from \citet{Arnold2014}, with centrally fast rotator galaxies remaining fast-rotating and centrally slow rotator galaxies remaining slow-rotating beyond $1\, R_{\mathrm{e}}$ \citep{Raskutti2014,Boardman2017}. The differences between these works could be due to differences in the sample selection criteria. For instance, the sample of ETGs from \citet{Boardman2017} contains ETGs with lower stellar masses than the SLUGGS sample, therefore the transition radius (which anti-correlates with the galaxy stellar mass; \citealt{Pulsoni2020}) in the stellar kinematics of the galaxies from \citet{Boardman2017} could simply be located at larger radii as compared to the SLUGGS galaxies. Nevertheless, all these results suggest the importance of studying the kinematic properties of larger samples of ETGs out to larger radii. This will not only help us to constrain the formation histories of ETGs, but also to understand the kinematic behaviour of the galaxy outskirts with respect to that of the galaxy central regions.

The stellar kinematic results obtained by \citet{Arnold2014} and \citet{Foster2016}, as part of SLUGGS, have shown the importance of studying the kinematic properties of the outer regions of ETGs, i.e. beyond $\sim2$-$3\, R_{\mathrm{e}}$.
In fact, the galaxy outskirts are characterized by longer dynamical timescales, therefore they better preserve the imprints of the galaxy formation processes in the stellar dynamic and population content.
However, the galaxy outskirts are difficult and time-consuming to observe due to their lower surface brightness as compared to the inner regions.

One way to overcome this issue and probe the outskirts of the galaxies is to use discrete kinematic tracers, such as GCs and planetary nebulae (PNe). 
GCs are observed in all galaxy types and numerous in the most massive ETGs. They are typically found to display a colour bimodality that reflects the two distinct sub-populations of the red, metal-rich and blue, metal-poor GCs, which are expected to trace the kinematic properties of the host ETGs differently \citep{Forbes1997,Strader2005,Brodie2006}. Specifically, the red, metal-rich GCs are expected to have formed largely in-situ together with the bulk of the host galaxy stars, while the blue, metal-poor GCs are expected to have formed mostly at early times of the galaxy formation process or subsequently accreted onto the host galaxy from tidally stripped dwarfs during the late evolutionary phase of the galaxy. Therefore, the red, metal-rich GCs should trace the kinematic properties of the stellar population of the bulge and spheroid in massive ETGs, while the blue, metal-poor GCs have typically a more extended spatial distribution than the red GCs and should trace the kinematic properties of the stellar population of the galaxy halo \citep{Forbes1997,Forbes2018}.
PNe are a late evolutionary stage of the stars with stellar masses $1\, \mathrm{M_{\odot}} < \mathrm{M}^{*}<8\, \mathrm{M_{\odot}}$ and are expected to trace the underlying stellar population in ETGs out to large radii (e.g. \citealt{Romanowsky2006,Buzzoni2006}).

Since GCs are bright star clusters and PNe are strong [OIII] emitters, they can be observed out to larger radii (i.e. $\sim8$-$10\, R_{\mathrm{e}}$) in ETGs, where the stellar component becomes too faint for spectroscopy studies. Some previous works have found that the PNe and GCs are generally good tracers of the kinematics of the underlying stellar population and stellar surface brightness profile for many ETGs (e.g. \citealt{Coccato2009,Pota2013,Pota2015}). Others adopted a novel approach to study the formation histories of S0 galaxies that combined the kinematics of the PNe and GCs with the photometric decomposition (i.e. bulge$+$disk) of the galaxy in order to associate the PNe and GCs to either the bulge or disk component using a maximum likelihood method (e.g. \citealt{Cortesi2011,Forbes2012,Cortesi2013,Cortesi2016,Zanatta2018}).

Recent simulations extending out to large galactocentric radii, i.e. $5\, R_{\mathrm{e}}$ \citep{Schulze2020}, have shown that different $V_{\mathrm{rot}}/\sigma$ profiles can be the result of merger events that occurred at different redshifts. These merger events are expected to influence not only the kinematics of the stars, but also that of both GC sub-populations depending on the mass-ratio and orbital configuration of the mergers \citep{Bekki2005,Cavanagh2020}.
Therefore, in this work, we study the kinematic profiles for a sample of $8$ selected S0 and E/S0 galaxies in the SLUGGS survey that are located mainly in low-density environments by combining the stellar with the GC and PNe kinematics in order to reach beyond $\sim2$-$3\, R_{\mathrm{e}}$, similarly to what has already been done in \citet{Dolfi2020} for the nearest S0 galaxy, NGC 3115.
If the PNe and GCs are tracing the kinematics of the underlying stellar population in the galaxies, then we can use them as proxies for the stars to study the galaxy outskirts beyond the radius probed by the stars. 
From the comparison of the overall $V_{\mathrm{rot}}/\sigma$ kinematic profiles of the galaxies out to large radii, i.e. $\sim5\, R_{\mathrm{e}}$, with the simulations, we can constrain the merger history of our galaxies with the most likely epoch when the merger event occurred that shaped the characteristic $V_{\mathrm{rot}}/\sigma$ profiles of our galaxies

The structure of the paper is as follows. In Sec. \ref{sec:galaxy_sample}, we describe the selected sample of ETGs with their stellar, GC and PNe kinematic data. In Sec. \ref{sec:results}, we present the kinematic results in the form of 2D kinematic maps and 1D kinematic profiles for all the kinematic tracers of each individual galaxy. In Sec. \ref{sec:discussion}, we discuss the likely formation histories of our sample of S0 galaxies from the comparison with simulations of galaxy formation. Appendix A gives a more detailed description of the kinematic properties of each one of our individual selected ETGs.

\section{The galaxy sample}
\label{sec:galaxy_sample}

    We select a sample of $8$ ETGs out of $25$ from the SLUGGS survey \citep{Brodie2014}), based on the availability of a sufficient number of both GCs and PNe with measured radial velocities in order to study their kinematic properties. 
    The GC and PNe spectroscopic catalogues come from the SLUGGS \citep{Forbes2017_GC} and the extended Planetary Nebula Spectrograph (ePN.S; \citealt{Pulsoni2018}) surveys, respectively. Photometry measurements are available for the GC systems of each one of our selected galaxies and they have been presented in \citet{Pota2013,Pota2015}.
    
\begin{figure*}\centering
        \includegraphics[width=1.0\textwidth]{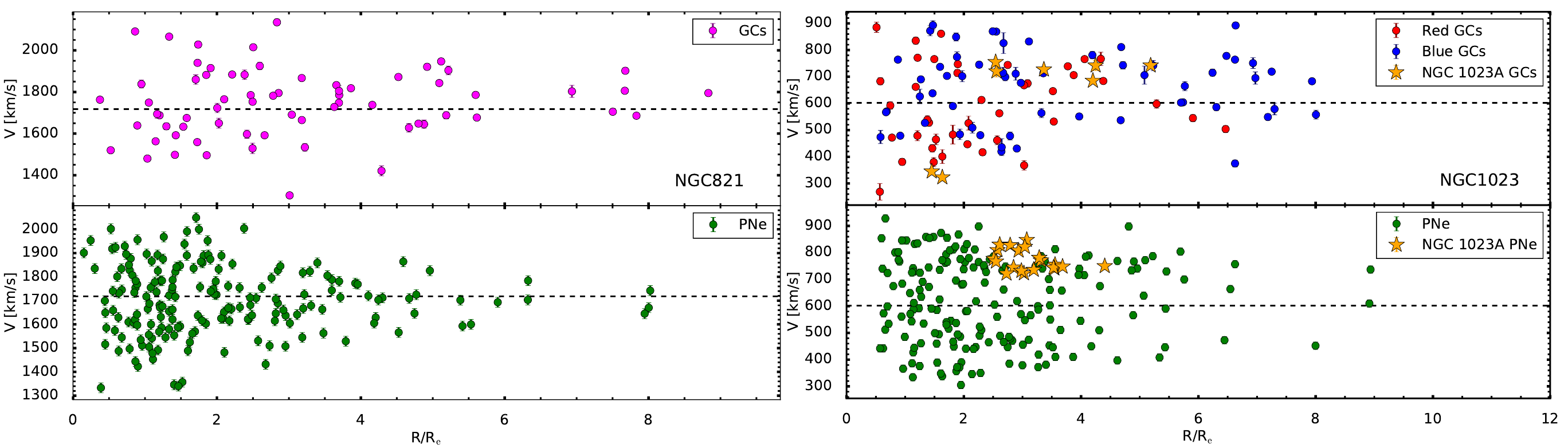}
        \includegraphics[width=1.0\textwidth]{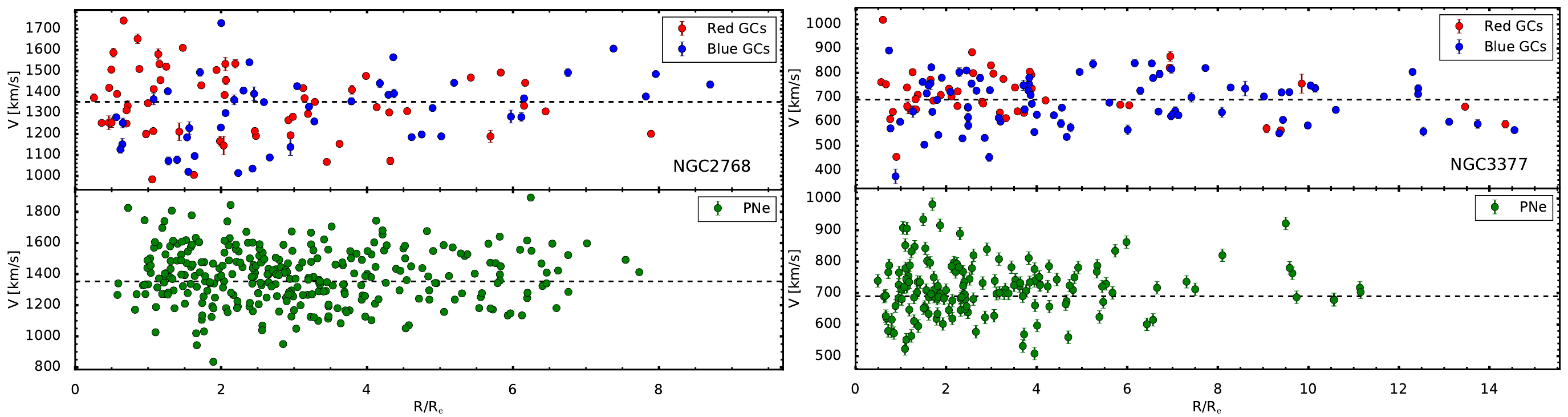}
        \includegraphics[width=1.0\textwidth]{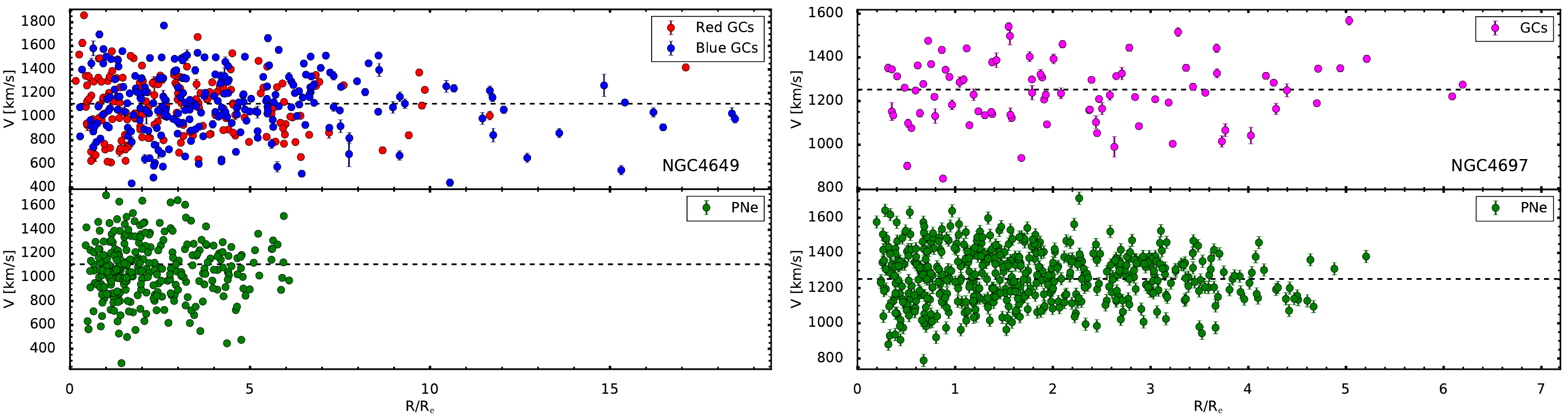}
        \includegraphics[width=1.0\textwidth]{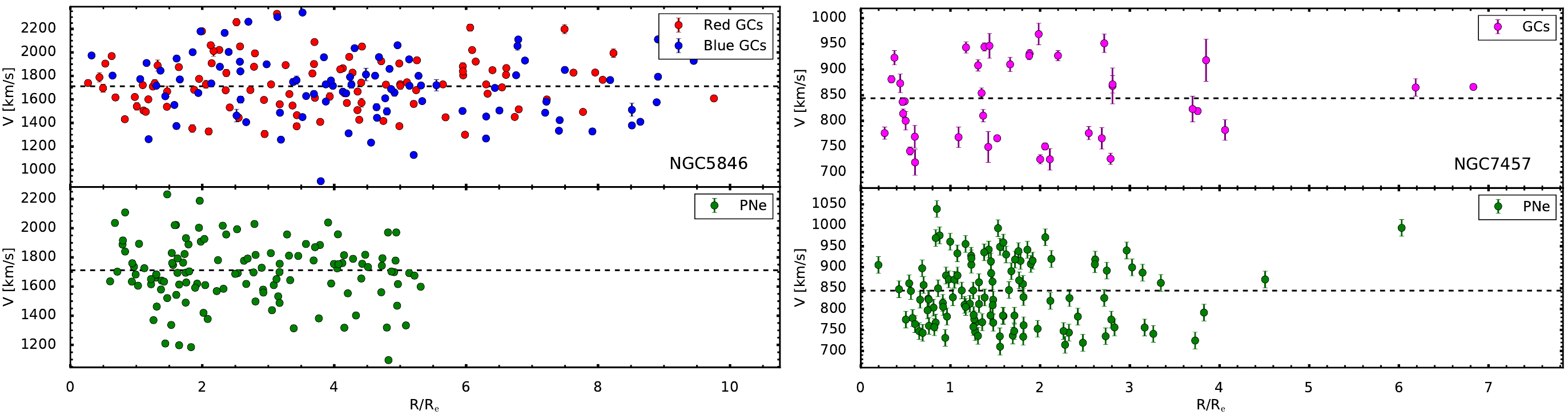}
        \caption{Phase space diagrams showing the measured radial velocities of the GC and PNe tracers of each galaxy as a function of the galactocentric radius, $R$, divided by the corresponding effective radius of the galaxy, $R_{\mathrm{e}}$ (see Table \ref{tab:sluggs_galaxies_summary}). The red, metal-rich and blue, metal-poor GC sub-populations are represented as red and blue dots, respectively, for all those galaxies for which we identify a colour bimodality distribution, otherwise the GCs are shown as magenta dots. The PNe are shown as green dots. For NGC 1023, we show the $8$ GCs and $20$ PNe (orange stars) associated to the companion galaxy, NGC 1023A, that we remove from the final catalogues as described in Sec. A1.1.}
        \label{fig:phase_space_diagrams}.
    \end{figure*}

    Table \ref{tab:sluggs_galaxies_summary} shows the list of the $8$ selected ETGs with a summary of some of their main characteristics. We also indicate the total number of spectroscopically confirmed GCs and PNe in each galaxy with the adopted colour-split to separate between the red, metal-rich and blue, metal-poor GC sub-populations, where applicable. The colour-splits to separate between the red and blue GCs were selected after visually inspecting the colour-magnitude diagrams, in the corresponding magnitude bands, of the photometric GC catalogues. We note that the GC colour-splits adopted in this work are consistent with those from \citet{Pota2013,Pota2015} and \citet{Cortesi2016}. NGC 3115 is also included in the list, but its study has been carried out in a separate paper \citep{Dolfi2020}. All selected ETGs have a total number of spectroscopically confirmed GCs or PNe greater than $100$ and Fig. \ref{fig:phase_space_diagrams} shows the measured radial velocities of the GC and PNe systems as a function of the galactocentric radius, $R/R_{\mathrm{e}}$, for each galaxy.
    
    The stellar kinematics for all our galaxies come from ATLAS$^{\mathrm{3D}}$ \citep{Cappellari2011_ATLAS3D}, MUSE \citep{Guerou2016} in the case of NGC 3115, and SLUGGS \citep{Arnold2014,Foster2016} surveys. The former provides spatially resolved 2D maps of the inner $1\, R_{\mathrm{e}}$ of ETGs, while the latter provides discrete stellar kinematic measurements around the galaxies that are obtained using the Stellar Kinematics from Multiple Slits (SKiMS) technique \citep{Proctor2009,Foster2009,Foster2011} and extend further out to $\sim2$-$4\, R_{\mathrm{e}}$. 
    We note here that all our selected ETGs, except for NGC 3115 and NGC 5846, show regular rotation in the 2D kinematic maps of their velocity fields within the central $\sim1\, R_{\mathrm{e}}$ and are, thus, classified as fast rotators \citep{Emsellem2004}. NGC 5846 is the only massive ETG in our selected sample classified as slow-rotator within the central $\sim1\, R_{\mathrm{e}}$ from ATLAS$^{\mathrm{3D}}$ \citep{Emsellem2004}. Finally, NGC 3115 was not observed as part of the ATLAS$^{\mathrm{3D}}$ \citep{Cappellari2011_ATLAS3D}, however the stellar kinematic results from MUSE, extending out to $\sim3\, R_{\mathrm{e}}$, show that NGC 3115 is also a fast rotator S0 galaxy \citep{Guerou2016}.
    
    We note that $5$ out of $9$ of our selected ETGs in Table \ref{tab:sluggs_galaxies_summary} are found in groups, while $3$ are found in the field and only one galaxy, NGC 4649, is in a cluster environment. 
    Additionally, we note that only three of the ETGs in Table \ref{tab:sluggs_galaxies_summary} are classified as pure S0s (i.e. NGC 1023, NGC 3115 and NGC 7457), while the remaining six ETGs are classified as E/S0s or Es. 
    However, even though these six ETGs are not classified as pure S0s based on their morphology (see Table \ref{tab:sluggs_galaxies_summary}), they have been all found to show evidence of a disk-like component from early kinematic or photometric studies (see Appendix A also for more specific details about the main properties of each individual galaxy in Table \ref{tab:sluggs_galaxies_summary} with the description of their GC and PNe systems).
    
    \begin{table*}
    \caption{Early-type galaxies selected for study in this work. The main properties of each galaxy, such as morphology, environment, distance, effective radius ($R_{\mathrm{e}}$), photometric position angle ($\mathrm{PA}_{\mathrm{phot}}$ measured from North towards East), photometric axial-ratio ($\mathrm{q}_{\mathrm{phot}}=1-\epsilon_{\mathrm{phot}}$), systemic velocity ($V_{\mathrm{sys}}$) and stellar velocity dispersion within $1\, \mathrm{kpc}$ ($\sigma_{1\, \mathrm{kpc}}$), are taken from \citep{Brodie2014}. The stellar masses are taken from \citet{Forbes2017} and were calculated from the total $3.6\, \mathrm{\mu m}$ magnitudes assuming a kroupa initial mass function \citep{Kroupa2002}. Additionally, we quote the total number of spectroscopically confirmed and unique GCs and PNe from the SLUGGS \citep{Forbes2017_GC} and ePN.S \citep{Pulsoni2018} surveys, respectively, for each galaxy. For the GCs, we specify the adopted colour-split to separate between the red, metal-rich and blue, metal-poor GC sub-populations for all the galaxies where a clear colour bi-modality is observed. For the PNe, we specify the mean error ($\Delta V$) associated with the PNe radial velocity measurements as given in \citet{Pulsoni2018}. The values for NGC3115 are those published in \citet{Dolfi2020}.}
    \centering
    \resizebox{18cm}{!}{
    \begin{tabular}{c|c|c|c|c|c|c|c|c|c|c|c|c|c|c|c|c|} \hline \hline
    Galaxy name & Morph. & Environm. & Dist. & $\log M^{*}$ & $R_{\mathrm{e}}$ & $\mathrm{PA}_{\mathrm{phot}}$ & $\mathrm{q}_{\mathrm{phot}}$ & $V_{\mathrm{sys}}$ & $\sigma_{\mathrm{1\, kpc}}$ & $\mathrm{N_{GC,tot}}$ & GC split & $\mathrm{N_{GC,red}}$ & $\mathrm{N_{GC,blue}}$ & $\mathrm{N_{PNe}}$ & $\Delta V$ \\ 
     & & & $\mathrm{(Mpc)}$ & $\mathrm{(M_{\odot})}$ & $\mathrm{(arcsec)}$ & (degree) & & $\mathrm{(kms^{-1})}$ & $\mathrm{(kms^{-1})}$ & & $\mathrm{(mag)}$ & & & & $\mathrm{\mathbf{(kms^{-1})}}$ \\ \hline \hline
    NGC 821  & E6/S0      & Field   & $23.4$ & $11.00$ & $40$ & $31.2$ & $0.65$ & $1718$ & $193$ & $68$  & no  & - & - & $186$ & $21$ \\ 
    NGC 1023 & SB0      & Group   & $11.1$ & $10.99$ & $48$ & $83.3$ & $0.37$ & $602$ & $183$ & $102$ & $(g-z)=1.13$  & $41$ & $61$ & $183$ & $14$ \\ 
    NGC 2768 & E6/S0   & Group   & $21.8$ & $11.21$ & $63$ & $91.6$ & $0.43$ & $1353$ & $206$ & $104$ & $(Rc-z)=0.57$  & $58$ & $46$ & $315$ & $20$ \\ 
    NGC 3115 & S0     & Field   & $9.4$  & $10.93$ & $35$ & $40$ & $0.34$ & $663$  & $248$ & $191$ & $(g-i)=0.93$  & $100$ & $91$ & $189$ & $20$ \\ 
    NGC 3377 & E5-6    & Group   & $10.9$ & $10.50$ & $36$ & $46.3$ & $0.67$ & $690$  & $135$ & $126$ & $(g-i)=0.925$ & $46$ & $80$ & $152$ & $20$ \\ 
    NGC 4649 & E2/S0   & Cluster & $16.5$ & $11.60$ & $66$ & $91.3$ & $0.84$ & $1110$ & $308$ & $430$ & $(g-z)=1.2$ & $179$ & $251$ & $298$ & $20$ \\ 
    NGC 4697 & E6      & Group   & $12.5$ & $11.15$ & $62$ & $67.2$ & $0.68$ & $1252$ & $180$ & $86$  & no & - & - & $531$ & $35$ \\ 
    NGC 5846 & E0-1/S0 & Group   & $24.2$ & $11.46$ & $59$ & $53.3$ & $0.92$ & $1712$ & $231$ & $192$ & $(g-i)=0.95$ & $102$ & $90$ & $124$ & $21$ \\ 
    NGC 7457 & S0      & Field   & $12.9$ & $10.13$ & $36$ & $124.8$ & $0.53$ & $844$  & $74$  & $40$  & no  & - & - & $113$ & $20$ \\ 
    \hline \hline
    \end{tabular}}
    \label{tab:sluggs_galaxies_summary}
    \end{table*}

\section{Results}
\label{sec:results}

    \subsection{2D Kinematic maps and 1D kinematic profiles}
    \label{sec:kinematic_results}
    In this section, we produce the 2D kinematic maps of the GC and PNe tracers, as well as stars, for each galaxy in Table \ref{tab:sluggs_galaxies_summary}. Similarly to \citet{Dolfi2020} and in previous literature works (e.g. \citealt{Proctor2009,Foster2013,Foster2016,Bellstedt2017}), we use the kriging spatial interpolation technique \citep{Pastorello2014}, which is an inverse-noise-weighted method, to generate the continuous 2D velocity and velocity dispersion maps shown in Fig. \ref{fig:NGC1023_2d_kinematic_maps} and A1-A8. 
    We fold the PNe kinematic catalogue only for NGC 4649 prior to producing the 2D maps in order to have full position angle ($\mathrm{PA}$) coverage around the galaxy, as described in Sec. A2.2.
    
    To estimate the velocity dispersion of the GC and PNe tracers of each galaxy, we calculate the standard deviation of the tracer velocities from the mean in spatial bins centered at the position of each tracer and containing a chosen number of nearest-neighbour objects. 
    In general, the higher is the number of nearest-neighbour objects, the more accurate the estimate of the velocity dispersion of the GCs and PNe would be in each spatial bin. However, we need to be careful when choosing the number of nearest-neighbour objects so that we do not average together GC and PNe tracers that are located in significantly different spatial regions around the galaxy.
    For this reason, we use $5$ nearest-neighbours for the GCs and $10$ nearest-neighbours for the PNe, since the latter have better sampled velocity fields, i.e. higher number of objects in the catalogues. We use $10$ nearest-neighbour for the GCs for only NGC 4649 as this galaxy has the largest spectroscopic GC catalogue (see Table \ref{tab:sluggs_galaxies_summary}). This procedure is the same as the one adopted in \citet{Dolfi2020} and it is not necessary for the stellar kinematics for which we have the integrated line-of-sight velocity dispersion.
    
    We calculate the 1D kinematic profiles of all our kinematic datasets, i.e. ATLAS$^{\mathrm{3D}}$ and SLUGGS stars, GCs and PNe (see Sec. \ref{sec:galaxy_sample}) using the \texttt{kinemetry}\footnote{\url{http://davor.krajnovic.org/idl/}} method \citep{Krajnovic2006}, which was initially used with 2D spatially resolved stellar kinematic maps of galaxies to recover the best-fitting moments of the Line-Of-Sight-Velocity-Distribution (LOSVD) as a function of the galactocentric radius (e.g. \citealt{Krajnovic2011}). Here, we adopt a similar procedure as in \citet{Proctor2009,Foster2011,Foster2016,Bellstedt2017,Dolfi2020} to recover the LOSVD moments for our non-homogeneous and sparse kinematic datasets, i.e. SLUGGS stars, GCs and PNe. 
    As for the 2D kinematic maps, we use the folded PNe catalogue of NGC 4649 (see Sec. A2.2) to derive its 1D kinematic profiles.
    Within \texttt{kinemetry}, the computation of the best-fitting moments of the LOSVD is performed at the different galactocentric radii by least-squares minimization of the cosine model function, described in equations $3$ and $4$ of \citet{Dolfi2020}, fitted to the observed kinematic data binned in concentric elliptical annuli. For a more detailed description of how we apply \texttt{kinemetry} to our kinematic datasets see section 6.2 in \citet{Dolfi2020}, while for the equations describing in details the chi-squared minimization procedure within \texttt{kinemetry} for recovering the best-fitting LOSVD moments of our kinematic data see \citet{Foster2011,Foster2016}.
    
    As previously done in \citet{Dolfi2020} for NGC 3115, we focus on recovering the first and second order moments of the LOSVD (i.e. $V_{\mathrm{rot}}$ and $\sigma$, respectively) for all the kinematic datasets of each one of our galaxies in Table \ref{tab:sluggs_galaxies_summary} and we fit for both the $\mathrm{PA}_{\mathrm{kin}}$ and $\mathrm{q}_{\mathrm{kin}}$ parameters, where we find that these can be reliably constrained (i.e. they do not excessively vary as a function of the galactocentric radius within the different elliptical annuli). If this is not the case, then we fix them to the corresponding photometric values, i.e. $\mathrm{PA}_{\mathrm{kin}}=\mathrm{PA}_{\mathrm{phot}}$ and $\mathrm{q}_{\mathrm{kin}}=\mathrm{q}_{\mathrm{phot}}$ (see Table \ref{tab:sluggs_galaxies_summary}), of the galaxies for more stable results during the fit. 
    We find this to be mostly the case for $\mathrm{q_{\mathrm{kin}}}$, which is generally hard to constrain even in the presence of well-sampled datasets \citep{Foster2016}, and has been, therefore, usually fixed to the photometric value in many previous works (e.g. \citealt{Foster2011,Foster2013,Foster2016,Bellstedt2017,Dolfi2020}). 
    Overall, we find that whether we fix or fit for $\mathrm{q}_{\mathrm{kin}}$, the shape of the derived rotation velocity and velocity dispersion profiles as a function of the galactocentric radius does not significantly change.
    
    The 1D kinematic profiles of the distinct tracers (i.e. GCs, PNe and stars) of each galaxy in Table \ref{tab:sluggs_galaxies_summary} are shown in Fig. \ref{fig:1D_kinematic_profiles}.
    Each sub-figure of Fig. \ref{fig:1D_kinematic_profiles} represents one individual galaxy, with the derived $\mathrm{PA}_{\mathrm{kin}}$ and $\mathrm{q}_{\mathrm{kin}}$ (where fitted), rotation velocity, $V_{\mathrm{rot}}$, velocity dispersion, $\sigma$, and rotation velocity to velocity dispersion, $V_{\mathrm{rot}}/\sigma$, profiles shown from the top to the bottom panel, respectively. 
    The $\mathrm{PA}_{\mathrm{kin}}$ is always measured from North towards East for consistency with the 2D kinematic maps in Fig. \ref{fig:NGC1023_2d_kinematic_maps} and A1-A8 and with the $\mathrm{PA}_{\mathrm{phot}}$ of the galaxy shown in Table \ref{tab:sluggs_galaxies_summary}. 
    The $1$-sigma errors of the $\mathrm{PA}_{\mathrm{kin}}$ and $\mathrm{q}_{\mathrm{kin}}$ profiles are calculated from running \texttt{kinemetry} on $500$ bootstrapped samples obtained from sampling with replacement the original kinematic datasets. On the other hand, the $1$-sigma errors of the rotation velocity and velocity dispersion profiles are the standard errors of the mean, calculated from the standard deviation of the velocity and velocity dispersion measurements of the data points in each elliptical annuli\footnote{These are the same concentric elliptical annuli used in the \texttt{kinemetry} algorithm to recover the 1D kinematic profiles at the different galactocentric radii.}, divided by the square root of the number of objects in each bin.
    We also include the 1D kinematic profiles of NGC 3115, where the $1$-sigma errors of the $V_{\mathrm{rot}}$ and $\sigma$ profiles are also the standard errors of the mean for consistency with the other galaxies, for the comparison. However, we refer the reader to \citet{Dolfi2020} for the detailed description of the main kinematic features of NGC 3115.
    
    \subsection{The kinematic behaviour of the galaxies}
    \label{sec:kinematic_behaviour}
\begin{figure*}
    \centering
        \includegraphics[width=1.0\textwidth]{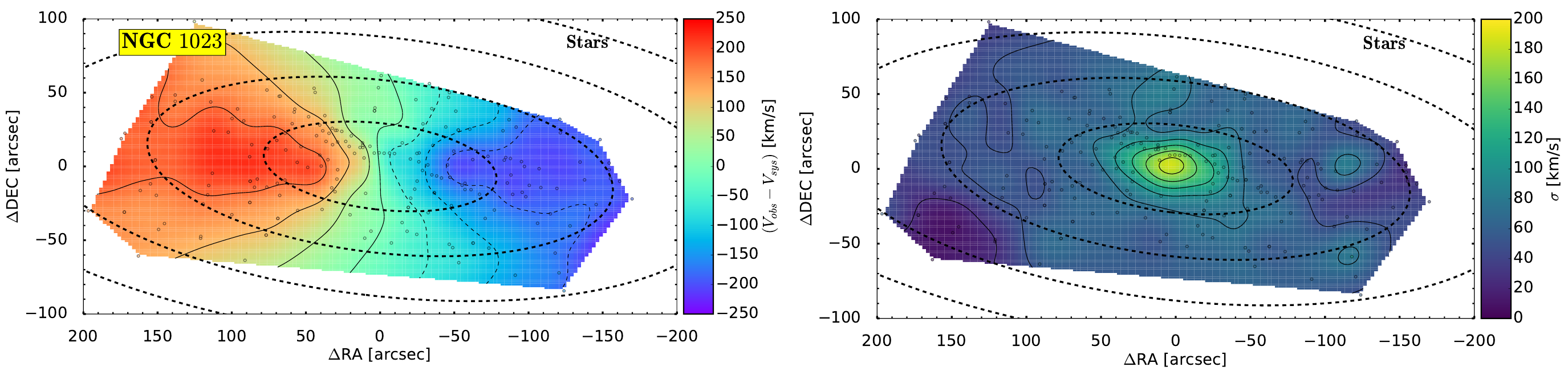}
        \includegraphics[width=1.0\textwidth]{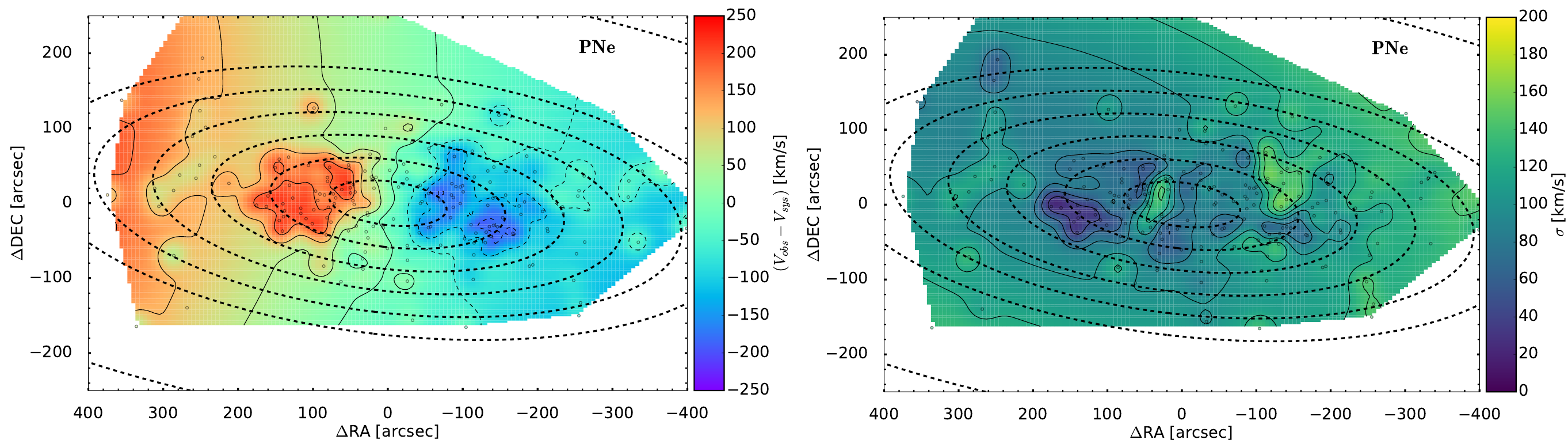}
        \includegraphics[width=1.0\textwidth]{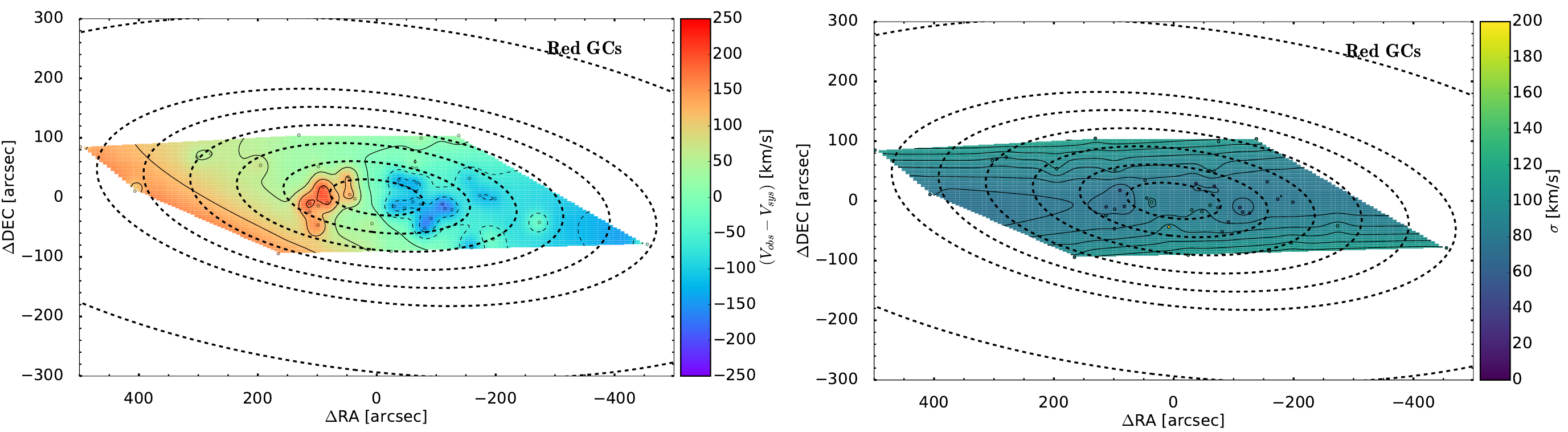}
        \includegraphics[width=1.0\textwidth]{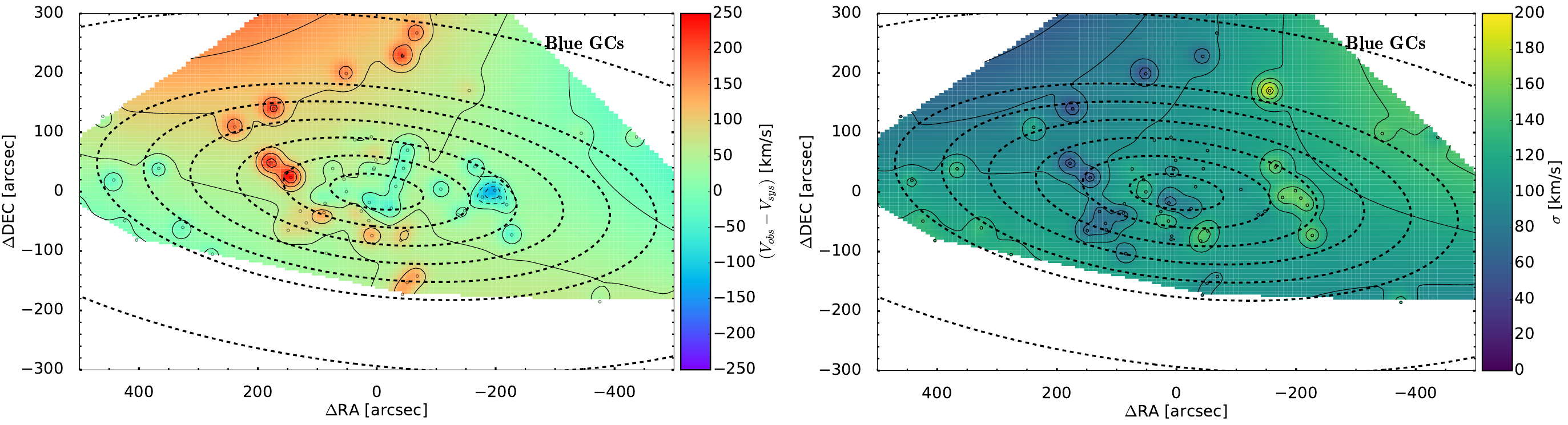}
        \caption{From top to bottom, 2D velocity (left-hand side) and velocity dispersion (right-hand side) maps of the SLUGGS stars, PNe, red and blue GCs for NGC 1023, shown here as an example (the 2D kinematic maps of the remaining eight galaxies in Table \ref{tab:sluggs_galaxies_summary} are shown in the Figures A1-A8). North is up, East is left. The small open circles indicate the positions of the tracers, the black solid lines indicate the iso-velocity contours and the black dashed lines represent the $1$-$6\, R_{\mathrm{e}}$ and $10\, R_{\mathrm{e}}$ photometric ellipses, which are inclined by the $\mathrm{PA}_{\mathrm{phot}}$ of the galaxy. The $\mathrm{PA}_{\mathrm{phot}}$ is measured from North towards East. The 2D maps are normalized with respect to the velocity scale bar on the right of each map.} 
    \label{fig:NGC1023_2d_kinematic_maps}
    \end{figure*}

    From the 2D kinematic maps in Fig. \ref{fig:NGC1023_2d_kinematic_maps} and A1-A8, and from the 1D kinematic profiles in Fig. \ref{fig:1D_kinematic_profiles}, we find that six galaxies (i.e. NGC 1023, NGC 2768, NGC 3115, NGC 3377, NGC 4697 and NGC 7457) show good kinematic alignment of the GC sub-populations and PNe with respect to the underlying stars (hereafter, \textit{aligned} galaxies). Specifically, four of the galaxies (i.e. NGC 1023, NGC 2768, NGC 3115, NGC 4697 and NGC 7457) have the $\mathrm{PA}_{\mathrm{kin}}$ of both the GCs and PNe that are closely aligned with the $\mathrm{PA}_{\mathrm{kin}}$ of the underlying stars (that is, in turn, closely aligned with the $\mathrm{PA}_{\mathrm{phot}}$ of the galaxy given in Table \ref{tab:sluggs_galaxies_summary}), as they are consistent within the $1$-sigma errors. Additionally, the $\mathrm{PA}_{\mathrm{kin}}$ of both the GCs and PNe in these four galaxies do not show significant scatter around the $\mathrm{PA}_{\mathrm{kin}}$ of the underlying stars, but they all display rotation along the photometric major-axis of the galaxy, as it can be seen also from the 2D kinematic maps in Fig. \ref{fig:NGC1023_2d_kinematic_maps}, A1, A3, A4 (and in fig. 6 of \citet{Dolfi2020} for NGC 3115). 
    
    An exception is NGC 3377, whose blue GC sub-population does not show evidence of a significant rotation (see Fig. A2 and Fig. \ref{fig:1D_kinematic_profiles}). Additionally, the PNe of NGC 3377 show a possible twist in the $\mathrm{PA}_{\mathrm{kin}}$ between $\sim3$-$5\, R_{\mathrm{e}}$ and evidence of such a kinematic twist of the PNe was also previously seen by \citet{Coccato2009}. However, the PNe show rotation that is closely aligned to that of the underlying stars, as well as red GCs, within the inner $\sim3\, R_{\mathrm{e}}$. At larger radii, we see that the rotation of the PNe is still overall consistent with the photometric major-axis of the galaxy from the 2D kinematic maps. Therefore, we also include NGC 3377 in the group of the \textit{aligned} galaxies.
    
    On the other hand, the remaining three galaxies show kinematic misalignments of the GC sub-populations and PNe with respect to the underlying stars, i.e. NGC 821 and NGC 4649, or no net rotation, i.e. NGC 5846 (hereafter, \textit{mis-aligned} galaxies). Specifically, NGC 4649 shows the largest scatter in the $\mathrm{PA}_{\mathrm{kin}}$ of the PNe and blue GC sub-population with respect to the underlying stars, with clear kinematic twists and minor-axis rotation beyond $\sim2\, R_{\mathrm{e}}$. On the other hand, the red GCs of NGC 4649 have $\mathrm{PA}_{\mathrm{kin}}$ that is slightly offset from that of the stars, but it does not show significant variations as a function of the radius (Fig. \ref{fig:1D_kinematic_profiles} and A7).
    Similarly, in NGC 821, the PNe are rotating more closely aligned to the photometric minor-axis of the galaxy and the GCs are also rotating along a $\mathrm{PA}_{\mathrm{kin}}$ that is offset from the photometric major-axis of the galaxy (Fig. A5) and consistent with the large $1$-sigma errors in the 1D kinematic profiles in Fig. \ref{fig:1D_kinematic_profiles}.
    
    Finally, in NGC 5846, we are not able to reliably recover the $\mathrm{PA}_{\mathrm{kin}}$ of the stars, PNe and GC sub-populations. However, from the 2D kinematic maps in Fig. A8, we do not see evidence of any ordered rotation around the galaxy and therefore, of any possible kinematic alignment between the distinct tracers. 
    For this reason, we include these three galaxies (i.e. NGC 821, NGC 4649 and NGC 5846) in the group of the \textit{mis-aligned} galaxies.

    \begin{figure*}
        \centering
        \includegraphics[width=0.25\textwidth]{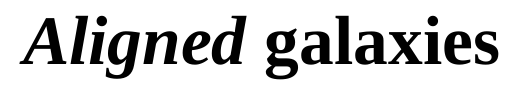}
        \includegraphics[width=1.10\textwidth]{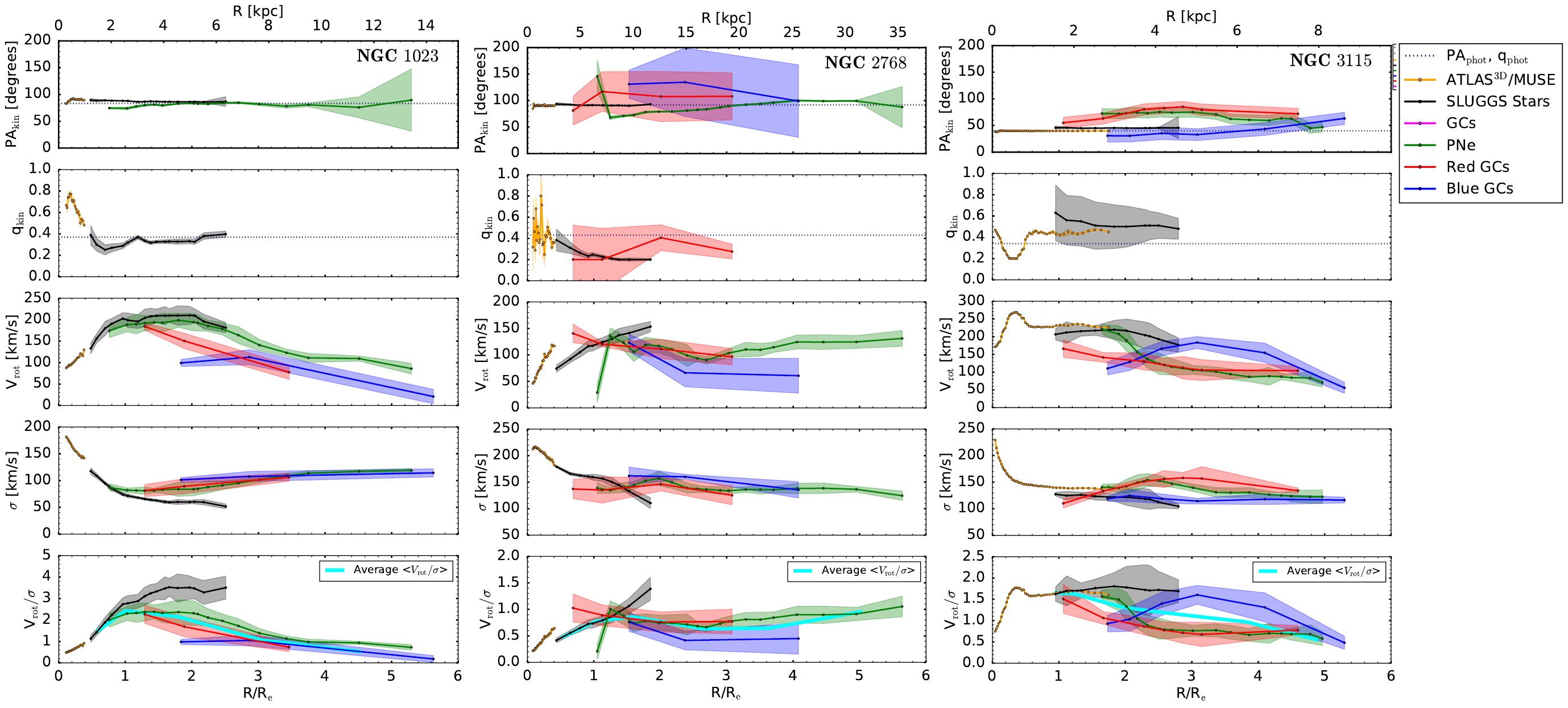}
        \includegraphics[width=1.0\textwidth]{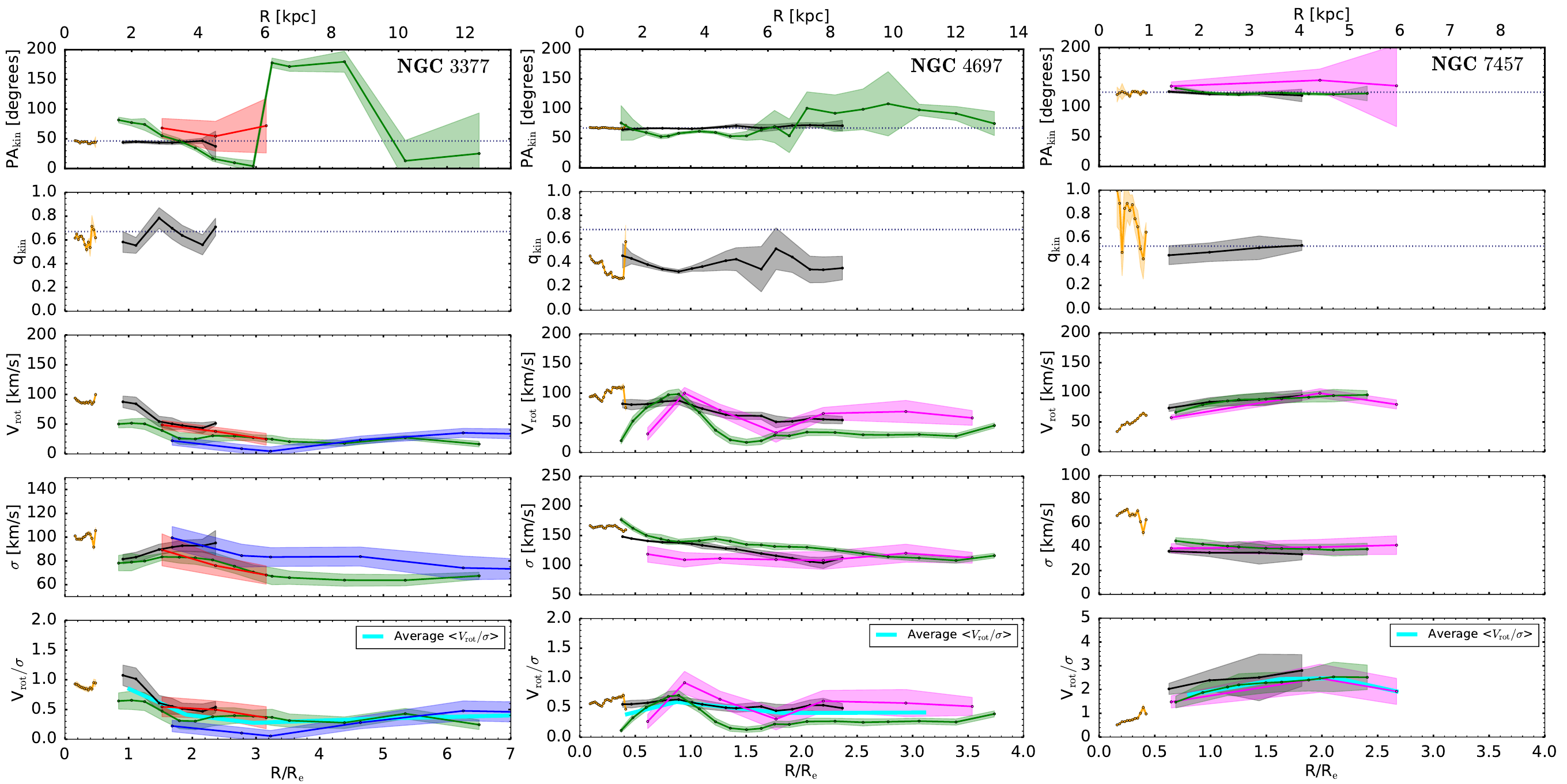}
        \caption{1D kinematic profiles for the ATLAS$^{\mathrm{3D}}$, or MUSE for NGC 3115 (orange line), and SLUGGS (black line) stars, PNe (green line), red and blue GCs (red and blue lines, respectively) and all GCs (magenta line) for those galaxies with no bimodal color distribution in the GC population (see Table \ref{tab:sluggs_galaxies_summary}). Each column represents one individual galaxy, which are being grouped based on whether they are \textit{aligned} or \textit{mis-aligned}. NGC 3115 is taken from \citet{Dolfi2020}. From top to bottom in each panel, we show the derived 1D profiles of the kinematic position angle, $\mathrm{PA}_{\mathrm{kin}}$, kinematic axial-ratio, $\mathrm{q}_{\mathrm{kin}}$, rotation velocity, $V_{\mathrm{rot}}$, velocity dispersion, $\sigma$, and rotation velocity to velocity dispersion, $V_{\mathrm{rot}}/\sigma$. For the \textit{aligned} galaxies, we also show the smooth $V_{\mathrm{rot}}/\sigma$ profile (cyan line), calculated by averaging the individual $V_{\mathrm{rot}}/\sigma$ profiles of the distinct kinematic tracers in the overlapping radii for each galaxy. For NGC 4649, we only fit for the $\mathrm{PA}_{\mathrm{kin}}$ and we keep $\mathrm{q}_{\mathrm{kin}}$ fixed to the photometric value, $\mathrm{q}_{\mathrm{phot}}$, of the galaxy (see Table \ref{tab:sluggs_galaxies_summary}) for all our kinematic datasets, while, for NGC 5846, we keep both $\mathrm{PA}_{\mathrm{kin}}$ and $\mathrm{q}_{\mathrm{kin}}$ fixed to the photometric values, $\mathrm{PA}_{\mathrm{phot}}$ and $\mathrm{q}_{\mathrm{phot}}$, of the galaxy.}
        \label{fig:1D_kinematic_profiles}
    \end{figure*}
    \begin{figure*}
    \ContinuedFloat
        \centering
        \includegraphics[width=0.25\textwidth]{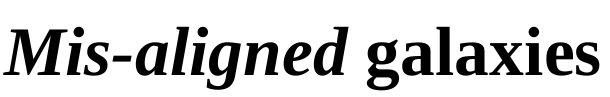}
        \includegraphics[width=1.0\textwidth]{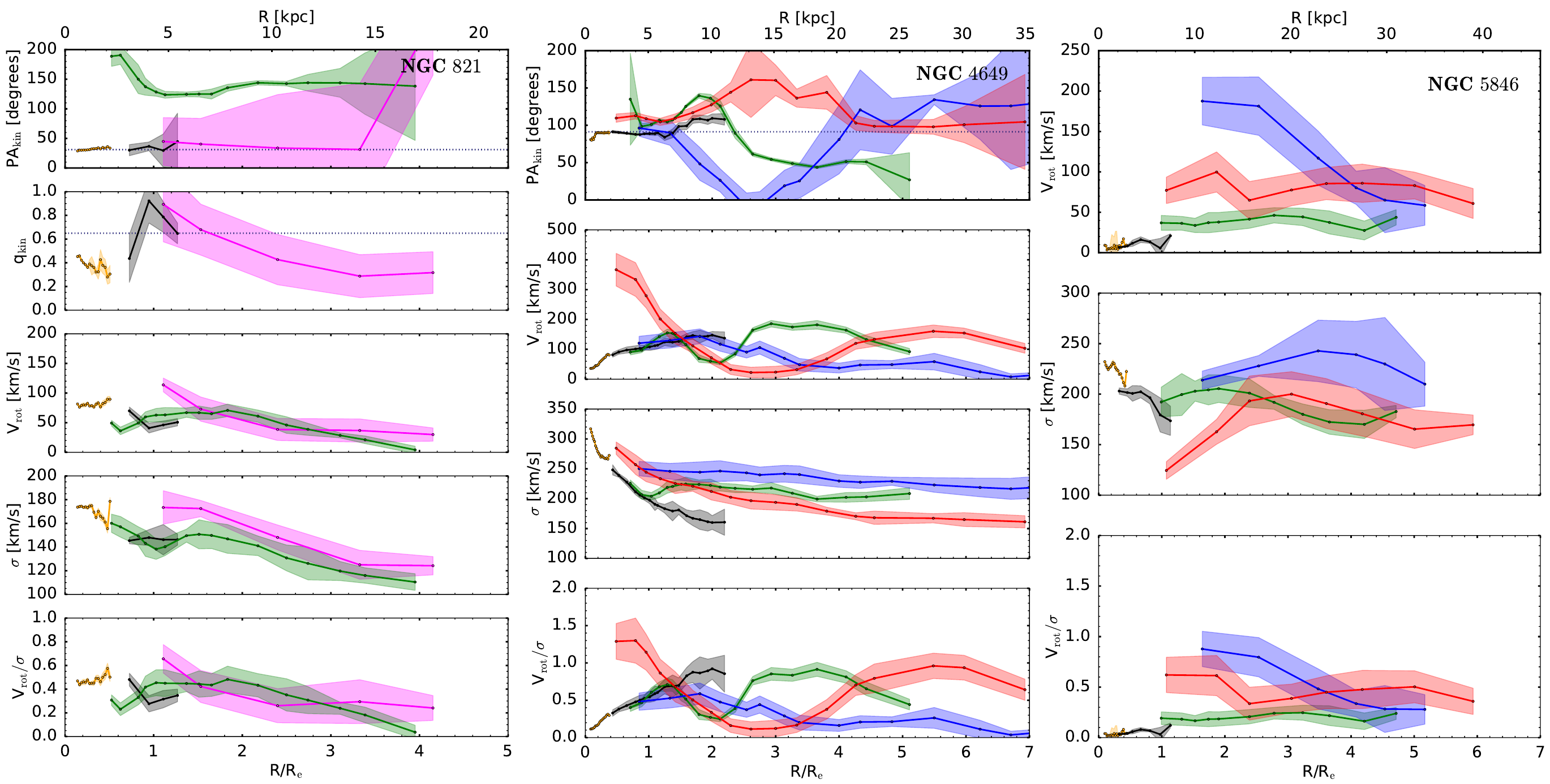}
        \caption*{\textit{Continued}.}
    \end{figure*}
    
    The different kinematic behaviours of the \textit{aligned} and \textit{mis-aligned} galaxies would suggest different formation histories for these two galaxy groups, with the latter being, possibly, the result of more recent interaction and merger events that have perturbed the dynamical equilibrium state of the galaxies.
    In Table \ref{tab:kinematic_alignment}, we summarize the kinematic alignment or misalignment of the GC and PNe tracers with respect to the underlying stars (from SLUGGS) and we indicate whether the red and blue GC sub-populations are kinematically aligned for each galaxy, based on the 2D and 1D kinematic results shown in the Fig. A1-A8 and in Fig. \ref{fig:1D_kinematic_profiles}, respectively, and described in more detail in the Appendix A for each individual galaxy. 
    
    \begin{table*}
    \caption{Summary table indicating whether the GC sub-populations and PNe are kinematically aligned to the underlying stars and whether the red and blue GC sub-populations are also kinematically aligned in each galaxy, i.e. their $\mathrm{PA}_{\mathrm{kin}}$ are consistent within the $1$-sigma errors based on the 2D kinematic maps and 1D kinematic profiles. If no colour bimodality distribution is observed, then all (red$+$blue) GCs are used. We assign "X" if there is no alignment between the pair of tracers that we are comparing in each galaxy and "$\checkmark$" if there is alignment. The results for NGC 3115 are taken from \citet{Dolfi2020}.}
    \centering
    \resizebox{15cm}{!}{
    \begin{tabular}{|c|c|c|c|c|c|} \hline \hline
    \bf{Galaxy} & \bf{Stars vs. PNe} & \bf{Stars vs. Red GCs} & \bf{Stars vs. Blue GCs} & \bf{Stars vs. GCs} & \bf{Red GCs vs. Blue GCs} \\ \hline \hline
    NGC 821  & X & - & - & X & - \\ 
    NGC 1023 & $\checkmark$ & $\checkmark$ & $\checkmark$ & - & $\checkmark$ \\ 
    NGC 2768 & $\checkmark$ & $\checkmark$ & $\checkmark$ & - & $\checkmark$ \\ 
    NGC 3115 & $\checkmark$ & $\checkmark$ & $\checkmark$ & - & X \\ 
    NGC 3377 & $\checkmark$ & $\checkmark$ & X & - & X \\ 
    NGC 4649 & X & $\checkmark$ & X & - & X \\ 
    NGC 4697 & $\checkmark$ & - & - & $\checkmark$ & - \\ 
    NGC 5846 & X & X & X & - & X \\ 
    NGC 7457 & $\checkmark$ & - & - & $\checkmark$ & - \\ 
    \hline \hline
    \end{tabular}}
    \label{tab:kinematic_alignment}
    \end{table*}
    
\begin{figure*}
        \centering
        \includegraphics[width=0.25\textwidth]{1D_KinematicProfiles/Title.pdf}
        \includegraphics[width=1.08\textwidth]{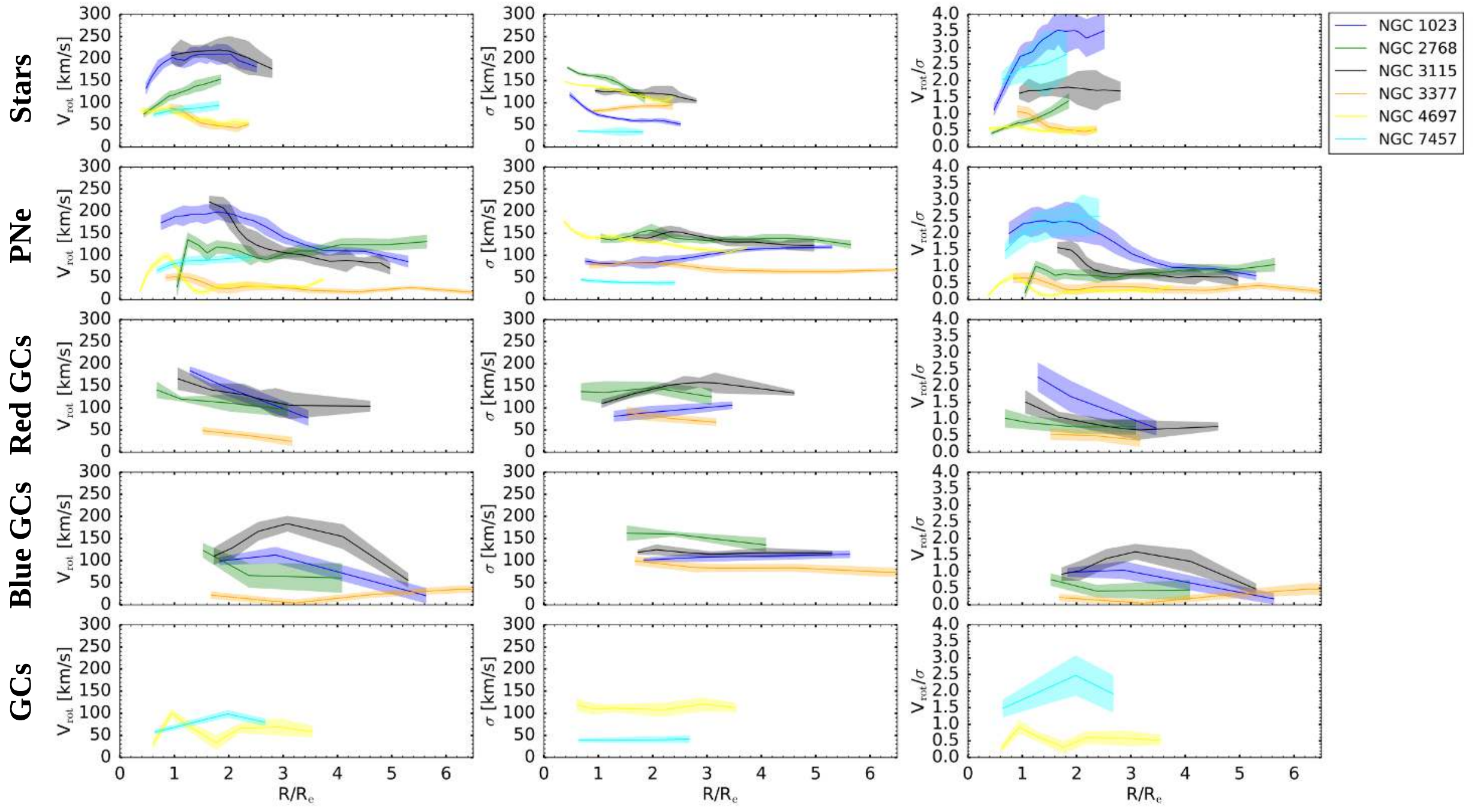}
        \includegraphics[width=0.25\textwidth]{1D_KinematicProfiles/Title2.pdf}
        \includegraphics[width=1.08\textwidth]{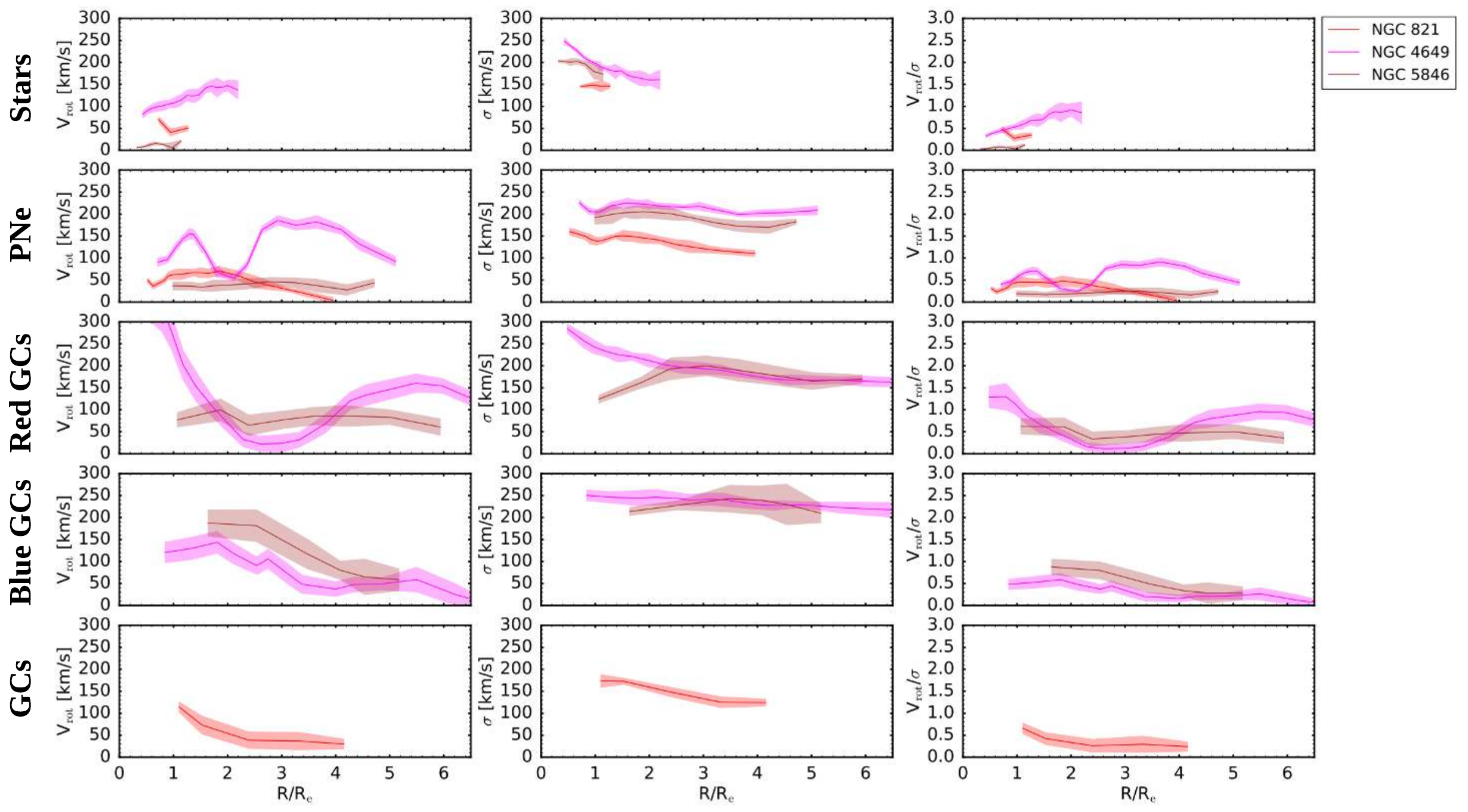}
        \caption{1D kinematic profiles of the \textit{aligned} (top) and \textit{mis-aligned} (bottom) galaxies compared together. The $V_{\mathrm{rot}}$, $\sigma$ and $V_{\mathrm{rot}}/\sigma$ profiles are shown from left to right, respectively, while the kinematic profiles of the stars (from SLUGGS), PNe, red GCs, blue GCs and all GCs (if no colour bimodality distribution is observed) are shown from the top to the bottom panel, respectively. The kinematic profiles are the same as in Fig. \ref{fig:1D_kinematic_profiles}, with the addition of NGC 3115 \citep{Dolfi2020} among the \textit{aligned} galaxies.}
        \label{fig:1D_kinematic_profiles_comparison}
    \end{figure*}

    In Fig. \ref{fig:1D_kinematic_profiles_comparison}, we compare together the 1D kinematic profiles of the stars, PNe and GC sub-populations of the \textit{aligned} (top) and \textit{mis-aligned} (bottom) galaxies, with the aim of identifying common trends in their behaviours. Therefore, in Fig. \ref{fig:1D_kinematic_profiles_comparison}, we show the $V_{\mathrm{rot}}$, $\sigma$ and $V_{\mathrm{rot}}/\sigma$ profiles (from left to right) for the stars (from SLUGGS), PNe, red GCs, blue GCs and all GCs (if no colour bimodality distribution is observed) from the top to the bottom panel, respectively. 
    We also include NGC 3115 for the comparison with the \textit{aligned} galaxies, with its kinematic profiles taken from \citet{Dolfi2020}.
    We notice that the $V_{\mathrm{rot}}/\sigma$ profiles of the \textit{aligned} and \textit{mis-aligned} galaxies have very similar shapes to the corresponding $V_{\mathrm{rot}}$ profiles. Therefore, we focus in this section on describing the common features of the $V_{\mathrm{rot}}/\sigma$ profiles of the different kinematic tracers of the galaxies.

    We notice that the velocity dispersion profiles of the \textit{aligned} and \textit{mis-aligned} galaxies in Fig. \ref{fig:1D_kinematic_profiles_comparison} all show weakly decreasing or overall flat behaviours out to large radii, with the exception of NGC 1023, whose PNe and GC sub-populations are characterized by slightly rising velocity dispersion profiles beyond $\sim2\, R_{\mathrm{e}}$.

    For the \textit{aligned} galaxies in Fig. \ref{fig:1D_kinematic_profiles_comparison}, we notice that their $V_{\mathrm{rot}}/\sigma$ profiles show clearly peaked (NGC 1023, NGC 3115) to flatter (NGC 2768, NGC 3377, NGC 4697) behaviours. NGC 7457 also show a peaked $V_{\mathrm{rot}}/\sigma$ profile, as seen from the GCs at $\sim2\, R_{\mathrm{e}}$, even though both the stars and PNe show rising $V_{\mathrm{rot}}/\sigma$ out to $\sim2$-$2.5\, R_{\mathrm{e}}$.
    It is worth noticing here that all \textit{aligned} galaxies have blue GCs that are rotating, with the exception of NGC 3377 that shows very low rotation in its blue GC sub-population, with $V_{\mathrm{rot}}/\sigma\lesssim0.5$ out to large radii.
    Additionally, while NGC 1023 and NGC 2768 have overall consistent $V_{\mathrm{rot}}/\sigma$ profiles between the red and blue GC sub-populations, the blue GCs of NGC 3115 are characterized by a peak of the $V_{\mathrm{rot}}/\sigma\sim1.5$ located at larger radii (i.e. $\sim3\, R_{\mathrm{e}}$) than that of the red GCs (i.e. $\sim1$-$1.5\, R_{\mathrm{e}}$.
    
    The \textit{mis-aligned} galaxies in Fig. \ref{fig:1D_kinematic_profiles_comparison} are, instead, characterized by more peculiar $V_{\mathrm{rot}}/\sigma$ profiles. 
    Only NGC 821 shows an overall flatter $V_{\mathrm{rot}}/\sigma$ profile out to large radii, similarly to NGC 2768, NGC 3377 and NGC 4697. However, in NGC 821, the rotation of the PNe and GCs occurs more closely aligned to the photometric minor-axis of the galaxy. 
    On the other hand, NGC 5846 shows spheroid-like kinematics (i.e. $V_{\mathrm{rot}}/\sigma<0.5$) out to large radii, while NGC 4649 shows the more interesting double-peaked $V_{\mathrm{rot}}/\sigma$ profiles, as seen from the PNe and red GCs (see Appendix A) for a more detailed description of the kinematic profiles of these two individual galaxies).
    
    Finally, we look for any mass or environment dependence in the 1D kinematic profiles of our galaxies. 
    Some previous works have found differences in the stellar kinematics of S0 galaxies in different environments, with S0s located in clusters and large groups being more rotationally supported than S0s found in isolation or in small groups (e.g. \citealt{Deeley2020,Coccato2020}). This might suggest different formation pathways for cluster and field S0s. However, we do not find any clear dependence on the environment in our kinematic profiles, nor more specifically in the $V_{\mathrm{rot}}/\sigma$ profiles. We argue that this is likely due to the small galaxy sample size that we are using and to the fact that our sample only contains one cluster (i.e. NGC 4649) and three field (i.e. NGC 821, NGC 3115 and NGC 7457) S0s, while the remaining galaxies are located in groups. Therefore, we are missing the high density environments of galaxy clusters.
    
    On the other hand, we do find a dependence of the velocity dispersion profiles on the stellar mass of the galaxies for all our kinematic tracers (i.e. stars, PNe and GCs), with the more massive galaxies showing higher velocity dispersion profiles than the less massive ones, as expected and also observed in previous works (e.g. \citealt{Foster2018}). At the same time, we also find a stellar mass dependence of the $V_{\mathrm{rot}}/\sigma$ profiles of the stars, PNe and GC sub-populations, with low mass galaxies showing higher $V_{\mathrm{rot}}/\sigma$ profiles than more massive ones. This is, again, consistent with the general picture from previous observational results, in which fast rotators are typically found among the population of low mass galaxies, while slow rotators are found among the class of rounder and more massive galaxies (e.g. \citealt{Cappellari2013}).

\section{Discussion}
\label{sec:discussion}

    \subsection{Do the PNe and GC sub-populations follow the rotation of the stars?}
    \label{sec:discrete_tracers_vs_stars}
    In Fig. \ref{fig:1D_kinematic_profiles}, we see that the PNe are overall good tracers of the kinematics of the stars in the overlapping radii out to $\sim2$-$3\, R_{\mathrm{e}}$ for the \textit{aligned} galaxies, suggesting that we can use the PNe as proxies of the dynamics of the stars out to larger radii (e.g. \citealt{Coccato2009,Coccato2013,Cortesi2016,Zanatta2018}). We see that the red GCs also show a good agreement with the kinematics of the stars out to similar radii.
    The blue GCs show overall consistent rotation with that of red GCs within the $1$-sigma errors in NGC 1023 and NGC 2768, suggesting that both GC sub-populations are likely tracing the kinematic properties of the same underlying stellar population in these two \textit{aligned} galaxies.
    The blue GCs are also rotating in NGC 3115, however their rotation velocity peaks at larger radii (i.e. $\sim3\, R_{\mathrm{e}}$) than that of the red GCs (i.e. $\sim1$-$2\, R_{\mathrm{e}}$), possibly suggesting differences in the formation history of NGC 3115 as compared to NGC 1023 and NGC 2768, e.g. different properties of the merger event, that we discuss in more details in Sec. \ref{sec:formation_of_aligned_galaxies}. 
    On the other hand, the blue GCs in NGC 3377 show very little amount of rotation at all radii as compared to the red GCs, suggesting that they are likely tracing the kinematic properties of the underlying stellar population of the galaxy halo, with some blue GCs that may also have an ex-situ origin (e.g. \citealt{Forbes1997,Forbes2018}).

    The \textit{mis-aligned} galaxies show more interesting kinematic behaviours, as both the PNe and GC sub-populations display minor-axis rotation and kinematic twists with respect to the underlying stars (i.e. NGC 821 and NGC 4649). NGC 5846 is also an example of a non-rotating galaxy. These kinematic behaviours suggest a more complex formation histories for the \textit{mis-aligned} galaxies, with, possibly, recent merger and accretion events.
    
    \subsection{How did the \textit{aligned} galaxies form?}
    \label{sec:formation_of_aligned_galaxies}
\begin{figure*}
        \centering
        \includegraphics[width=0.75\textwidth]{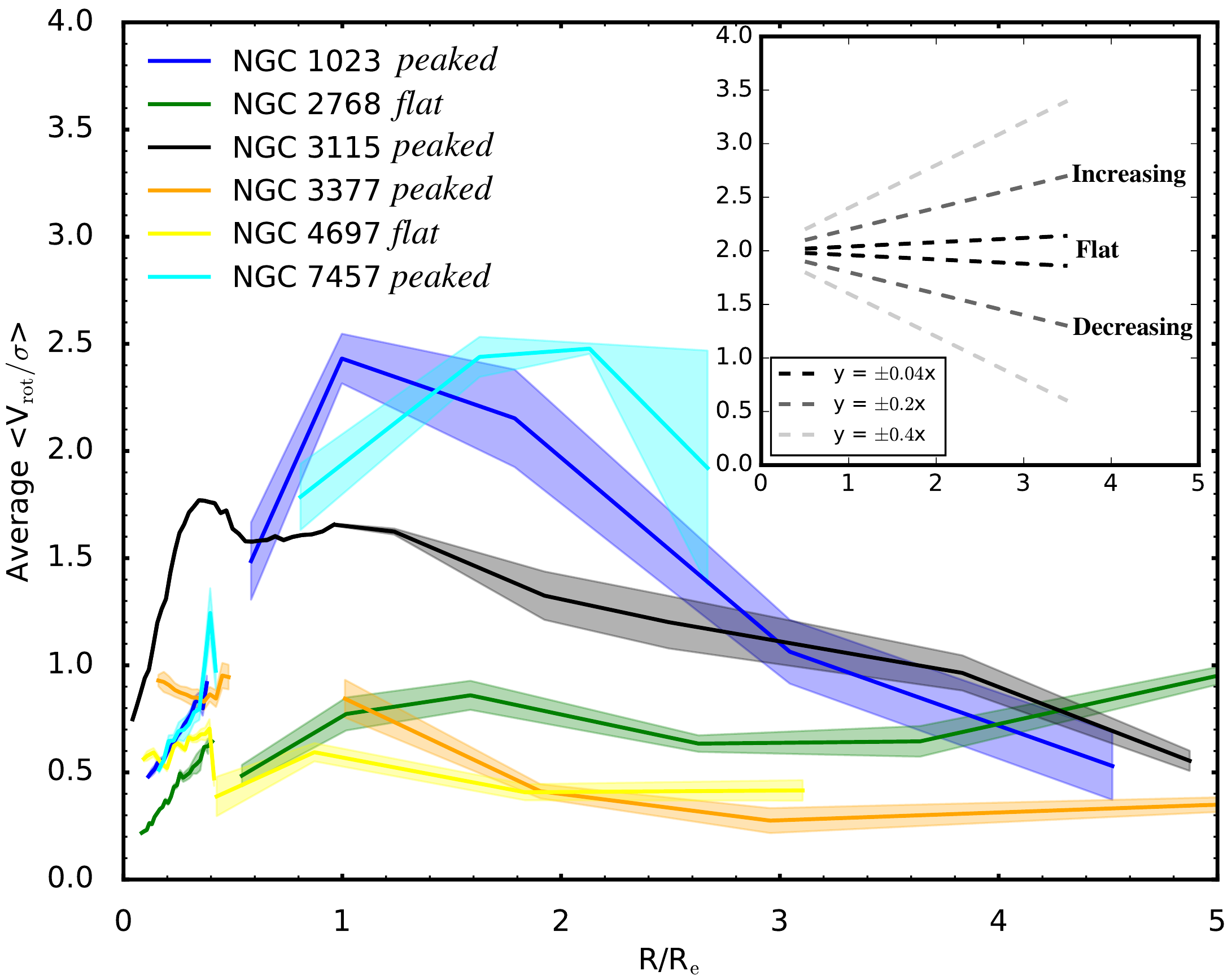}
        \caption{Smooth $V_{\mathrm{rot}}/\sigma$ profiles for the six \textit{aligned} galaxies, calculated by averaging the individual $V_{\mathrm{rot}}/\sigma$ profiles of the distinct kinematic tracers (i.e. stars, PNe and GCs) of each galaxy in the overlapping radii. The shaded areas represent the corresponding $1$-sigma standard errors of the mean at the radii where the average $V_{\mathrm{rot}}/\sigma$ profile is calculated. The inset plot shows six linear functions characterized by different slopes, i.e. $\pm0.04$, $\pm0.2$, $\pm0.4$. The $\pm0.04$ slope corresponds to the $V_{rot}/\sigma$ gradient adopted by \citet{Schulze2020} to separate the \textit{flat} $V_{rot}/\sigma$ profiles from the \textit{increasing} ($V_{\mathrm{rot}}/\sigma>0.04$) and \textit{peaked} ($V_{\mathrm{rot}}/\sigma<-0.04$) $V_{rot}/\sigma$ profiles. The ATLAS$^{\mathrm{3D}}$ data are also shown in this plot, however they typically do not reach beyond $\sim0.5\, R_{\mathrm{e}}$ and, therefore, do not overlap in radius with the other kinematic tracers (with the exception of the MUSE data of NGC 3115).}
        \label{fig:1D_stellar_Vsigma_vs_simulation}
    \end{figure*}

    We have seen that the six \textit{aligned} galaxies show good kinematic alignment of both the PNe and GC tracers with respect to the underlying stars in the overlapping radii. 
    In Fig. \ref{fig:1D_stellar_Vsigma_vs_simulation}, we now calculate their $V_{\mathrm{rot}}/\sigma$ profiles by averaging the contributions of the individual $V_{\mathrm{rot}}/\sigma$ profiles of the distinct kinematic tracers (i.e. stars, PNe and GCs) of each galaxy in the overlapping radii. The shaded areas represent the $1$-sigma standard errors of the mean. The inset plot shows the gradients, as adopted in the \textit{Magneticum} simulations by \citet{Schulze2020}, to classify their simulated ETGs as \textit{flat} (i.e. $-0.04 < V_{\mathrm{rot}}/\sigma < 0.04$), \textit{increasing} (i.e. $V_{\mathrm{rot}}/\sigma>0.04$) and "peaked and decreasing" (hereafter, \textit{peaked}; i.e. $V_{\mathrm{rot}}/\sigma<-0.04$) from the stellar kinematics extending out to $\sim5\, R_{\mathrm{e}}$.
    From the comparison of the observed $V_{\mathrm{rot}}/\sigma$ profiles of the six \textit{aligned} galaxies in Fig. \ref{fig:1D_stellar_Vsigma_vs_simulation} with the gradients from \citet{Schulze2020}, we classify NGC 1023, NGC 3115, NGC 3377 and NGC 7457 as \textit{peaked}, while NGC 2768 and NGC 4697 as \textit{flat}.

    \subsubsection{Galaxies with \textit{peaked} $V_{\mathrm{rot}}/\sigma$ profiles}
    \label{sec:peaked_galaxies}
    The four galaxies (i.e. NGC 1023, NGC 3115, NGC 3377, NGC 7457) with \textit{peaked} $V_{\mathrm{rot}}/\sigma$ profiles should share similar accretion histories in more recent times (i.e. $z\lesssim1$) according to \citet{Schulze2020}, dominated by minor (1:10 $<$ mass-ratio $<$ 1:4) and mini (mass-ratio $<$ 1:10) mergers that enhanced the velocity dispersion of the galaxy outer regions without disrupting its central disk-like kinematics. Some of these \textit{peaked} galaxies may have also experienced a major merger (mass-ratio $>$ 1:4), however this is most likely to have occurred at early times, i.e. more than $5\, \mathrm{Gyr}$ ago. In any case, \citet{Schulze2020} find that more than half of the population of the \textit{peaked} galaxies (i.e. $\sim60\%$) is likely to have experienced no major mergers in its past evolutionary history. 
    
    Using the Illustris TNG100 simulations, \citet{Pulsoni2020} studied the accretion histories of a sample of ETGs in the stellar mass range $10^{10.3} \leq \mathrm{M}^{*}/\mathrm{M}_{\odot} \leq 10^{12}$, reaching out to $15\, R_{\mathrm{e}}$. They find overall consistent results with those from \citet{Schulze2020}, where galaxies with \textit{peaked} $V_{\mathrm{rot}}/\sigma$ profiles have typically lower stellar masses and have evolved mostly passively after having, possibly, experienced a gas-rich major merger at early times (i.e. $z>1$). Although a significant fraction (i.e. $\sim60\%$) of the galaxies with \textit{peaked} $V_{\mathrm{rot}}/\sigma$ profiles may have not experienced any major merger and have only evolved through the accretion of low-mass galaxies in minor and mini mergers that influenced the outer galaxy regions, the gas-rich nature of these merger events may still play an important role. Indeed, \citet{Pulsoni2020} find that the peak of rotation in galaxies with \textit{peaked} $V_{\mathrm{rot}}/\sigma$ profiles becomes less prominent as the fraction of ex-situ stars accreted through major mergers increases. However, the peak in rotation can re-establish itself if new ($z\leq1$) stars are formed within the galaxy from its cold gas, either in-situ or recently accreted through mergers. This is true for both fast and slow rotator galaxies (see their figure 8).
    
    Finally, even though mini mergers do not largely affect the central peak of rotation of the galaxies with \textit{peaked} $V_{\mathrm{rot}}/\sigma$ profiles, they can increase the stellar rotation of the galaxies at larger radii if high fractions of the stellar mass (i.e. $10$-$30\%$) are accreted through this channel \citep{Pulsoni2020}. This mild increase of the stellar $V_{\mathrm{rot}}/\sigma$ at larger radii could be a consequence of the orbital configuration of the merger, as \citet{Karademir2019} have shown with simulations that mini mergers can increase the size of the host galaxy disk if they occur along the disk plane. Therefore, these mini mergers could produce galaxies with extended stellar disks characterized by, possibly, higher rotation out to larger radii.
    
    NGC 1023 and NGC 3115 are more massive than NGC 3377 and NGC 7457. For this reason, NGC 1023 and NGC 3115 should be more similar to the Class II galaxies of \citet{Pulsoni2020}, that are in-situ dominated in the inner regions and ex-situ dominated in the outskirts of the galaxies. On the other hand, NGC 3377 and NGC 7457 should have more similar properties to the Class I galaxies of \citet{Pulsoni2020}, that are in-situ dominated at all radii.
    Therefore, this suggests that NGC 3377 and NGC 7457 have, on average, accreted larger (i.e. up to $\sim50$-$60\%$) gas fractions than NGC 1023 and NGC 3115 (i.e. up to $\sim30$-$40\%$), possibly through an early gas-rich major merger, that contributed to the in-situ star formation of the galaxies. 
    This merger event is expected to have occurred at earlier times, i.e. $z\geq2$, for NGC 3377 and NGC 7457 as compared to NGC 1023 and NGC 3115, i.e. $1<z<2$, while no major mergers have likely happened in the evolution histories of all four galaxies since $z=1$ \citep{Pulsoni2020,Schulze2020}.
    
    However, NGC 1023, NGC 3115 and NGC 7457 are rotationally supported S0 galaxies, with $V_{\mathrm{rot}}/\sigma>1$ out to $\sim3\, R_{\mathrm{e}}$, as seen in Fig. \ref{fig:1D_stellar_Vsigma_vs_simulation}. 
    Following \citet{Bournaud2005}, a mean $V_{\mathrm{rot}}/\sigma>1$ is more compatible with a minor merger rather than a major one, which would produce hotter disk kinematics with mean $V_{\mathrm{rot}}/\sigma<1$. For this reason, NGC 1023 and NGC 7457 are more consistent with having experienced at most a minor merger with mass-ratio 1:10, due to their high $V_{\mathrm{rot}}/\sigma>2$ \citep{Bournaud2005}. On the other hand, NGC 3115 is more consistent with a minor merger with mass-ratio 1:7 that resulted in its $1<V_{\mathrm{rot}}/\sigma<2$ \citep{Bournaud2005} as seen in Fig. \ref{fig:1D_stellar_Vsigma_vs_simulation}.
    Indeed, for NGC 3115, \citet{Dolfi2020} have proposed a scenario in which the galaxy was originally a spiral that experienced an early (i.e. $\sim9\, \mathrm{Gyr}$ ago) gas-rich minor merger in the mass-ratio range 1:4-1:10, which shaped its current embedded kinematically cold disk. In more recent times, NGC 3115 has evolved mostly passively, exhausting its own gas through in-situ star formation and accreting low-mass dwarf galaxies in mini mergers that produced the spheroid-like kinematics of the outer regions of the galaxy.   
    This formation scenario for NGC 3115 is consistent with the stellar population results from \citet{Poci2019}, who found a dominant old (i.e.  $\sim9\, \mathrm{Gyr}$), metal-rich stellar population in the kinematically cold disk component of NGC 3115, therefore ruling out the occurrence of recent major or minor merger events that would likely destroy this disk structure. At the same time, the old and metal-poor stellar population with hotter kinematics found in the stellar halo of NGC 3115 suggests a late accretion of low-mass dwarf galaxies in dry minor and mini mergers onto the outer regions of the galaxy.
    
    For NGC 1023, we propose a similar formation history as for NGC 3115, characterized by an early (i.e. $1<z<2$) gas-rich minor merger with mass-ratio 1:10. This gas-rich minor merger would be consistent with the high stellar $V_{\mathrm{rot}}/\sigma>2$ of NGC 1023 (see Fig. \ref{fig:1D_kinematic_profiles}), according to \citet{Bournaud2005}, and would shape the embedded kinematically cold disk of NGC 1023, according to \citet{Naab2014}. Moreover, NGC 1023 is classified as an SB0 galaxy from previous photometric studies that detected the presence of a barred bulge \citep{Barbon1975}. 
    From simulations, \citet{Cavanagh2020} have shown that bars can be formed from both major and minor mergers and that the orbital configuration of the merger plays a decisive role in determining whether the bar can survive after the merger. Specifically, they found that minor mergers with mass-ratio 1:10 and closely aligned spin angles are most conducive to the bar formation.
    Therefore, this seems to reinforce our proposed scenario that NGC 1023 formed through an early gas-rich minor merger with mass-ratio 1:10, where the merging galaxies had similar orientations of their spin angles. 
    The $V_{\mathrm{rot}}/\sigma$ profiles of the red and blue GCs that show little rotation at larger radii (see Fig. \ref{fig:1D_kinematic_profiles}) would also be more consistent with a minor merger scenario, according to \citet{Bekki2005}.
    \citet{Cortesi2016} have also found that the majority of the red GCs (and a small fraction of the blue GCs) are associated with the disk of NGC 1023 and, therefore, show a similar disk-like kinematics as the stars and PNe. These GCs could have, thus, formed together with the disk of NGC 1023 at early times (i.e. $z\sim2$) and, possibly, at the epoch of the early gas-rich minor merger.
    At late times, NGC 1023 has likely continued its evolution history through the accretion of dwarf galaxies in mini mergers \citep{Schulze2020}. Indeed, NGC 1023 is currently ongoing an interaction with its companion dwarf galaxy, NGC 1023A, suggesting that NGC 1023 is still experiencing its late mini merger accretion. The HI gas cloud detected around NGC 1023, with its asymmetric spatial distribution \citep{Sancisi1984,Capaccioli1986}, could be the result of the gas that was stripped from the dwarf galaxies accreted by NGC 1023 in mini mergers, as previously suggested also by \citet{Cortesi2016}.
    Finally, \citet{Corsini2016} detected the presence of a nuclear disk in NGC 1023, which is characterized by a younger ($\sim3.4\, \mathrm{Gyr}$) and more metal-rich ([Fe/H]$=0.50\, \mathrm{dex}$) stellar population than the host galaxy bulge, suggesting that it may have formed from pre-processed, metal-enriched gas within NGC 1023. 
    Therefore, the presence of the young nuclear disk would seem to agree with a recent (and, possibly, still ongoing) mini merger accretion in NGC 1023, where the gas stripped from the dwarf galaxies may have been funnelled towards the galaxy central regions, facilitated by the presence of the bar, to form the young and metal-rich nuclear disk of NGC 1023.
    
    Overall, the formation histories of NGC 1023 and NGC 3115 should be very similar, with the main difference being, possibly, in the orbital configuration of the mergers. In Fig. \ref{fig:1D_kinematic_profiles}, we notice that the blue GCs of NGC 3115 have a $V_{\mathrm{rot}}/\sigma$ peak located at larger radii (i.e. $\sim3\, R_{\mathrm{e}}$) than the red GCs (i.e. $\sim1\, R_{\mathrm{e}}$). On the other hand, the red and blue GCs of NGC 1023 have similar $V_{\mathrm{rot}}/\sigma$ profiles. 
    According to \citet{Bekki2005}, minor mergers can produce a rotation velocity difference between the blue and red GCs, as we see for NGC 3115 in Fig. \ref{fig:1D_kinematic_profiles}. The fact that we do not see this rotation velocity difference in the red and blue GCs of NGC 1023, that also formed through a minor merger of similar mass-ratio, could suggest different orbital configurations of the minor mergers in NGC 1023 and NGC 3115.
    Alternatively, \citet{Karademir2019} showed that mini mergers occurring along the disk plane of the galaxy can increase the size of the host galaxy disk. Therefore, this could suggest that the mini mergers in NGC 3115 have occurred along the plane of the disk, causing the size of the disk to increase. As a result, we see a more extended disk-like kinematics in NGC 3115 than in NGC 1023.
    Additional simulations of GC kinematics would be required in order to understand whether a specific orbital configuration of the minor and mini mergers can, indeed, produce the observed rotation velocity difference in the red and blue GCs of NGC 3115, as compared to NGC 1023. 

    Similarly to NGC 1023, NGC 7457 should have also formed through a gas-rich minor merger with mass-ratio 1:10, which would be more consistent with its high stellar $V_{\mathrm{rot}}/\sigma>2$ (see Fig. \ref{fig:1D_kinematic_profiles}), according to \citep{Bournaud2005}. 
    However, according to \citet{Pulsoni2020}, low-mass galaxies, such as NGC 7457, should be in-situ dominated at all radii, with very large gas fractions (i.e. $\sim50$-$60\%$) that were accreted at early times (i.e. $z>2$) and that contributed to the in-situ star formation of the galaxy until more recently. 
    These galaxies are also expected to have accreted only very few ex-situ stars onto their outskirts from mini mergers. 
    Therefore, this would suggest that the vast majority of the gas fraction in NGC 7457 already belonged to the galaxy (i.e. in-situ origin).
    From the stellar kinematics, \citet{Bellstedt2017} showed that the high $V_{\mathrm{rot}}/\sigma$ of NGC 7457 was more consistent with that of a disk progenitor galaxy that evolved passively through secular evolution, consuming its own in-situ gas.
    \citet{Zanatta2018} also found that $\sim70\%$ of the GCs of NGC 7457 belong to the disk of the galaxy and closely follow the kinematics of the stars and PNe (as we have also seen in Fig. \ref{fig:1D_kinematic_profiles}), therefore suggesting mostly an in-situ origin for the GCs and a secular evolution for NGC 7457.
    According to \citet{Schulze2020}, around half (i.e. $\sim40\%$) of the population of galaxies with \textit{peaked} $V_{\mathrm{rot}}/\sigma$ profiles are found to have experienced no minor mergers. However, they are still likely to have experienced, even multiple (i.e. up to $4$), mini mergers in their late (i.e. $z<1$) evolution histories, which contributed to the few ex-situ stars in the galaxy outskirts.
    These mini mergers are expected to enhance the dispersion of the galaxy, and of its GCs \citep{Bekki2005}, in the outer regions, without altering the central $V_{\mathrm{rot}}/\sigma$ profiles of NGC 7457's GCs and stars. In Fig. \ref{fig:1D_kinematic_profiles}, we see that there is a hint that the $V_{\mathrm{rot}}/\sigma$ profile of the GCs starts decreasing beyond $\sim2\, R_{\mathrm{e}}$ and such behaviour, even if weak, is also visible from the $V_{\mathrm{rot}}/\sigma$ profile of the GCs in \citet{Zanatta2018}. 
    Finally, NGC 7457 shows the youngest ($\sim6\, \mathrm{Gyr}$) stellar population \citep{McDermid2015,Forbes2017}, with respect to the other galaxies in our sample, suggesting that the galaxy continued to form new stars from its large in-situ gas reservoir during its passive evolution until more recently. Indeed, according to \citet{Pulsoni2020}, Class I galaxies, such as NGC 7457, are characterized by the largest fraction ($\sim50\%$) of in-situ formed stars at $z\leq1$, with respect to the other classes.
    Therefore, NGC 7457 seems to be consistent with mostly a secular evolution during which it consumed its own in-situ gas. At late times, NGC 7457 has likely accreted dwarf galaxies in mini mergers that caused the rotation velocity of the GCs to, possibly, decrease beyond $\sim2\, R_{\mathrm{e}}$. It would be interesting to extend the stellar and GC kinematics beyond $2\, R_{\mathrm{e}}$, in order to confirm whether the $V_{\mathrm{rot}}/\sigma$ profiles are, indeed, decreasing, consistent with a spheroid-like kinematics at larger radii produced by the late mini merger accretion.
    
    Similarly to NGC 7457, NGC 3377 should also be mostly in-situ dominated at all radii, with a somewhat larger fractions of ex-situ stars that are accreted through minor and mini mergers \citep{Pulsoni2020}. 
    The stellar $V_{\mathrm{rot}}/\sigma\sim1$ of NGC 3377 (see Fig. \ref{fig:1D_kinematic_profiles}) would be consistent with a minor merger with mass-ratio 1:4.5, according to \citet{Bournaud2005}. This minor merger was likely gas-rich and occurred early (i.e. $1<z<2$), in order to shape the embedded kinematically cold disk of NGC 3377, according to \citet{Naab2014,Schulze2020}.
    Additionally, the red GCs are following the rotation of the stars (see Fig. \ref{fig:1D_kinematic_profiles}) in the overlapping radii and they show little rotation at larger radii, which is also consistent with a minor merger scenario \citep{Bekki2005}. 
    However, it is interesting to note that the red GCs in NGC 3377 are not as old as the red GCs in NGC 3115, as found by \citep{Usher2019}. This would suggest a more recent formation for the red GCs of NGC 3377, as compared to the red GCs of NGC 3115, possibly after the epoch of the gas-rich minor merger that produced the host galaxy disk. 
    A fraction of the GCs in NGC 3377 could also have an ex-situ origin. Indeed, the results from \citet{Usher2019} showed that the GCs of NGC 3377 display a wide range of metallicities, including few very young and metal-poor GCs, suggesting the contribution from recently accreted dwarf galaxies.
    According to \citet{Pulsoni2020} and \citet{Schulze2020}, the late (i.e. $z<1$) evolution history of NGC 3377 is likely dominated by the accretion of dwarf galaxies in mini mergers, which may have contributed to ex-situ stars as well as to ex-situ GCs in the outer regions of the galaxy. 
    Specifically, the blue GCs in NGC 3377 show very little rotation out to large radii, as compared to the other kinematic tracers in NGC 3377 and to the other galaxies in our sample (see Fig. \ref{fig:1D_kinematic_profiles}), suggesting that some blue GCs have an ex-situ origin. 
    
    \subsubsection{Galaxies with \textit{flat} $V_{\mathrm{rot}}/\sigma$ profiles}
    \label{sec:flat_galaxies}
    The two galaxies with \textit{flat} $V_{\mathrm{rot}}/\sigma$ profiles (i.e. NGC 2768, NGC 4697) should share similar accretion histories according to \citet{Schulze2020} and should be dominated by minor and mini mergers in more recent times (i.e. $z\lesssim1$), similarly to the \textit{peaked} galaxies. However, galaxies with \textit{flat} $V_{\mathrm{rot}}/\sigma$ profiles are more likely to have experienced also a late (i.e. $z\lesssim1$) major merger that destroyed the central disk-like kinematics of the galaxy \citep{Schulze2020}. These results are also consistent with those from \citet{Pulsoni2020}, who find that the $V_{\mathrm{rot}}/\sigma$ of the galaxies decreases as the ex-situ fraction increases. Therefore, galaxies with \textit{flat} $V_{\mathrm{rot}}/\sigma$ profiles should have typically higher stellar masses, since they have likely experienced more recent ($z\leq1$) and more gas poor mergers that brought in larger fractions of ex-situ stars, without much gas, flattening the central disk-like kinematics of the galaxy \citep{Pulsoni2020}.
    NGC 2768 and NGC 4697 are also more massive than the four \textit{peaked} galaxies, therefore they should be characterized by similar in-situ and ex-situ fractions in the inner regions and be ex-situ dominated in the outskirts of the galaxies (i.e. Class III of \citealt{Pulsoni2020}).
    
    NGC 2768 shows a $V_{\mathrm{rot}}/\sigma\sim1$ out to large radii from the stars and PNe (see Fig. \ref{fig:1D_kinematic_profiles}), which would be consistent with a minor merger with mass-ratio 1:4.5, according to \citet{Bournaud2005}. 
    However, this merger is likely to have occurred more recently at $z\lesssim1$ \citep{Schulze2020,Pulsoni2020} and to be more gas-poor, as compared to the mergers experienced by the \textit{peaked} galaxies that are typically gas-rich \citep{Pulsoni2020}.
    \citet{Naab2014} showed that a late ($z<2$) gas-poor major merger can spin-up the stars in the merger remnant and produce a rising $\lambda_{\mathrm{R}}$ profile, as seen in figure 5 of \citet{Naab2014}, which resembles the rising stellar $V_{\mathrm{rot}}/\sigma$ profile, shown in Fig. \ref{fig:1D_kinematic_profiles}. However, a gas-poor major merger with mass-ratio $>$1:4 would produce a galaxy remnant with hotter disk kinematics, i.e. $V_{\mathrm{rot}}/\sigma<1$ \citep{Bournaud2005}, which would not be entirely consistent with the $V_{\mathrm{rot}}/\sigma$ of NGC 2768. 
    Therefore, we suggest that NGC 2768 likely formed through a late (i.e. $z\lesssim1$) gas-poor minor merger with mass-ratio $\sim$1:4.5, as was previously suggested by \citet{Zanatta2018}. 
    This late gas-poor minor merger was likely responsible for producing the kinematic misalignment between the stars and gas in the centre of the galaxy \citep{Sarzi2006} and the \textit{flat} $V_{\mathrm{rot}}/\sigma$ profile of the stars and PNe out to larger radii (see Fig. \ref{fig:1D_stellar_Vsigma_vs_simulation}). Additionally, the gas-poor minor merger also likely contributed to a fraction of the ex-situ stars, even though a significant fraction of ex-situ stars (i.e. $\sim20\%$) is expected to come from the mini mergers that also shaped the late ($z<1$) evolution history of the galaxy \citep{Pulsoni2020,Schulze2020}.
    The kinematics of the red and blue GC sub-populations, which show low $V_{\mathrm{rot}}/\sigma\lesssim1$ out to large radii, is also more consistent with a minor merger scenario \citep{Bekki2005}. 
    Moreover, \citet{Forbes2012} have shown that the red GCs follow the radial density distribution of the bulge PNe and stars and are, therefore, mostly associated with the spheroidal component of the galaxy. This could be consistent with NGC 2768 having undergone a more gas-poor accretion history \citep{Pulsoni2020} that has produced the spheroid-like kinematics of the GC sub-populations, a fraction of which could likely have an ex-situ origin.
    
    NGC 4697 shows a $V_{\mathrm{rot}}/\sigma\sim0.5$ (see Fig. \ref{fig:1D_stellar_Vsigma_vs_simulation}), which would be consistent with a major merger with mass-ratio 1:3, according to \citet{Bournaud2005}.
    However, the GCs show a $V_{\mathrm{rot}}/\sigma$ that is decreasing at larger radii and this kinematic behaviour is more consistent with a minor merger \citep{Bekki2005}.
    \citet{Bournaud2005} showed that multiple minor mergers in the mass-ratio range 1:4.5-1:10 can, ultimately, produce an elliptical galaxy with stellar kinematics and morphology that are similar to those of an elliptical galaxy produced through an individual major merger.
    Therefore, we suggest that NGC 4697 has likely formed through multiple minor mergers (at least two) in the mass-ratio range 1:4.5-1:10. These minor mergers likely occurred more recently (i.e. $z\lesssim1$) and were gas-poor, as for NGC 2768, therefore contributing to some fraction of the ex-situ stars at all radii \citep{Pulsoni2020,Schulze2020}. At late times (i.e. $z<1$), NGC 4697 has likely continued its evolution through the accretion of dwarf galaxies in mini mergers that contributed to the remaining fraction of ex-situ stars \citep{Pulsoni2020,Schulze2020}.
    An early work has detected the presence of a young (i.e. $\sim1\, \mathrm{Gyr}$ old) stellar population as well as of low levels of ongoing star formation in the central regions of NGC 4697 \citep{Ford2013}. According to \citet{Pulsoni2020}, galaxies in Class III, such as NGC 4697, may also have accreted some small gas fractions, i.e. $\sim30\%$, from which new stars can be born. If this gas was funnelled towards the central regions of the galaxy at the time of the minor mergers, then it could have triggered a more recent star formation event that formed the young stars detected by \citet{Ford2013} in the galaxy central regions. Alternatively, a more recent star formation event could have been induced from the remaining in-situ gas of NGC 4697 that was funnelled towards the central regions of the galaxy at the epoch of the late gas-poor minor and mini mergers. 
    This young stellar population could, possibly, be associated with the sub-population of bright PNe that was identified by \citet{Sambhus2006}, as theoretical studies suggests that such a bright PNe sub-population in ETGs would require high-mass, young (i.e. $\leq1\, \mathrm{Gyr}$ old) stars \citep{Kalirai2008,Ciardullo2010}.
    \citet{Sambhus2006} showed that the bright PNe sub-population had an asymmetric distribution and was concentrated near the central regions of the galaxy, but it was not tracing the kinematics of the underlying stars \citep{Sambhus2006}. 
    Therefore, this could be consistent with a scenario in which this small fraction of young stars in the central regions of the galaxy has formed from the gas, of either in-situ or ex-situ origin, that induced a recent star formation event during the late minor and mini merger accretion of NGC 4697.

    \subsection{How did the \textit{mis-aligned} galaxies form?}
    \label{sec:formation_of_misaligned_galaxies}
    The three \textit{mis-aligned} galaxies show kinematic twists and misalignment of both the PNe and GC sub-populations with respect to the underlying stars. For this reason, they have more peculiar kinematic profiles that cannot be easily linked to one of the three characteristic profile shapes (i.e. \textit{peaked}, \textit{flat}, \textit{increasing}) investigated by \citet{Schulze2020}. This suggests, therefore, that these galaxies had a more complex formation history that involved a larger number of recent merger and interaction events.
    
    NGC 821 is an isolated elliptical galaxy that should share a similar accretion history to the Class III galaxies of \citet{Pulsoni2020}, characterized by more recent and gas-poor mergers. 
    The stellar $V_{\mathrm{rot}}/\sigma\sim0.5$ would be consistent with a major merger of mass-ratio 1:3 \citep{Bournaud2005}. 
    This gas-poor major merger would have likely occurred more recently, i.e. $z\lesssim1$, and contributed to the accretion of a fraction of the ex-situ stars \citep{Pulsoni2020}. 
    However, \citet{Proctor2005} found evidence of strong metallicity gradients in NGC 821, with the central regions showing young ($\sim4\, \mathrm{Gyr}$ old) ages and high ([Z/H]$=0.5\, \mathrm{dex}$) metallicities. Outside the central $1\, R_{\mathrm{e}}$ of the galaxy, the ages steeply rise to $\sim12\, \mathrm{Gyr}$ and the metallicities drop to $\sim1/3$ of the central value. 
    These steep metallicity gradients rule out a recent major merger in the formation history of the galaxy that would produce a mixing of the stellar population and flatter metallicity gradients, as also shown with simulations \citep{Cook2016}.
    Therefore, it seems more likely that NGC 821 has experienced multiple minor mergers rather than a single major merger. Indeed, \citet{Bournaud2005} showed that multiple minor mergers in the mass-ratio range 1:4.5-1:10 have the same effect of a single major merger, ultimately producing an elliptical galaxy with hotter kinematics.
    For NGC 821, we, thus, propose that the galaxy formed through multiple gas-poor minor (at least two) mergers in the mass-ratio range 1:4.5-1:10 in more recent times (i.e. $z\lesssim1$), similarly to NGC 4697.
    This formation scenario for NGC 821 would be in agreement with the conclusions reached by \citet{Proctor2005}, who suggested that the most likely explanation for the central young and metal-rich stellar population in NGC 821 was the accretion of a low-mass galaxy that triggered a centrally concentrated burst of star formation from the in-situ gas of the galaxy.
    The decreasing $V_{\mathrm{rot}}/\sigma$ profile of the GCs (see Fig. \ref{fig:1D_kinematic_profiles}) also suggests a minor merger scenario that is expected to produce little rotation of the GCs at larger radii \citep{Bekki2005}.
    Finally, the minor-axis rotation of the PNe at all radii as well as the kinematic misalignment shown by the GCs with respect to the underlying stars (see Fig. \ref{fig:1D_kinematic_profiles} and Appendix A6) could, possibly, be a result of the late gas-poor accretion history of NGC 821. \citet{Naab2014} found that late gas-poor major mergers can produce slowly rotating galaxies with kinematic twists and misalignment. In the case of NGC 821, the multiple (and late) gas-poor minor mergers could have had the effect of a gas-poor major merger, producing the minor-axis rotation of the PNe.
    \citet{Coccato2009} found similar misalignment in the rotation velocity of the PNe with respect to the underlying stars and \citet{Pota2013} found that the blue GCs are rotating along the photometric minor-axis of the galaxy, consistently with the PNe, while some outer red GCs are counter-rotating with respect to the underlying stars. \citet{Coccato2009} suggested that the disagreement between the stars and PNe may be due to a rapid change of the major kinematic axis of rotation of NGC 821 just outside $\sim1\, R_{\mathrm{e}}$, beyond which the stellar kinematics do not reach. 
    It would be interesting to extend the stellar kinematics beyond $1\, R_{\mathrm{e}}$ with new observations in order to confirm whether the major kinematic axis of rotation of the stars changes beyond $\sim1\, R_{\mathrm{e}}$, in agreement with the PNe data. 
    
    NGC 4649 is the only cluster galaxy in our sample and the most massive together with NGC 5846. Therefore, NGC 4649 should be ex-situ dominated at all radii (Class IV galaxy; \citealt{Pulsoni2020}). This galaxy is expected to have had a very gas-poor accretion history, with only very small fraction of accreted gas (i.e. $<20\%$), and to have experienced at least one gas-poor major merger in recent times, i.e. $z<1$, that contributed to a large fraction (up to $\sim40\%$) of the ex-situ stars \citep{Pulsoni2020}. 
    \citet{Naab2014} found that a late gas-poor major merger can spin up the stars in the merger remnant, producing a fast rotator galaxy with rising $\lambda_{\mathrm{R}}$ profile out to $\sim2\, R_{\mathrm{e}}$, which roughly resembles the rising stellar $V_{\mathrm{rot}}/\sigma$ profile out to $\sim2\, R_{\mathrm{e}}$ of NGC 4649 in Fig. \ref{fig:1D_kinematic_profiles}.
    According to \citet{Bournaud2005}, a major merger with mass-ratio 1:3 can produce elliptical galaxies with less disky isophotes (low ellipticity) and stellar $V_{\mathrm{rot}}/\sigma<1$, which would also be consistent with our kinematic results in Fig. \ref{fig:1D_kinematic_profiles}.
    Both the red and blue GCs sub-populations show rotation in the inner regions of the galaxy within $\sim2\, R_{\mathrm{e}}$, consistent with the photometric major-axis of the galaxy. The red GCs also show rotation at larger radii, i.e. beyond $\sim4\, R_{\mathrm{e}}$, which is overall aligned with the galaxy photometric major-axis (see Fig. \ref{fig:1D_kinematic_profiles}). 
    We note here that we do not detect rotation at large radii for the blue GCs, as for the red GCs, from our 1D kinematic profiles in Fig. \ref{fig:1D_kinematic_profiles}. However, \citet{Pota2015} found that the blue GCs in NGC 4649 also show significant rotation at larger radii, similarly to the red GCs. However, \citet{Pota2015} found that the rotation of the blue GCs increases beyond $\sim30\, \mathrm{kpc}$. 
    Our 1D kinematic profiles of NGC 4649 in Fig. \ref{fig:1D_kinematic_profiles} do not reach beyond $\sim35\, \mathrm{kpc}$, therefore we are not detecting the rotation of the blue GCs at larger radii. 
    If the blue GCs are also rotating at larger radii as seen by \citet{Pota2015}, then a late (i.e. $z<1$) gas-poor major merger formation scenario would be consistent with the kinematic results shown in Fig. \ref{fig:1D_kinematic_profiles}. In fact, according to \citet{Bekki2005}, a major merger is expected to spin up both the red and blue GC sub-populations at larger radii, with the blue GCs having larger velocity dispersion than the red GCs, as we have seen in Fig. \ref{fig:1D_kinematic_profiles}.
    We notice that the PNe also show a dip in the rotation velocity at $\sim2\, R_{\mathrm{e}}$, beyond which the rotation increases again but it is more closely aligned to the photometric minor-axis of the galaxy, as was also seen in \citet{Pota2015} and \citet{Coccato2013}. 
    The kinematic twist of the PNe could, possibly, also be a result of the late gas-poor major merger \citep{Naab2014}.
    \citet{Forbes2017_GC} have found a large number of ultra-compact dwarfs (UCDs) that are associated with NGC 4649. \citet{Strader2004} also identified a very massive UCD in NGC 4649, which is located at $\sim2\, R_{\mathrm{e}}$ towards South-West, in correspondence of the rotation velocity dip of the PNe and red GCs (see Fig. \ref{fig:1D_kinematic_profiles}).
    Therefore, these UCDs could be the remnant nuclei of dwarf galaxies (e.g. \citealt{Drinkwater2000}) that have been tidally stripped by NGC 4649 during its late (i.e. $z<1$) gas-poor minor and mini merger accretion events that are expected to contribute in roughly equal amounts to the fraction of the ex-situ stars \citep{Pulsoni2020}.
    
    Similarly to NGC 4649, NGC 5846 is also expected to be ex-situ dominated at all radii due to its large stellar mass. Therefore, its accretion history should be characterized by more very gas-poor merger events that happened very recently, i.e. $z<1$.
    NGC 5846 is a massive slow rotator galaxy with $V_{\mathrm{rot}}/\sigma\lesssim0.2$ out to large radii, as seen from both the stars and PNe in Fig. \ref{fig:1D_kinematic_profiles}. 
    According to \citet{Naab2014}'s simulations, the kinematic properties of NGC 5846 would be more consistent with a formation history dominated by late gas-poor minor mergers, as a late gas-poor major merger is, instead, expected to produce a weakly rising stellar rotation out to $\sim2\, R_{\mathrm{e}}$.
    Indeed, multiple minor, as well as mini, mergers are expected to, ultimately, produce an elliptical slowly rotating galaxy, with similar properties as if it had undergone one single major merger \citep{Bournaud2005}. Finally, both the red and blue GC sub-populations do not show any strong rotation at large radii, consistent with a minor merger formation pathway \citep{Bekki2005}.
    We note here that the rotation of the red and blue GCs in the inner regions of the galaxy within $\sim2$-$3\, R_{\mathrm{e}}$ in Fig. \ref{fig:1D_kinematic_profiles} may not be real but due to the contamination of the GC system of NGC 5846 with outliers from its nearby companion galaxies, e.g. NGC 5846A, as also previously pointed out by \citet{Pota2013}. We also mention this issue in the observed rotation of the red and blue GCs in the inner regions of NGC 5846 in Sec. A2.3.1.
    Therefore, we propose that NGC 5846 has likely formed through a series of multiple gas-poor minor mergers in recent times (i.e. $z<1$), which may have also contributed ex-situ GCs.
    
\section{Summary and conclusions}
\label{sec:conclusions}
In this work, we have studied the kinematic properties of $9$ S0 and E/S0 galaxies (including NGC 3115 from \citealt{Dolfi2020}) by combining the stars from ATLAS$^{\mathrm{3D}}$ and SLUGGS with the discrete PNe and GC tracers, in order to probe the outer regions of the galaxies out to $\sim4$-$6\, R_{\mathrm{e}}$.

From 2D kinematic maps (see Fig. A1-A8) and 1D kinematic profiles (see Fig. \ref{fig:1D_kinematic_profiles}), we find that:
\begin{itemize}
    \item For NGC 1023, NGC 2768, NGC 3115, NGC 3377, NGC 4697 and NGC 7457 (\textit{aligned} galaxies), the kinematics of their PNe and GC systems are well aligned with the kinematics of the stars in the overlapping radii. Overall, the rotation is also consistent with the photometric major-axis of the galaxies. This suggests that both the PNe and GC sub-populations are tracing the main underlying stellar population in these six galaxies out to large galactocentric radii;
    \item For NGC 821, NGC 4649 and NGC 5846 (\textit{mis-aligned} galaxies), the kinematics of their PNe and GC systems are not aligned with the kinematics of the stars in the overlapping radii. Specifically, the PNe and GCs show rotation more closely aligned with the photometric minor-axis of the galaxy in NGC 821, while the PNe and blue GCs show a kinematic twist in the major-axis of rotation beyond $\sim2\, R_{\mathrm{e}}$ in NGC 4649 and, finally, no clear rotation is detected from all kinematic tracers in NGC 5846. This suggests more complex formation histories for these three galaxies that have likely involved more numerous and more recent mergers.
\end{itemize}
From the comparison with the simulations of \citet{Schulze2020}, we suggest that four of the \textit{aligned} galaxies (i.e. NGC 1023, NGC 3115, NGC 3377 and NGC 7457) show the characteristic \textit{peaked} $V_{\mathrm{rot}}/\sigma$ profile shape, while the remaining two galaxies (i.e. NGC 2768 and NGC 4697) show the characteristic \textit{flat} $V_{\mathrm{rot}}/\sigma$ profile shape.
However, in Fig. \ref{fig:1D_stellar_Vsigma_vs_simulation}, we see that all six \textit{aligned} galaxies show similar kinematic behaviours at large radii, i.e. beyond $\sim2$-$3\, R_{\mathrm{e}}$, characterized by low $V_{\mathrm{rot}}/\sigma$.
Therefore, we suggest two main times in the assembly history of the six \textit{aligned} galaxies. At late ($z\lesssim1$) times, all six \textit{aligned} galaxies had similar accretion histories likely characterized by the accretion of dwarf galaxies in mini mergers that enhanced the velocity dispersion of the galaxies in the outer regions (i.e. beyond $\sim2$-$3\, R_{\mathrm{e}}$) without altering their central disk-like kinematics. 
However, from the differences in the $V_{\mathrm{rot}}/\sigma$ profiles of the galaxies in the central regions (i.e. within $\sim2$-$3\, R_{\mathrm{e}}$), we suggest that the two \textit{flat} galaxies had, possibly, experienced also a late ($z\lesssim1$), and relatively more gas-poor, minor merger that flattened the $V_{\mathrm{rot}}/\sigma$ profile in the central regions of the galaxy \citep{Schulze2020,Pulsoni2020}. On the other hand, we suggest that the four galaxies with \textit{peaked} $V_{\mathrm{rot}}/\sigma$ profiles had, possibly, experienced an early ($z>1$) gas-rich minor merger with some larger accreted gas fractions that preserved the amplitude of the $V_{\mathrm{rot}}/\sigma$ peak in the central regions of the galaxies \citep{Pulsoni2020}, or had evolved mostly passively through secular evolution as it seems to be the case for NGC 7457.

Among the four \textit{peaked} galaxies, we also note differences in the 1D kinematic profiles of their tracers (see Fig. \ref{fig:1D_kinematic_profiles}), therefore suggesting differences in their individual assembly histories that in some cases may, possibly, be a result of the specific orbital configuration of the mergers (see Sec. \ref{sec:peaked_galaxies}). 
Specifically, we suggest that the orbital configuration of the minor and mini merger events may have induced the formation of the bar in NGC 1023, i.e. a minor merger of mass-ratio 1:10 with closely aligned spin angles \citep{Cavanagh2020}, as well as the higher rotation velocity amplitude of the blue GCs than the red GCs at $\sim3\, R_{\mathrm{e}}$ in NGC 3115 \citep{Bekki2005}. In fact, the late accretion of low-mass dwarf galaxies along a preferential direction may have caused the more extended disk-like kinematics in NGC 3115 than in NGC 1023, i.e. mini mergers occurring along the disk plane of the galaxy \citep{Karademir2019}.
However, further simulations of GC kinematics that explore a wider range of parameters in the orbital configuration of the mergers would be required in order to explain these observed differences.

Finally, all three \textit{mis-aligned} galaxies had also overall similar assembly histories, characterized by late (i.e. $z<1$) mergers. These mergers were likely minor and occurred more than once in the case of NGC 821 and NGC 5846, therefore enhancing the velocity dispersion of the kinematic tracers at all radii. On the other hand, NGC 4649 is likely the result of a major merger (i.e. mass-ratio $>$1:4) that spun-up both the red and blue GC sub-populations at large radii, with minor and mini merger events that contributed to its population of ultra compact dwarfs (UCDs).

In this work, we find that the formation histories of S0 galaxies in low-density environments can be quite complex with merger events playing a crucial role in shaping the observed distinct $V_{\mathrm{rot}}/\sigma$ profiles of the different kinematic tracers in our galaxies out to large radii, depending mainly on their time of occurrence, specific orbital configuration and mass-ratio.

\section*{Acknowledgements}
We thank the anonymous referee for their very constructive comments and suggestions that helped improving this paper.
We thank the SLUGGS team for the collection, reduction and initial analysis of data from the SLUGGS survey used in this paper.
We would like to thank Adebusola Alabi for providing the spectroscopic PNe catalogues of some of the galaxies studied in this work. We thank Michael Drinkwater for useful discussions that have helped to improve this paper. DF, WC, KB and AD acknowledge support from the Australian Research Council under Discovery Project 170102344.
AFM has received financial support through the Postdoctoral Junior Leader Fellowship Programme from `La Caixa' Banking Foundation (LCF/BQ/LI18/11630007).
AJR was supported by National Science Foundation grant AST-1616710
and as a Research Corporation for Science Advancement Cottrell Scholar.

\section*{Data Availability}
Original Keck data are available from the Keck Observatory Archive (\url{https://www2.keck.hawaii.edu/koa/public/koa.php}).



\appendix

\section{Description of the individual galaxies}
\label{sec:notes_individual_galaxies}
In this section, we describe in more details the main properties and kinematic features of each individual galaxy listed in Table 1. 
As previously mentioned in Sec. 2, all galaxies listed in Table 1 show evidence of a rotating disk-like component from early studies. Apart from NGC 1023, NGC 3115 and NGC 7457 that are morphologically classified as S0s (see Table 1), NGC 821 shows evidence of a rapidly rotating stellar disk \citep{Emsellem2004,Proctor2009}, whose presence was also suggested from early photometric studies (e.g. \citealt{Michard1994,Goudfrooij1994,Ravindranath2001}) as well as from stellar population studies \citep{Proctor2005}. 
Similarly, evidence of a rotating stellar disk was also found in NGC 2768 (e.g. \citealt{Forbes2012}), NGC 3377 \citep{Halliday2001} and NGC 4697 \citep{Spiniello2015,Foster2016} from stellar kinematic or population studies. The presence of a disk is not clear in NGC 4649 \citep{Pota2015} or in NGC 5846, which has conflicting morphological classification (i.e. E0-1 in RC2 and S0 in RSA; \citealt{Michard1994}).
Therefore, apart from these two not clearly defined galaxies (i.e. NGC 4649 and NGC 5846), all our selected ETGs in Table 1 show evidence of a disk-like component, supporting our choice of including them in this current study of the assembly history of S0 galaxies.

\begin{figure*}
    \centering
        \includegraphics[width=1.0\textwidth]{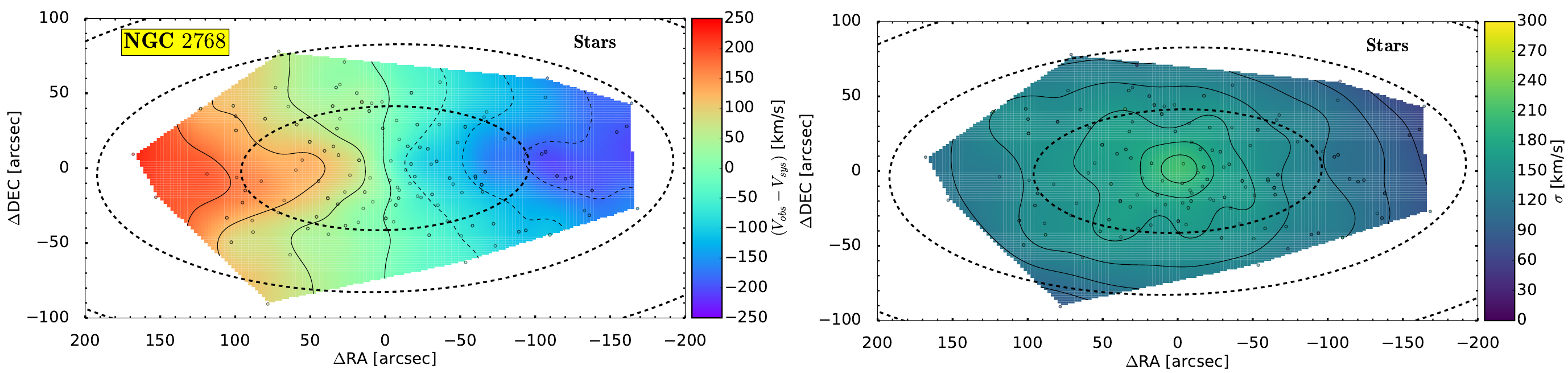}
        \includegraphics[width=1.0\textwidth]{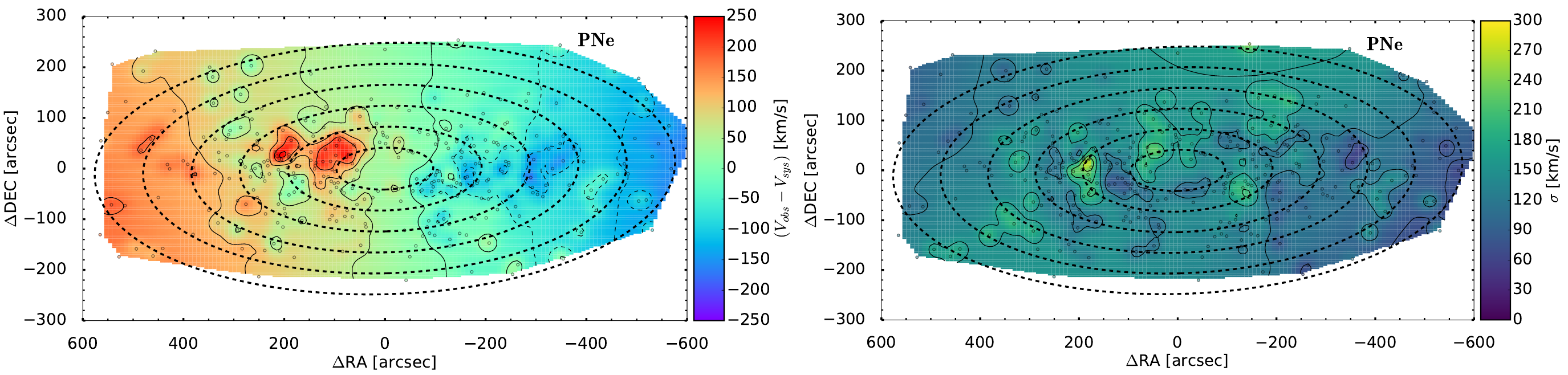}
        \includegraphics[width=1.0\textwidth]{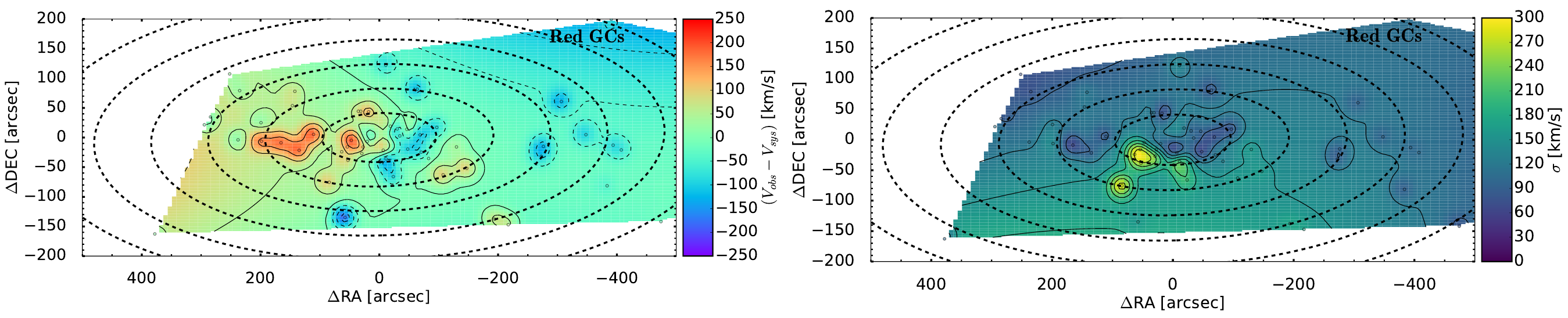}
        \includegraphics[width=1.0\textwidth]{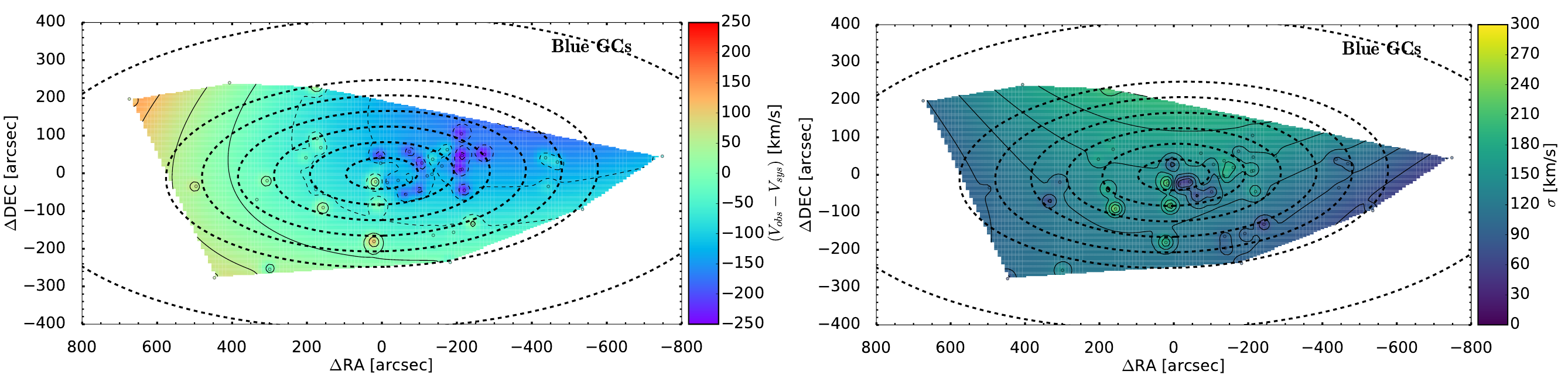}
        \caption{From top to bottom, 2D velocity (left-hand side) and velocity dispersion (right-hand side) maps of the SLUGGS stars, PNe, red and blue GCs for NGC 2768. The description is as in Fig. 2.}
    \label{fig:NGC2768_2d_kinematic_maps}
    \end{figure*}

\begin{figure*}
    \centering
        \includegraphics[width=0.85\textwidth]{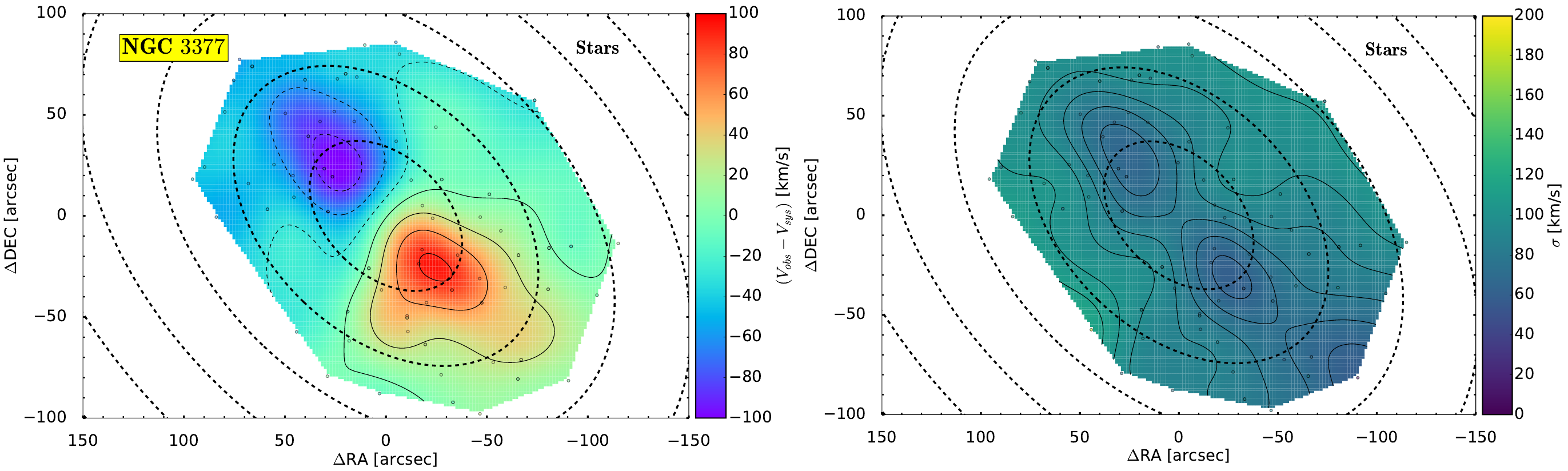}
        \includegraphics[width=0.85\textwidth]{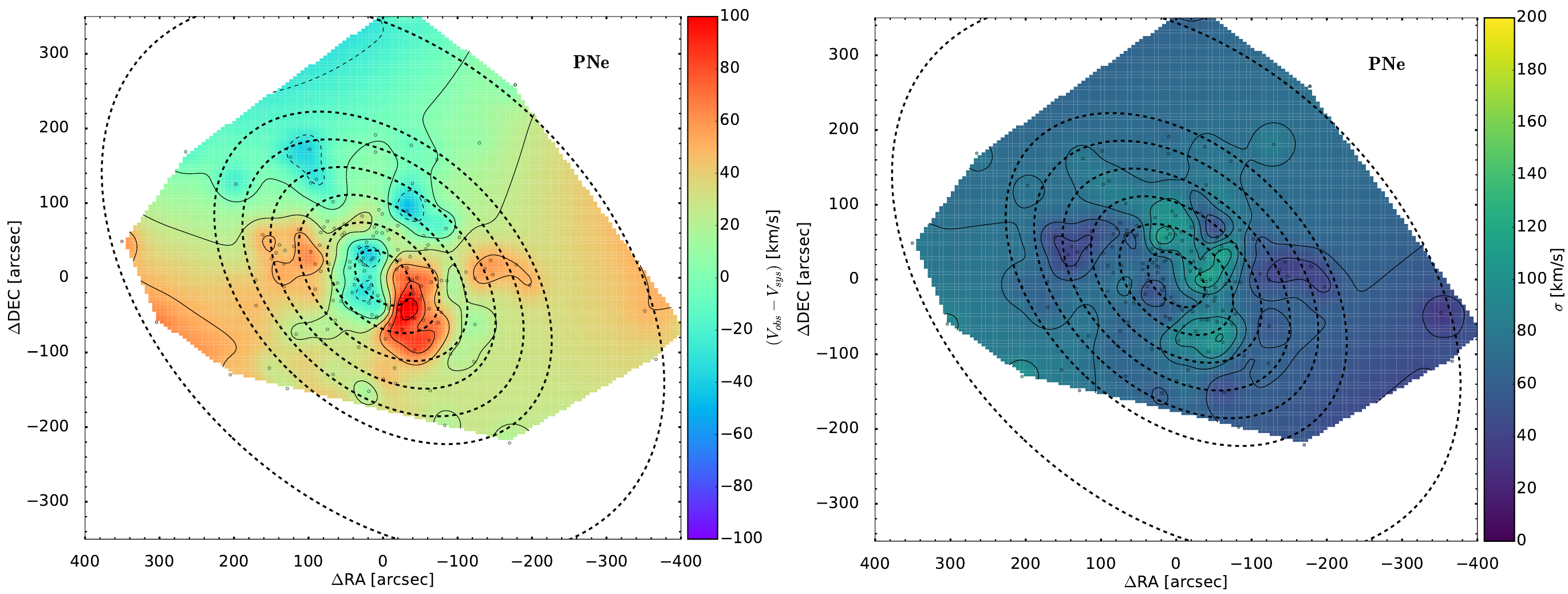}
        \includegraphics[width=0.85\textwidth]{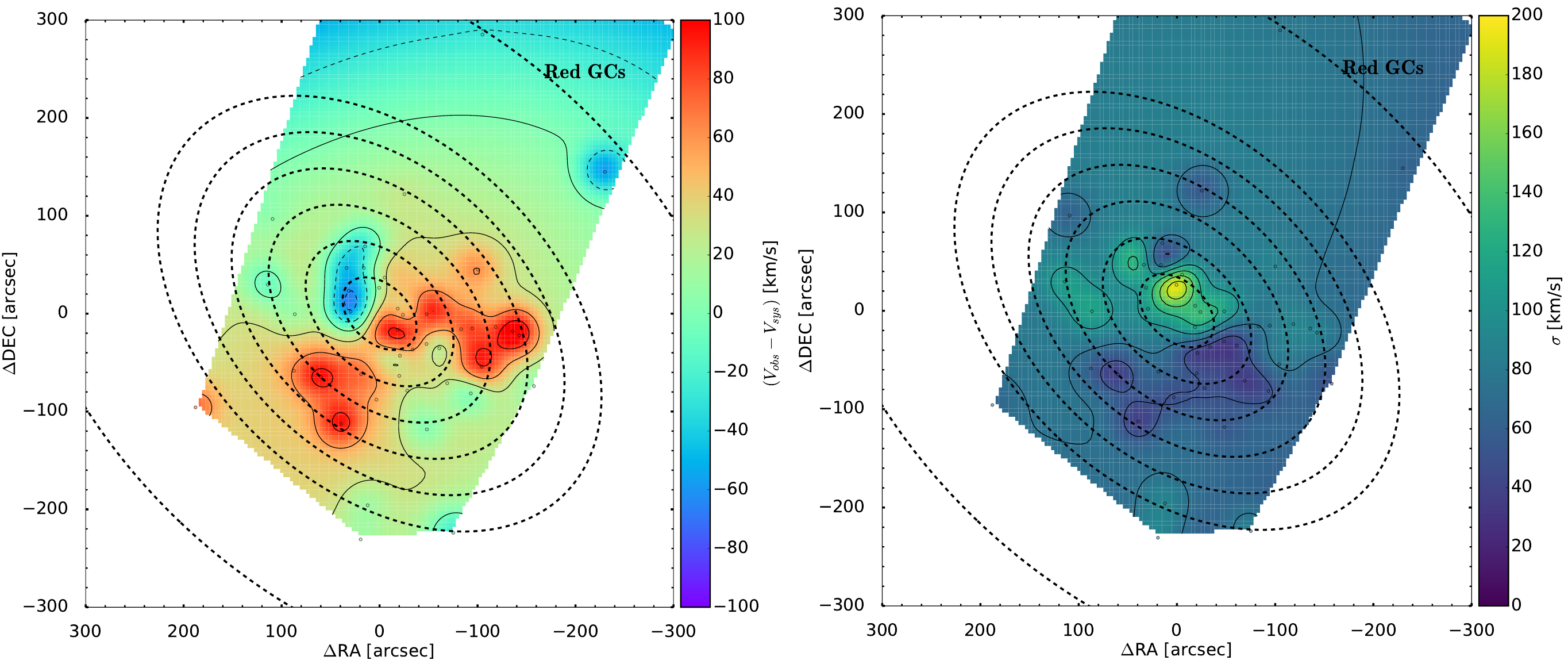}         \includegraphics[width=0.85\textwidth]{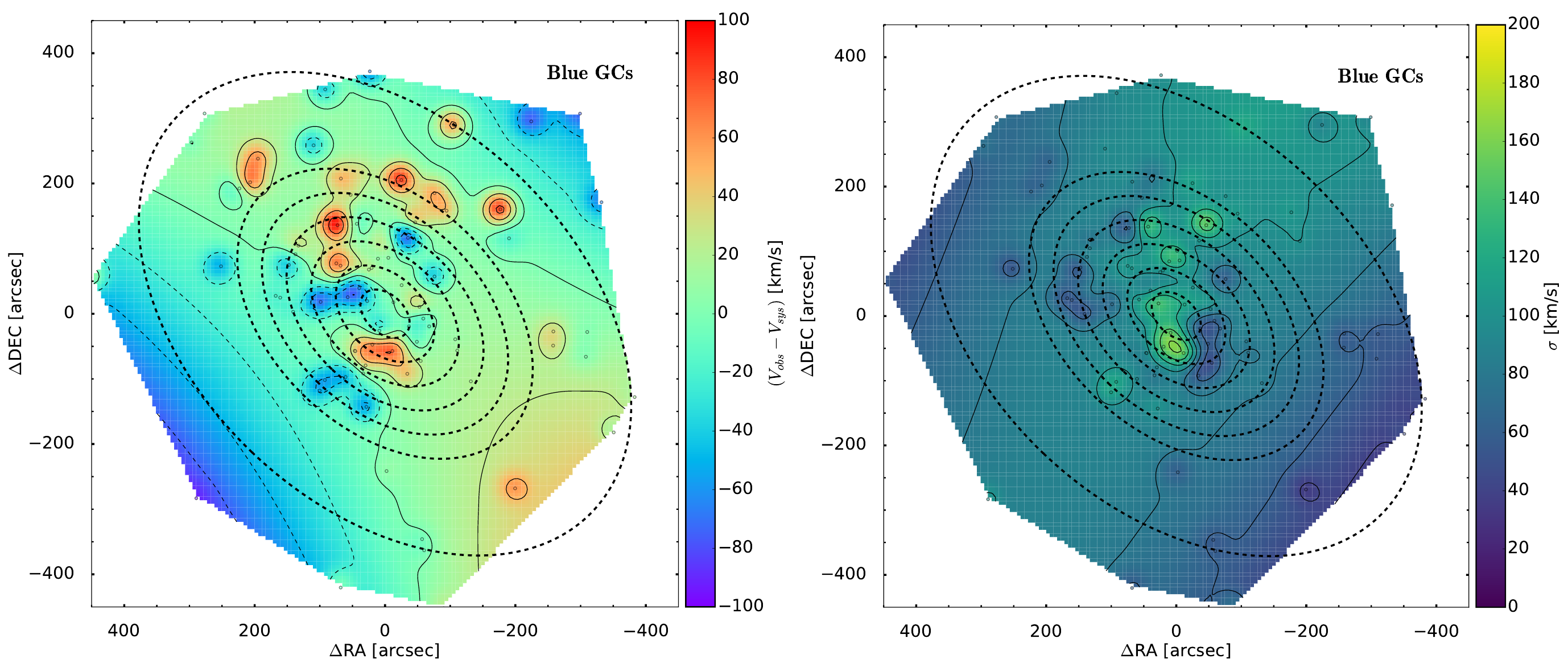}
        \caption{From top to bottom, 2D velocity (left-hand side) and velocity dispersion (right-hand side) maps of the SLUGGS stars, PNe, red and blue GCs for NGC 3377. The description is as in Fig. 2.}
    \label{fig:NGC3377_2d_kinematic_maps}
    \end{figure*}

\begin{figure*}
    \centering
        \includegraphics[width=1.0\textwidth]{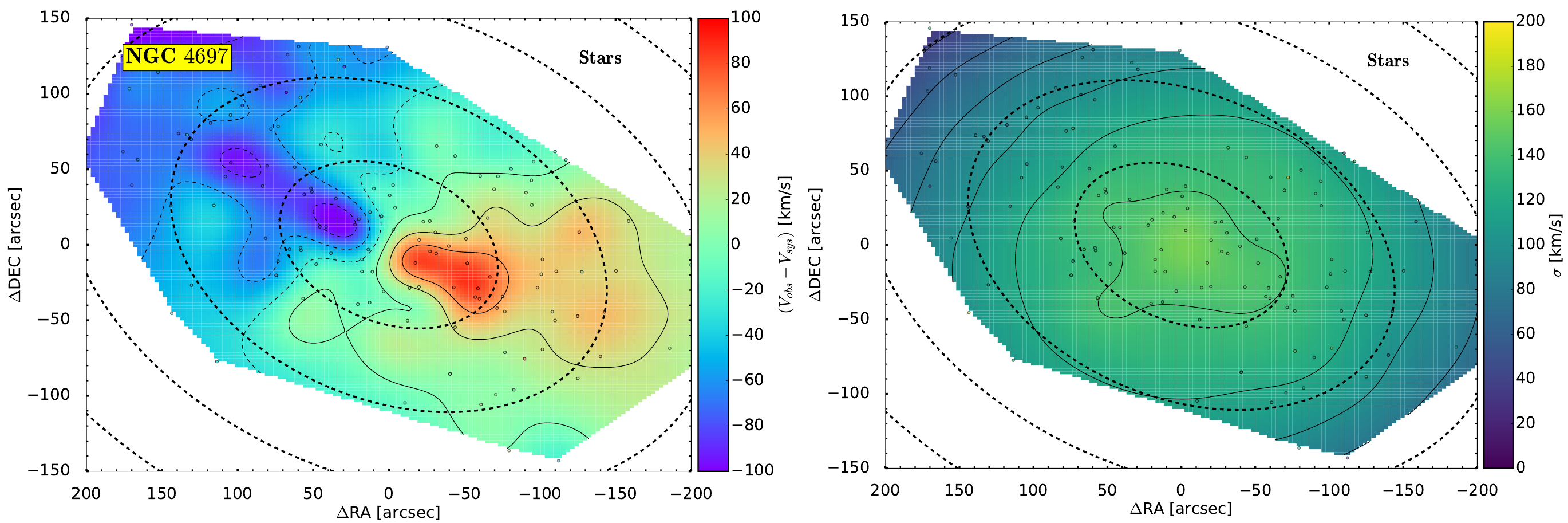}
        \includegraphics[width=1.0\textwidth]{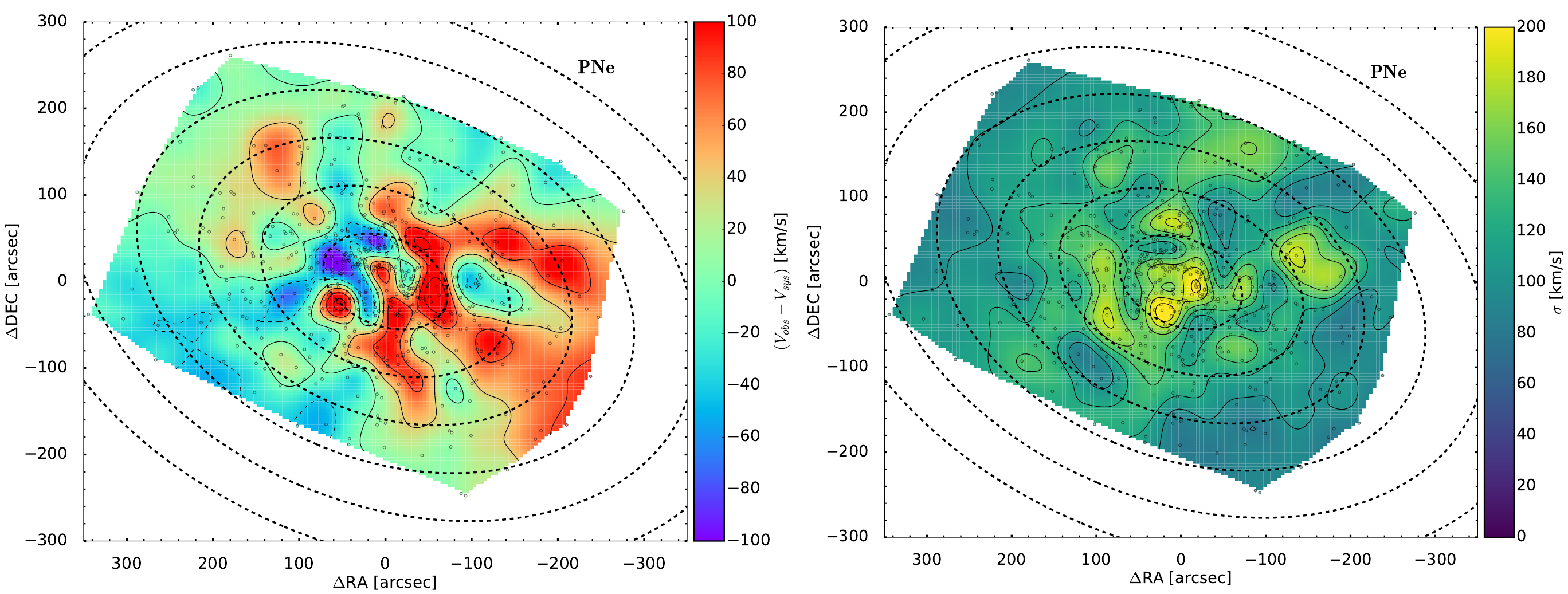}
        \includegraphics[width=1.0\textwidth]{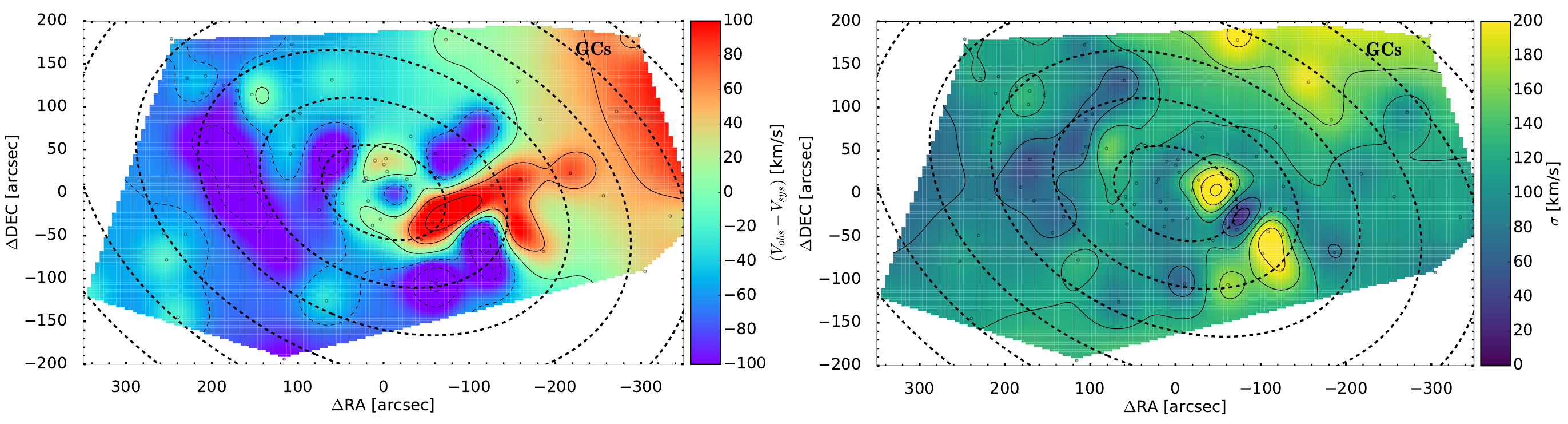}
        \caption{From top to bottom, 2D velocity (left-hand side) and velocity dispersion (right-hand side) maps of the SLUGGS stars, PNe and GCs for NGC 4697. The description is as in Fig. 2.}
    \label{fig:NGC4697_2d_kinematic_maps}
    \end{figure*}

\begin{figure*}
    \centering
        \includegraphics[width=1.0\textwidth]{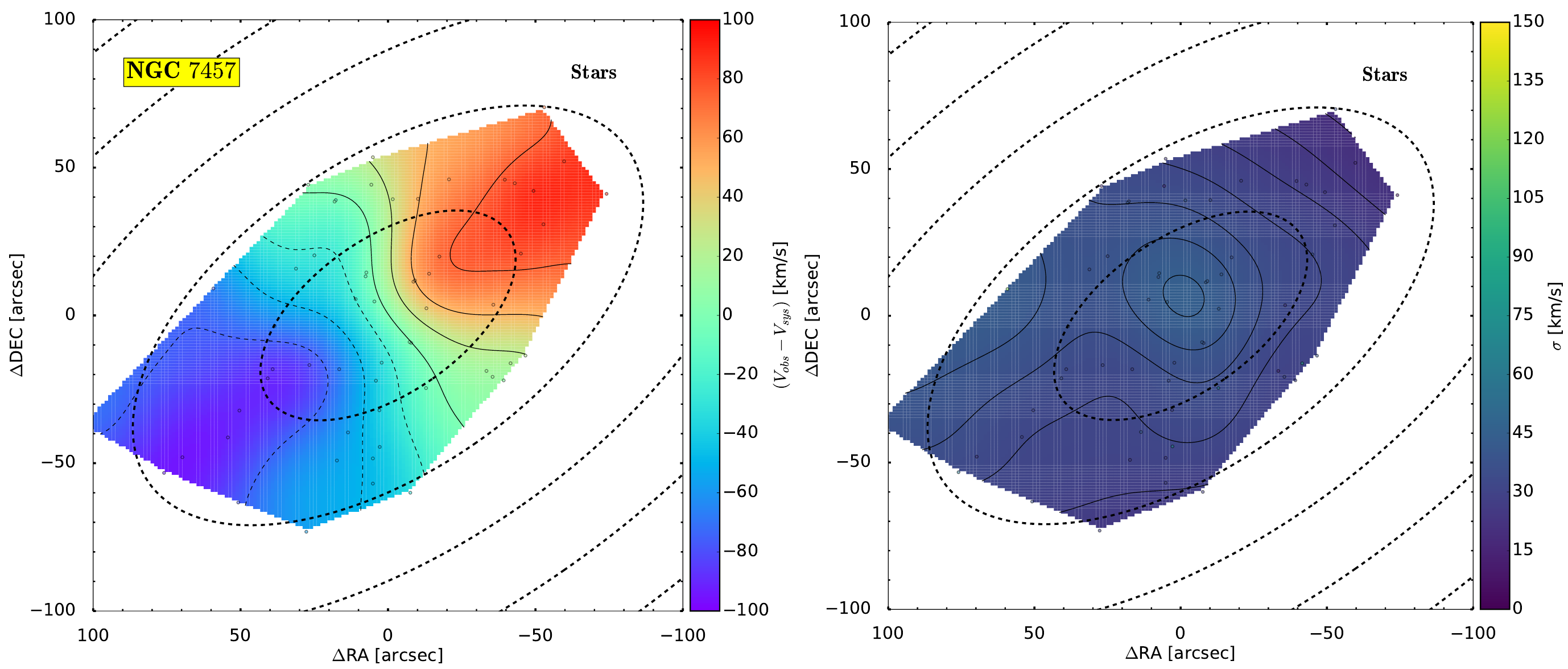}
        \includegraphics[width=1.0\textwidth]{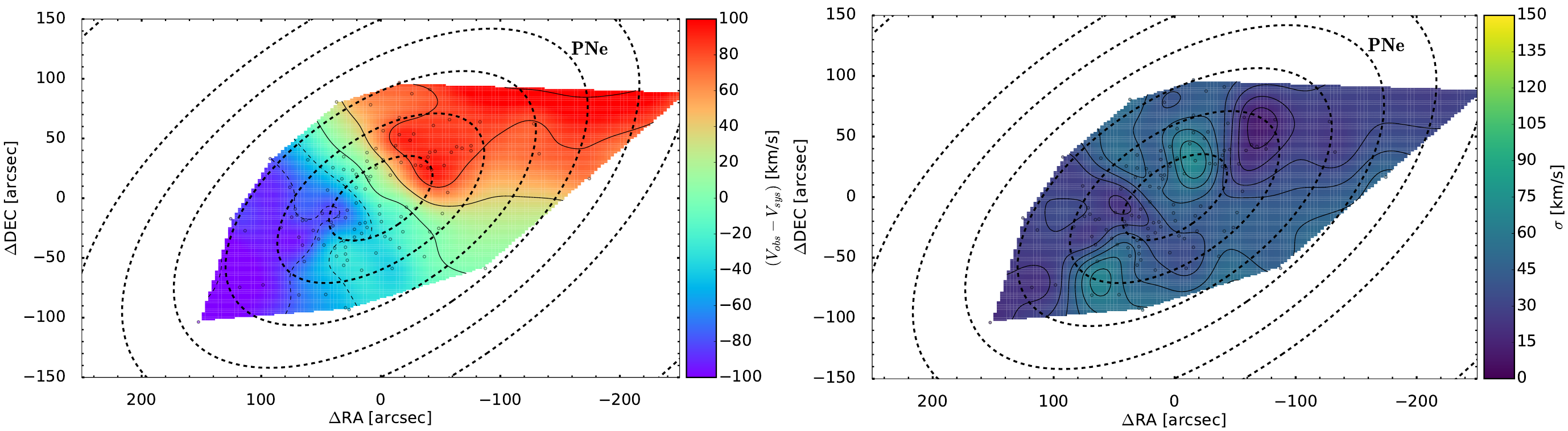}
        \includegraphics[width=1.0\textwidth]{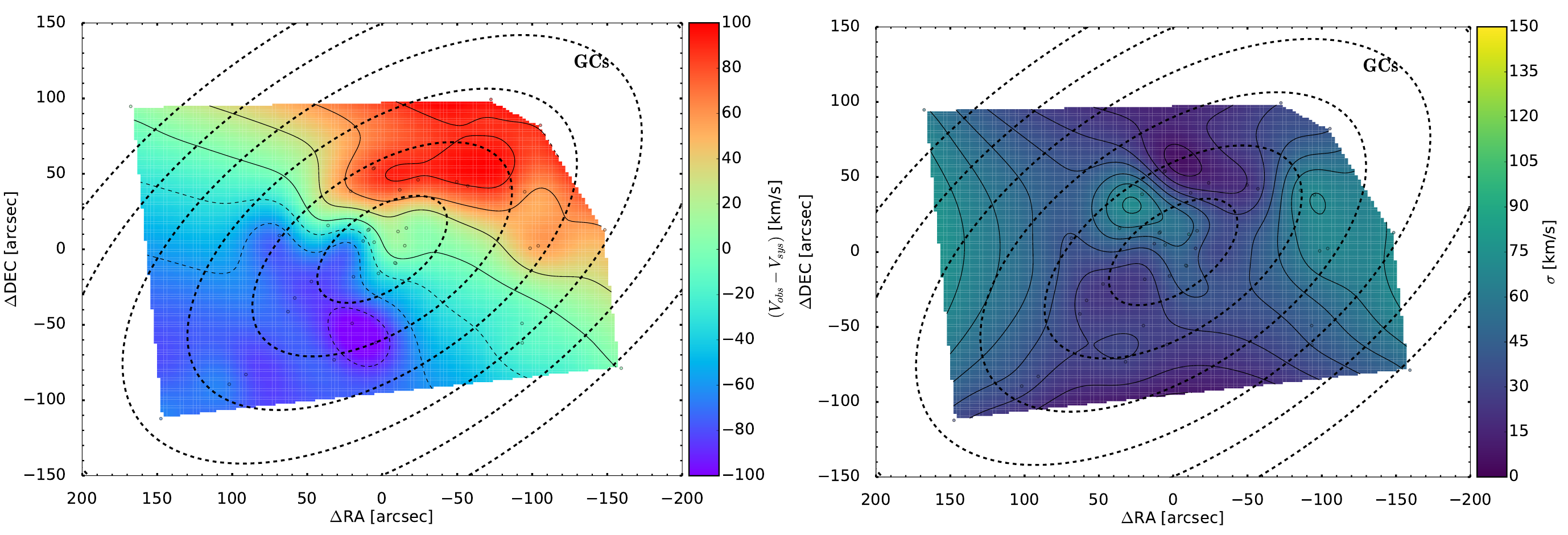}
        \caption{From top to bottom, 2D velocity (left-hand side) and velocity dispersion (right-hand side) maps of the SLUGGS stars, PNe and GCs for NGC 7457. The description is as in Fig. 2.}
    \label{fig:NGC7457_2d_kinematic_maps}
    \end{figure*}

\begin{figure*}
    \centering
        \includegraphics[width=1.0\textwidth]{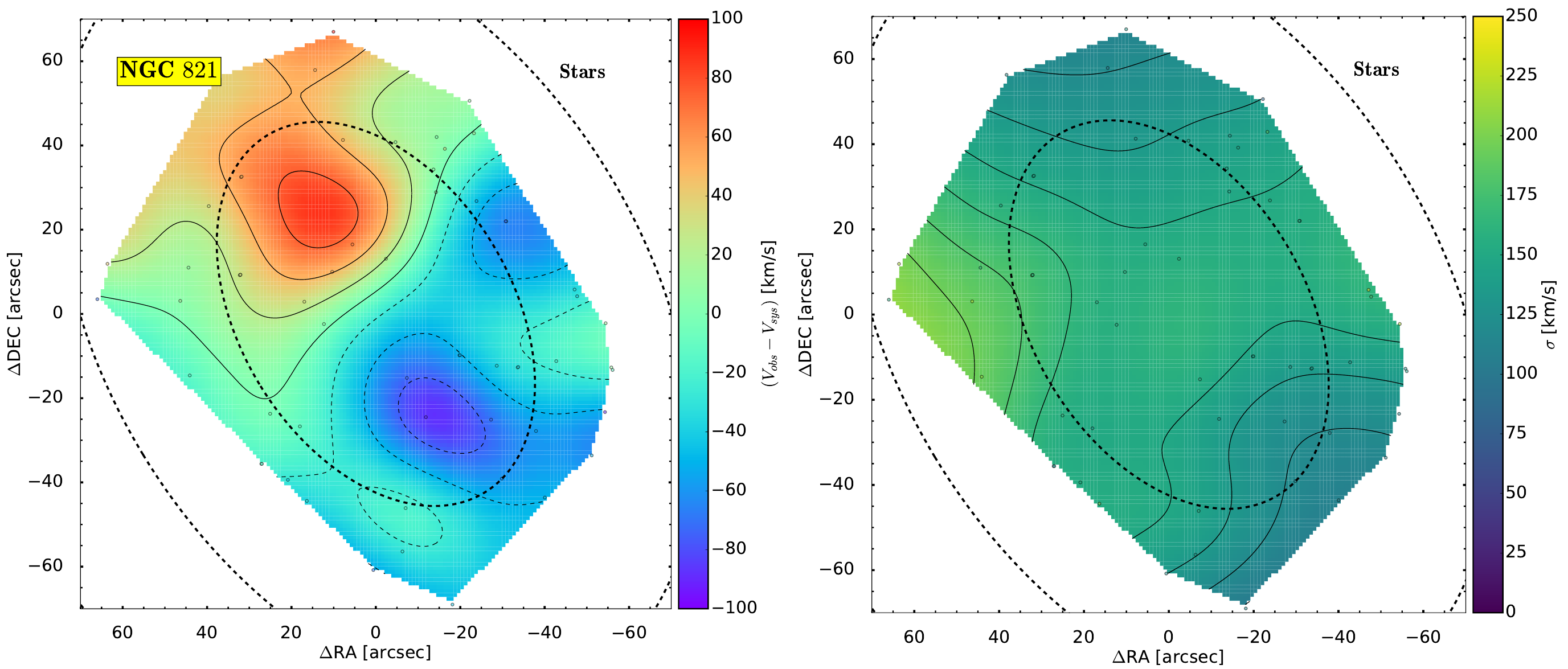}
        \includegraphics[width=1.0\textwidth]{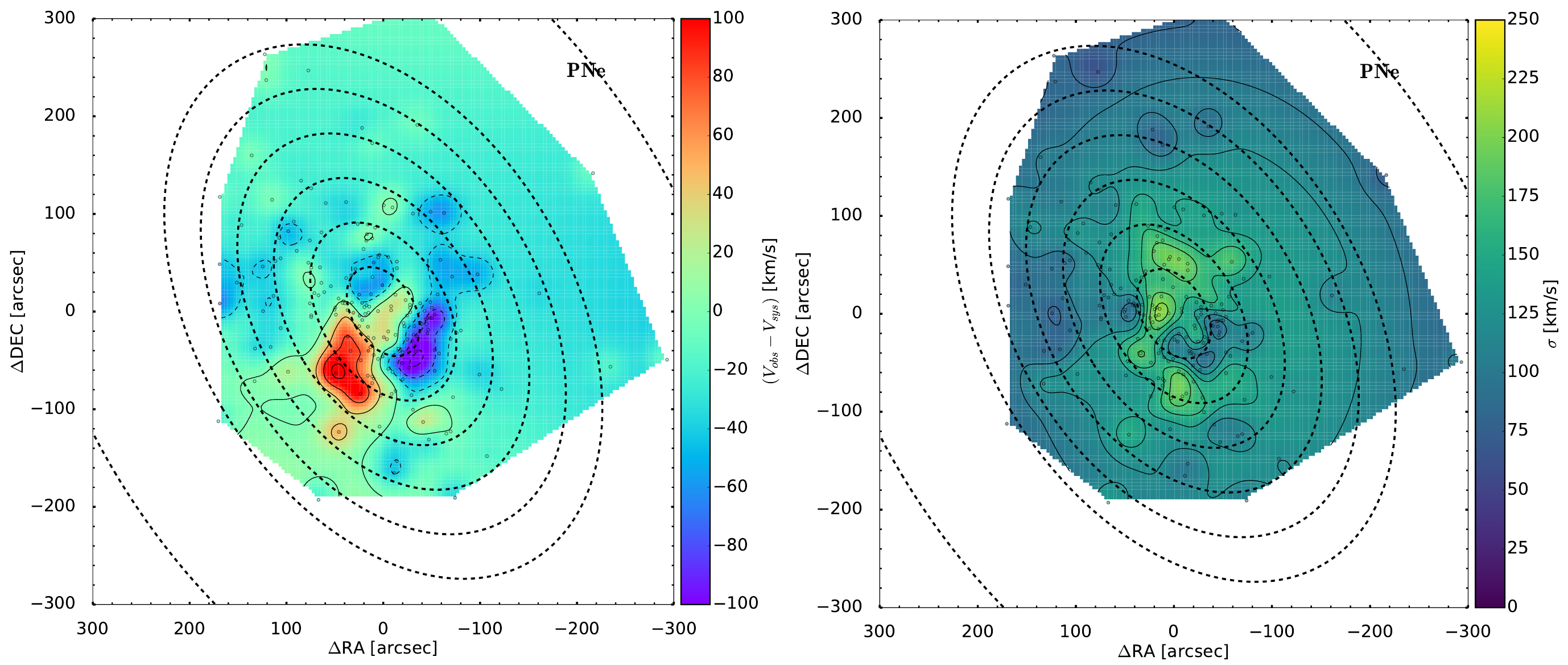}
        \includegraphics[width=1.0\textwidth]{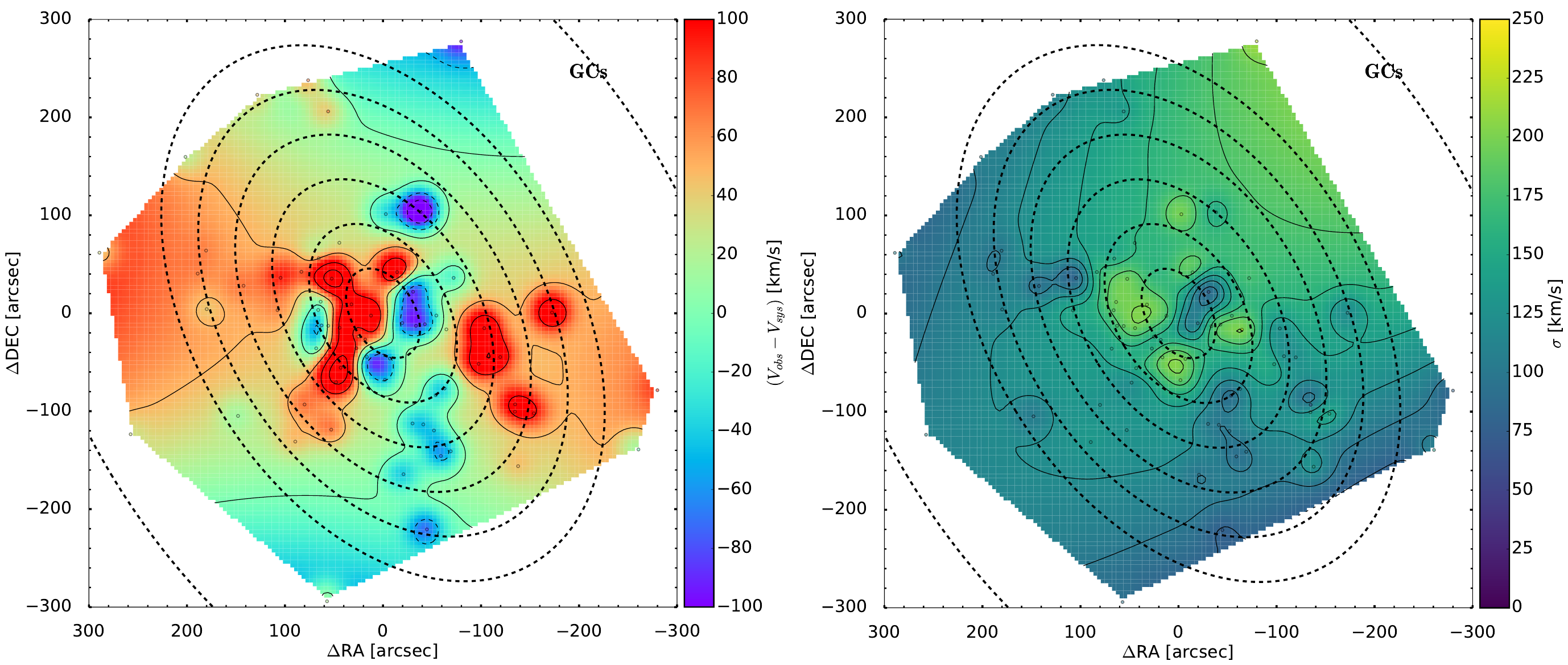}
        \caption{From top to bottom, 2D velocity (left-hand side) and velocity dispersion (right-hand side) maps of the SLUGGS stars, PNe and GCs for NGC 821. The description is as in Fig. 2.}
    \label{fig:NGC821_2d_kinematic_maps}
    \end{figure*}

The description of the kinematic results of all the distinct tracers (i.e. stars, PNe and GC sub-populations) for each galaxy is based on the 2D kinematic maps and 1D kinematic profiles shown in Fig. 2 and \ref{fig:NGC2768_2d_kinematic_maps}-\ref{fig:NGC5846_2d_kinematic_maps} and in Fig. 3, respectively, that are obtained as described in Sec. 3.1. We describe separately the \textit{aligned} and \textit{mis-aligned} galaxies introduced in Sec. 3.2.

\subsection{\textit{Aligned} galaxies}
\label{sec:aligned_galaxies}

    \subsubsection{NGC 1023}
    \label{sec:N1023_galaxy}
    NGC 1023 is the brightest member of a small galaxy group in Perseus (namely the NGC 1023 group), which is composed of $13$ bound galaxies \citep{Tully1980}. Specifically, NGC 1023 is classified as an SB0 galaxy from early photometric studies that confirmed the presence of a barred bulge (e.g. \citealt{Barbon1975}), which is found to rotate rapidly \citep{Kormendy1982}. 
    NGC 1023 has also a companion dwarf satellite galaxy, NGC 1023A, located South-East of its nucleus (see figure 2 of \citealt{Barbon1975}). The possible interaction between these two systems has been discussed in the literature as a potential explanation for the origin and peculiar distribution of the neutral hydrogen HI mass cloud observed around NGC 1023 \citep{Sancisi1984,Capaccioli1986}. 
    
    The GC system contains $110$ objects, six of which are likely associated with the nearby dwarf galaxy NGC 1023A \citep{Cortesi2016}. \citet{Cortesi2016} also removed some additional GCs that deviate strongly from the general rotation pattern of NGC 1023, however we choose to only remove two of these GCs (both are blue) that strongly counter-rotate at $\sim2\, R_{\mathrm{e}}$. The final GC catalogue contains $102$ objects that display a clear colour-bimodality distribution for $(g - z)=1.13$. Therefore, we split between the red, metal-rich and blue, metal-poor GC sub-populations that contain $41$ and $61$ objects, respectively.
    
    The PNe system is publicly available in \citet{Noordermeer2008} and it contains $203$ PNe, twenty of which are likely to belong to the nearby NGC 1023A dwarf galaxy \citep{Noordermeer2008}. After excluding this group of $20$ PNe, we obtain a final sample of $183$ PNe for NGC 1023. 
    
    \myparagraph{Kinematic profiles} 
    NGC 1023 is characterized by an increasing stellar rotation velocity profile from the innermost regions within $\sim0.5\, R_{\mathrm{e}}$ (ATLAS$^{\mathrm{3D}}$) out to $\sim1\, R_{\mathrm{e}}$, beyond which the rotation remains roughly constant out to $\sim2\, R_{\mathrm{e}}$. There is, possibly, evidence that the stellar rotation velocity starts decreasing beyond $\sim2\, R_{\mathrm{e}}$ (see Fig.3). The rotation velocity profile of the PNe shows excellent agreement with that of the stars in the overlapping radii between $\sim1$-$2.5\, R_{\mathrm{e}}$. At larger radii (i.e. beyond $\sim2\, R_{\mathrm{e}}$), the PNe rotation velocity decreases out to $\sim5\, R_{\mathrm{e}}$. 
    The red and blue GC sub-populations show decreasing rotation velocity profiles out to $\sim3.5\, R_{\mathrm{e}}$ and $\sim5.5\, R_{\mathrm{e}}$, respectively, which are overall consistent within the $1$-sigma errors. We see that the red GC rotation velocity profile is also consistent with both that of the stars and PNe, with similar amplitude at $\sim1.5\, R_{\mathrm{e}}$. On the other hand, the blue GC rotation velocity profile has somewhat smaller amplitude than the stars, PNe and red GCs at $\sim2\, R_{\mathrm{e}}$, while it shows good agreement with the PNe and red GCs at larger radii. 
    \citet{Cortesi2016} found that the blue GCs also display strong rotation velocity within the inner $\sim2.5\, R_{\mathrm{e}}$, with similar amplitude to that of the stars, PNe and red GCs. 
    We argue that the observed difference between our rotation velocity profile of the blue GCs with respect to that in \citet{Cortesi2016} at $\sim2\, R_{\mathrm{e}}$ may be the result of the removal of additional blue GCs deviating strongly from the regular rotation pattern of NGC 1023, as well as of the larger elliptical annuli used in \citet{Cortesi2016} than in this work to derive the corresponding 1D kinematic profiles. 
    
    The stellar velocity dispersion profile peaks at $\sim200\, \mathrm{kms^{-1}}$ within $\sim0.5\, R_{\mathrm{e}}$ (ATLAS$^{\mathrm{3D}}$) and, subsequently, steeply decreases out to $\sim1\, R_{\mathrm{e}}$, beyond which it remains roughly constant to $\lesssim100\, \mathrm{kms^{-1}}$ out to $\sim2\, R_{\mathrm{e}}$. The small region with very low ($\sim20\, \mathrm{kms^{-1}}$) velocity dispersion, seen at $\sim3\, R_{\mathrm{e}}$ towards South-East in Fig. 2, is likely associated with the presence of the companion galaxy, NGC 1023A, given the closely overlapping spatial locations (e.g. \citealt{Cortesi2016}).    
    The PNe show a weakly increasing velocity dispersion profile as a function of galactocentric radius, with higher values displayed at larger radii beyond $\sim2\, R_{\mathrm{e}}$. The red and blue GC sub-populations have similar increasing velocity dispersion profiles as a function of galactocentric radius, consistent with that of the PNe within the $1$-sigma errors. This increasing behaviour of the velocity dispersion profiles of the PNe, red and blue GCs was also previously observed in \citet{Cortesi2016}. 
    
    The stellar $V_{\mathrm{rot}}/\sigma$ profile rises steeply to $V_{\mathrm{rot}}/\sigma\sim3$ within the inner $\sim1\, R_{\mathrm{e}}$ and, then, it remains roughly flat out to $\sim2.5\, R_{\mathrm{e}}$. Similarly, the PNe show high values of the $V_{\mathrm{rot}}/\sigma\sim2$ between $\sim1$-$2\, R_{\mathrm{e}}$, consistent with the stars within the $1$-sigma errors. At larger radii, the $V_{\mathrm{rot}}/\sigma$ profile of the PNe decreases down to values $\lesssim1$ out to $\sim5\, R_{\mathrm{e}}$. The red and blue GCs show decreasing $V_{\mathrm{rot}}/\sigma$ profiles out to large radii, overall consistent with that of the PNe within the $1$-sigma errors. The red GCs show similar $V_{\mathrm{rot}}/\sigma\sim2$ amplitude to the PNe at $\sim1.5\, R_{\mathrm{e}}$, while the blue GCs have lower $V_{\mathrm{rot}}/\sigma\sim1$ than the PNe at $\sim2\, R_{\mathrm{e}}$, but consistent with the red GCs within the $1$-sigma errors.
    
    The $\mathrm{PA}_{\mathrm{kin}}$ profiles show good kinematic alignment between the stars, PNe, red and blue GC sub-populations, which are all consistent with rotation along the photometric major-axis of the galaxy at all radii within the $1$-sigma errors. The $\mathrm{q}_{\mathrm{kin}}$ profile of the stars shows that the inner regions within $\sim0.5\, R_{\mathrm{e}}$ are characterized by lower ellipticity (high $\mathrm{q}_{\mathrm{kin}}$) than the outer regions of the galaxy beyond $\sim0.5\, R_{\mathrm{e}}$, where the ellipticity is consistent with the photometric value out to $\sim2.5\, R_{\mathrm{e}}$.

    \subsubsection{NGC 2768}
    \label{sec:N2768_galaxy}
    NGC 2768 is classified as an E6/S0 group galaxy \citep{Sandage1994}. This galaxy is found to be characterized by red colours with a dust lane running along its photometric minor axis \citep{Kim1989}. Ionized gas has also been detected within $\sim20\arcsec$ from the central regions of the galaxy \citep{Kim1989}, and, interestingly, this gas is found to be kinematically misaligned with respect to the stars due to its rotation along the photometric minor-axis of the galaxy \citep{Wiklind1995,Sarzi2006}.
    
    The GC system contains $109$ objects, displaying a clear colour bimodality distribution for $(g-i)\sim1.02$ \citep{Pota2013}. Adopting this colour-split, we identify $37$ red, metal-rich and $44$ blue, metal-poor GCs.
    The PNe system is publicly available in \citet{Cortesi2013a} and it contains 315 PNe.

    The stellar, GC and PNe kinematics of NGC 2768 were previously studied by \citet{Forbes2012}, who found overall good agreement between the kinematic profiles of the distinct tracers and significant rotation for the disk component of the galaxy with rising $V_{\mathrm{rot}}/\sigma$ profile out to $\sim4\, R_{\mathrm{e}}$. A similar rising trend was also found for the $V_{\mathrm{rot}}$ profile of the PNe by \citet{Zanatta2018}. However, they found that both the red and blue GC sub-populations display lower amount of rotation, more consistent with spheroid-like kinematics. 
    
    \myparagraph{Kinematic profiles}
    NGC 2768 is characterized by an increasing stellar rotation velocity profile out to $\sim2\, R_{\mathrm{e}}$. We find a rotation amplitude offset of $\sim40$-$50\, \mathrm{kms^{-1}}$ between ATLAS$^{\mathrm{3D}}$ and SLUGGS at $\sim0.4\, R_{\mathrm{e}}$, with ATLAS$^{\mathrm{3D}}$ showing a steeper increase of the rotation velocity than SLUGGS. On the other hand, the stellar velocity dispersion profile does not show any offset at $\sim0.4\, R_{\mathrm{e}}$ and is continuously decreasing out to $\sim2\, R_{\mathrm{e}}$. 
    
    The PNe show rotation velocity consistent with the stars out to $\sim2\, R_{\mathrm{e}}$ along the photometric major-axis of the galaxy, with a peak in the rotation velocity between $\sim1.2$-$1.4\, R_{\mathrm{e}}$ and a subsequent small dip between $\sim2$-$3\, R_{\mathrm{e}}$. Beyond this radius, the rotation velocity profile of the PNe smoothly increases again out to $\sim6\, R_{\mathrm{e}}$.  
    Clear rotation is also seen for both the red and blue GC sub-populations between $\sim1$-$3\, R_{\mathrm{e}}$ along the photometric major-axis of the galaxy, similarly to the stars and PNe.
    The red GCs show a slightly decreasing rotation velocity profile out to $\sim3\, R_{\mathrm{e}}$, consistent with the PNe. The blue GCs also show a decreasing rotation velocity profile out to $\sim2.5\, R_{\mathrm{e}}$, beyond which it remains constant out to $\sim4\, R_{\mathrm{e}}$. 
    Overall, the PNe, red and blue GC rotation velocity profiles are consistent with each other within the $1$-sigma errors out to $\sim3\, R_{\mathrm{e}}$. They also show consistency with the rotation velocity of the stars at $\sim1.5\, R_{\mathrm{e}}$, but we are unable to confirm the kinematic agreement beyond this radius due to the limited extension of the SLUGGS stellar kinematic data.
    
    The velocity dispersion profiles of the PNe, red and blue GCs are consistent with each other within the $1$-sigma errors and show smoothly decreasing trends out to large radii along the photometric major-axis of the galaxy. From the 2D kinematic maps in Fig. A1, we see that the velocity dispersion of the PNe, red and blue GCs show higher values along the photometric minor-axis of the galaxy. They also show consistency with the stellar velocity dispersion profile at $\sim1.5\, R_{\mathrm{e}}$, but, as for the rotation velocity, we are unable to confirm this at larger radii. 
    
    The stellar $V_{\mathrm{rot}}/\sigma$ profile rises up to $\sim1.5$ within $2\, R_{\mathrm{e}}$. The PNe and red GCs show similar $V_{\mathrm{rot}}/\sigma\sim1$ profiles, which remains roughly flat and are consistent with the stars at $\sim1.5\, R_{\mathrm{e}}$. The blue GCs show a decreasing $V_{\mathrm{rot}}/\sigma$ profile out to $\sim2.5\, R_{\mathrm{e}}$, beyond which it remains overall flat out to large radii and consistent with the $V_{\mathrm{rot}}/\sigma$ profiles of both the PNe and red GCs within the $1$-sigma errors.
    
    The $\mathrm{PA}_{\mathrm{kin}}$ profiles show that the stars, PNe, red and blue GCs are kinematically aligned and are rotating along the photometric major-axis of the galaxy at all radii. The $\mathrm{q}_{\mathrm{kin}}$ profile of the stars is consistent with the photometric value within the inner $\sim0.5\, R_{\mathrm{e}}$, while it decreases out to $\sim2\, R_{\mathrm{e}}$. The $\mathrm{q}_{\mathrm{kin}}$ profile of the red GCs is consistent with that of the stars between $\sim1$-$2\, R_{\mathrm{e}}$ and it increases up to the photometric value at larger radii.
    However, we note that the $\mathrm{q}_{\mathrm{kin}}$ profile of the red GCs is overall consistent with the photometric value at all radii within the $1$-sigma errors.
    
    \subsubsection{NGC 3377}
    \label{sec:N3377_galaxy}
    NGC 3377 is a bright elliptical galaxy of E5-6 morphological type \citep{Sandage1984}, characterized by a small disk embedded in a boxy halo \citep{Halliday2001}. This galaxy is a member of the Leo I group with the smaller NGC 3377A being its nearest companion galaxy.
    
    The GC system contains $126$ objects, characterized by a bimodal colour distribution as was already found by \citet{Pota2013}. We study the kinematics of $46$ red, metal-rich and $80$ blue, metal-poor GCs adopting the colour-split $(g-i)=0.925$.
    
    The PNe system contains $152$ objects. The PNe kinematics for this galaxy were initially studied by \citet{Coccato2009}, who found significant rotation in the inner regions of the galaxy, i.e. within $\sim 100\arcsec$, which is characterized by a twist in the kinematic position angle ($\mathrm{PA}_{\mathrm{kin}}$) between $60\arcsec$ and $200\arcsec$.
    
    \myparagraph{Kinematic profiles}
    NGC 3377 shows roughly constant ($\sim80$-$100\, \mathrm{kms^{-1}}$) stellar rotation velocity out to $\sim1\, R_{\mathrm{e}}$, followed by a subsequent decrease out to $\sim2$-$2.5\, R_{\mathrm{e}}$. ATLAS$^{\mathrm{3D}}$ has shown that the stellar rotation velocity is also high in the central $\sim0.5\, R_{\mathrm{e}}$, consistent with the SLUGGS stellar kinematics. The PNe also show a decreasing rotation velocity profile out to $\sim2$-$3\, R_{\mathrm{e}}$, which is overall consistent with that of the stars within $\sim1$-$2\, R_{\mathrm{e}}$ despite an offset of $\sim40$-$50\,\mathrm{kms^{-1}}$ at $\sim1\, R_{\mathrm{e}}$. At larger radii, i.e. beyond $\sim3\, R_{\mathrm{e}}$, the rotation velocity of the PNe shows evidence of slightly increasing again towards North (see Fig. A2), where the rotation, possibly, occurs along a different $\mathrm{PA}_{\mathrm{kin}}$ than that of the inner $\sim2\, R_{\mathrm{e}}$ (see Fig. 3).
    The red GCs show decreasing rotation velocity profile out to $\sim3\, R_{\mathrm{e}}$, which is consistent with both that of the stars and PNe in the overlapping radii. Some red GCs seem to be rotating along the galaxy photometric minor-axis between $\sim3$-$4\, R_{\mathrm{e}}$ towards South-East (see Fig. A2). On the other hand, the blue GCs do not show evidence of any clear rotation at all radii.
    
    The stellar velocity dispersion profile slightly increases from $\sim80$ to $\sim100\, \mathrm{kms^{-1}}$ between $\sim1$-$2.5\, R_{\mathrm{e}}$. This behaviour is likely mimicking the increase of the stellar velocity dispersion at large radii, moving away from the photometric major-axis of the galaxy where the rotating disk-like structure is observed, as seen from the 2D kinematic map in Fig. A2 and also previously shown by \citet{Foster2016}. ATLAS$^{\mathrm{3D}}$ also show consistent results, with lower velocity dispersion along the photometric major-axis of the galaxy. However, ATLAS$^{\mathrm{3D}}$ shows higher velocity dispersion within $\sim0.5\, R_{\mathrm{e}}$ with an offset of $\sim20\, \mathrm{kms^{-1}}$ from SLUGGS at $\sim1\, R_{\mathrm{e}}$. The PNe and red GCs show very consistent, and decreasing, velocity dispersion profiles between $\sim2$-$3\, R_{\mathrm{e}}$. At larger radii, the velocity dispersion profile of the PNe remains roughly constant out to $\sim6\, R_{\mathrm{e}}$. The blue GCs also show decreasing velocity dispersion profile between $\sim2$-$3\, R_{\mathrm{e}}$, consistent with the PNe and red GCs, beyond which it remains roughly constant out to $\sim6\, R_{\mathrm{e}}$. The blue GCs show higher velocity dispersion along the photometric minor-axis of the galaxy, similarly to the stars (see Fig. A2). Overall, the velocity dispersion profiles of the PNe, red and blue GCs also show consistency with that of the stars in the overlapping radii at $\sim2\, R_{\mathrm{e}}$, but we are unable to confirm such agreement at larger radii due to the limited extension of the stellar kinematic data.
    
    The stellar $V_{\mathrm{rot}}/\sigma$ profile steeply decreases between $\sim1$-$2\, R_{\mathrm{e}}$. The PNe show a similar decreasing $V_{\mathrm{rot}}/\sigma$ profile out to large radii, i.e. $\sim5\, R_{\mathrm{e}}$, which is consistent with that of the stars in the overlapping radii within $\sim2\, R_{\mathrm{e}}$. The red GCs also show a smoothly decreasing $V_{\mathrm{rot}}/\sigma$ profile, consistent with both that of the stars and PNe. On the other hand, the blue GCs are not rotating, with $V_{\mathrm{rot}}/\sigma<0.5$ at all radii.
    
    The $\mathrm{PA}_{\mathrm{kin}}$ profile of the PNe shows that the rotation is overall kinematically aligned with that of the stars along the photometric major-axis of the galaxy within $\sim3\, R_{\mathrm{e}}$. At larger radii, i.e. between $\sim3$-$5\, R_{\mathrm{e}}$, we see a twist of the $\mathrm{PA}_{\mathrm{kin}}$, which shows rotation along an $\sim180\degr$ kinematic position angle. 
    We have seen this, possible, twist of the kinematic axis of rotation of the PNe beyond $\sim3\, R_{\mathrm{e}}$ in the 2D kinematic maps of Fig. A2 and it was also previously shown in \citet{Coccato2009}. The $\mathrm{q}_{\mathrm{kin}}$ profile of the stars is roughly constant at all radii and consistent with the photometric value.

    \subsubsection{NGC 4697}
    \label{sec:N4697_galaxy}
    NGC 4697 is a disky elliptical galaxy classified as E6 \citep{Michard1994}, which belongs to a galaxy group located in the Virgo Southern Extension \citep{Blakeslee2009}. Stellar kinematic studies have classified this galaxy as a fast-rotator (e.g. \citealt{Spiniello2015,Foster2016}), while stellar population analyses have found an older and metal-poorer stellar population in the bulge than in the disk \citep{Spiniello2015}.

    The GC system of NGC 4697 contains $86$ spectroscopically confirmed objects. We do not find evidence of any clear colour bimodality in the GC population, therefore we do not split between the two red, metal-rich and blue, metal-poor GC sub-populations for studying the GC kinematics of this galaxy.
    
    The PNe system of NGC 4697 is publicly available in \citet{Mendez2001,Mendez2008} and it contains a total of $531$ objects. The kinematics of the PNe system were studied by \citet{Coccato2009}, who found evidence of rotation within $\sim1\, R_{\mathrm{e}}$ along the photometric major-axis consistent with the stars.
    In a previous work, \citet{Sambhus2006} studied a PNe sub-sample containing $320$ objects, defined by all PNe with apparent magnitudes brighter than $\mathrm{[OIII]}(\lambda=5007\angstrom) = 27.6\, \mathrm{mag}$ outside a central elliptical region with semi-major axis length $\mathrm{a} = 60\arcsec$ \citep{Mendez2001}. They found a correlation between the PNe velocities and their $\mathrm{[OIII]}(\lambda=5007\angstrom)$ magnitudes, with faint PNe displaying co-rotation and bright PNe (i.e. $\mathrm{[OIII]}(\lambda=5007\angstrom) < 26.2\, \mathrm{mag}$) counter-rotation.
    
    Therefore, \citet{Sambhus2006} studied separately the faint (i.e. $\mathrm{[OIII]}(\lambda=5007\angstrom) > 26.9\, \mathrm{mag}$), intermediate (i.e. $26.2\, \mathrm{mag} < \mathrm{[OIII]}(\lambda=5007\angstrom) < 26.9\, \mathrm{mag}$) and bright (i.e. $\mathrm{[OIII]}(\lambda=5007\angstrom) < 26.2\, \mathrm{mag}$) PNe sub-populations, where the magnitude cuts were chosen to have equal-width bins with $\Delta \mathrm{[OIII]}(\lambda=5007\angstrom) = 0.7\, \mathrm{mag}$. 
    The azimuthal distributions of the bright and faint PNe sub-populations were only found to have $2\%$ probabilities of being consistent, while the intermediate and faint PNe were more consistent with a $28\%$ probability and the intermediate and bright PNe were only consistent with a $7\%$ probability from the Kolmogorov-Smirnov (KS) test.
    
    We consider the PNe catalogue of $526$ objects with both radial velocity and magnitude measurements, and we compare the PNe velocities with respect to the underlying 2D stellar velocity field. For the comparison, we use the 2D stellar kinematics from the SLUGGS survey \citep{Arnold2014,Foster2016}. The aim is to test whether the PNe show a significantly different kinematic behaviour with respect to the underlying stars and, if that is the case, whether such different behaviour shows any dependance as a function of the PNe apparent magnitudes, $\mathrm{[OIII]}(\lambda=5007\angstrom)$. 
    
    Therefore, we calculate the PNe velocities and velocity dispersion relative to the underlying 2D stellar velocity field at the PNe location, i.e. $(V_{\mathrm{PNe}} - V_{\mathrm{stars}})$ and $(\sigma_{\mathrm{PNe}} - \sigma_{\mathrm{stars}})$, as a function of the apparent PNe magnitudes, $\mathrm{[OIII]}(\lambda=5007\angstrom)$. Here, $V_{\mathrm{PNe}}$ refers to the PNe velocity already subtracted by the $V_{\mathrm{sys}}$ of NGC 4697 and the velocity dispersion is estimated at the location of each PNe from the standard deviation of the PNe velocities in bins containing the $10$ nearest objects. We use all $526$ PNe in the catalogue and we calculate the average velocities and velocity dispersion in magnitude bins containing $30$ objects.
    
    We find that there is a large scatter of the PNe velocities and velocity dispersion relatively to the underlying 2D stellar velocity and velocity dispersion fields. However we do not find evidence of a strong trend of the PNe velocities and velocity dispersion as a function of the PNe magnitudes, $\mathrm{[OIII]}(\lambda=5007\angstrom)$, which are overall consistent with the underlying stellar kinematics. Therefore, based on these results, we choose not to split the PNe catalogue of NGC 4697 into the magnitude bins adopted in \citet{Sambhus2006}, but we will study the kinematics of the PNe population as a whole.
    
    \myparagraph{Kinematic profiles}
    NGC 4697 shows a smoothly decreasing stellar rotation velocity profile out to $\sim2.5\, R_{\mathrm{e}}$. The ATLAS$^{\mathrm{3D}}$ and SLUGGS stellar rotation velocity profiles are consistent with each other, despite the offset in rotation of $\sim20\, \mathrm{kms^{-1}}$ at $\sim0.5\, R_{\mathrm{e}}$. The PNe and GCs show consistent rotation velocity profiles, with a peak of rotation occurring at $\sim1\, R_{\mathrm{e}}$ and a subsequent sharp decrease out to $\sim1.5$-$2\, R_{\mathrm{e}}$, beyond which the rotation velocity remains roughly constant out to $\sim3.5\, R_{\mathrm{e}}$. Some GCs show evidence of counter-rotation at $\sim2\, R_{\mathrm{e}}$ towards South-West as seen in Fig. A3.
    
    The stellar velocity dispersion smoothly decreases out to $\sim2.5\, R_{\mathrm{e}}$ along the photometric major-axis of the galaxy, while it remains roughly constant along the photometric minor-axis of the galaxy (see Fig. A3). The PNe show a similar decreasing velocity dispersion behaviour, as the stars, out to $\sim3.5\, R_{\mathrm{e}}$ along the photometric major-axis of the galaxy. On the other hand, the GCs have a roughly constant velocity dispersion profile at all radii, whose amplitude is consistent with that of the stars and PNe beyond $\sim2\, R_{\mathrm{e}}$. 
    
    The stellar $V_{\mathrm{rot}}/\sigma\sim0.5$ profile remains roughly flat out to large radii, i.e. $\sim2.5\, R_{\mathrm{e}}$. The PNe and GCs have consistent $V_{\mathrm{rot}}/\sigma$ profiles, with a peak at $\sim1\, R_{\mathrm{e}}$ and a subsequent sharp decrease out to large radii, similarly to the behaviour of the corresponding $V_{\mathrm{rot}}$ profiles.
    
    The $\mathrm{PA}_{\mathrm{kin}}$ profiles show good kinematic alignment between the stars and PNe out to $\sim2.5\, R_{\mathrm{e}}$, with the rotation occurring along the photometric major-axis of the galaxy. Beyond $\sim2.5\, R_{\mathrm{e}}$, the PNe, possibly, show some kinematic misalignment from the photometric major-axis, but they are overall consistent with it within the $1$-sigma errors. Additionally, we do not clearly see evidence of a possible kinematic misalignment beyond $\sim2.5\, R_{\mathrm{e}}$ from the 2D kinematic maps in Fig. A3. The $\mathrm{q}_{\mathrm{kin}}$ profile of the stars is roughly constant at all radii out to $\sim2.5\, R_{\mathrm{e}}$ and lower than the photometric value, implying higher flattening for the stars.

    \subsubsection{NGC 7457}
    \label{sec:N7457_galaxy}
    NGC 7457 is a disk-dominated S0 galaxy found in isolation with the possible evidence of a bar-like structure \citep{Michard1994}. Early studies have identified the presence of a small, compact nucleus within the innermost galaxy regions with a young ($\sim2$-$2.5\, \mathrm{Gyr}$ old) stellar population, which displays counter-rotation with respect to the galaxy bulge and disk \citep{Silchenko2002}. Evidence for a kinematically distinct core (KDC) characterized by young stellar ages in NGC 7457 has been also found by \citet{Kuntschner2006,Kuntschner2010}, who measured a strong $\mathrm{H}_{\beta}$ absorption.
    
    The GC system of NGC 7457 contains $40$ objects without evidence of any color bi-modality distribution. Therefore, we do not adopt any colour-split to separate between the two red, metal-rich and blue, metal-poor GC sub-populations.
    
    The PNe system of NGC 7457 contains $113$ objects. The combined kinematics of both the PNe and GC systems of NGC 7457 were previously studied by \citet{Zanatta2018}, who found this galaxy to be rotationally supported with $V_{\mathrm{rot}}/\sigma > 2$ at all radii.
    
    \myparagraph{Kinematic profiles}
    NGC 7457 shows an increasing stellar rotation velocity profile out to $\sim2\, R_{\mathrm{e}}$. Such behaviour is observed from both ATLAS$^{\mathrm{3D}}$ (within $\sim0.5\, R_{\mathrm{e}}$) and SLUGGS, which show good agreement in the rotation velocity. Both PNe and GCs also show a rising rotation velocity profile out to $\sim2\, R_{\mathrm{e}}$, very consistent with the stars. Beyond $\sim2\, R_{\mathrm{e}}$, the rotation velocity profile of the GCs shows evidence of starting to decrease (see Fig. 3).
    
    The stellar velocity dispersion from SLUGGS is very low (i.e. $\sim30$-$45\, \mathrm{kms^{-1}}$) at all radii with, possibly, a mild central peak of $\sim45$-$50\, \mathrm{kms^{-1}}$ (see Fig. A4), as previously shown by \citet{Foster2016} and \citet{Bellstedt2017}. The stellar velocity dispersion from ATLAS$^{\mathrm{3D}}$ shows inconsistency with SLUGGS in the inner $\sim0.5\, R_{\mathrm{e}}$, being characterized by higher values of $\sim75\, \mathrm{kms^{-1}}$, as also previously shown in \citet{Bellstedt2017}. It is not clear what causes this offset.
    The velocity dispersion of the PNe and GCs are higher than that of the stars in some localized regions, mostly along the photometric minor-axis of the galaxy (see Fig. A4). However, the velocity dispersion profiles of the PNe and GCs are overall very consistent with each other, as well as with that of the stars, within the $1$-sigma errors and show very low (i.e. $\sim40\, \mathrm{kms^{-1}}$) values at all radii (see Fig. 3). 
    
    The $V_{\mathrm{rot}}/\sigma$ profiles of the stars, PNe and GCs are very consistent with each other, rising up to $V_{\mathrm{rot}}/\sigma\sim2$-$3$ within $\sim2$-$2.5\, R_{\mathrm{e}}$. 
    
    The $\mathrm{PA}_{\mathrm{kin}}$ profiles show good kinematic alignment between the stars, PNe and GCs, which are rotating along the photometric major-axis of the galaxy at all radii. The $\mathrm{q}_{\mathrm{kin}}$ profile of the stars is consistent with the photometric value of the galaxy between $\sim0.5$-$2\, R_{\mathrm{e}}$, while it is higher (low ellipticity) in the central regions within $\sim0.5\, R_{\mathrm{e}}$.
    
\subsection{\textit{Mis-aligned} galaxies}
\label{sec:misaligned_galaxies}

    \subsubsection{NGC 821}
    \label{sec:N821_galaxy}
    NGC 821 is an isolated elliptical galaxy classified as E6 according to NED\footnote{The NASA/IPAC Extragalactic Database (NED) is operated by the Jet Propulsion Laboratory, California Institute of Technology, under contract with the National Aeronautics and Space Administration.}. However, \citet{Emsellem2004} found evidence of a rapidly rotating disk-like component from stellar kinematic studies, as part of the SAURON project. Further evidence of the presence of a stellar disk in NGC 821 comes also from earlier photometric studies, e.g. \citet{Michard1994,Goudfrooij1994,Ravindranath2001} as well as from stellar population and kinematic studies \citep{Proctor2005,Proctor2009}. Specifically, \citet{Proctor2009} found evidence of a rotating disk extending out to $\sim1\, R_{\mathrm{e}}$, while \citet{Proctor2005} found strong age and metallicity gradients within the inner $\sim1\, R_{\mathrm{e}}$ of the galaxy, with a young ($\sim1$-$4\, \mathrm{Gyr}$ old) and metal-rich ($[Z/H]\sim0.5\, \mathrm{dex}$) central stellar population that transitions to an older ($\sim10\, \mathrm{Gyr}$) and more metal-poor ($[Z/H]\sim-0.5\, R_{\mathrm{e}}$) one at $\sim1\, R_{\mathrm{e}}$.

    The photometrically-identified GC system of NGC 821 contains $320 \pm 45$ objects and it displays a colour bimodality distribution at $(B - Rc) \sim 1.3$ \citep{Spitler2008}. This colour bimodality is also visible in the spectroscopic GC catalogue containing $68$ objects. \citet{Pota2013} studied the kinematics of the two GC sub-populations separately in this galaxy, finding minor-axis rotation for the blue GCs, as was also observed for the PNe by \citet{Coccato2009}. 
    In this work, we choose not to split the GC population of NGC 821, as that would result in $14$ red and $35$ blue GCs, but we study the kinematics of the two GC sub-populations combined. In this way, we can work with a higher number of GCs to recover the kinematic profiles more reliably. 
    
    The PNe system contains $186$ unique objects with measured radial velocities and it was obtained by combining the individual spectroscopic catalogues of \citet{Coccato2009} and \citet{Teodorescu2010}. The former catalogue was observed as part of the ePN.S survey \citep{Pulsoni2018} and contains $127$ PNe, while the latter one was observed with the Subaru/FOCAS spectrograph and contains $155$ PNe candidates ($145$ with measured radial velocities). We cross-match these PNe catalogues and find $84$ objects in common in the two surveys, similarly to \citet{Teodorescu2010}. The combination of the $84$ common PNe with the $43$ PNe from ePN.S and the $71$ PNe from Subaru/FOCAS gives us a total of $188$ PNe with measured radial velocities. After the removal of two objects from this sample that are outliers in the phase-space diagram, we end up with the final spectroscopic catalogue of $186$ PNe with mean error on the radial velocity measurements of $\Delta V = 21\, \mathrm{km/s}$. 
    
    \myparagraph{Kinematic profiles}
    NGC 821 has a roughly flat stellar rotation velocity profile within $\sim0.5\, R_{\mathrm{e}}$ (from ATLAS$^{\mathrm{3D}}$), which decreases out to $\sim1\, R_{\mathrm{e}}$ and, then, it remains roughly constant (from SLUGGS). The PNe show rotation between $\sim1$-$2\, R_{\mathrm{e}}$, followed by a subsequent decrease out to large radii, i.e. $\sim4\, R_{\mathrm{e}}$. The rotation velocity profile of the PNe is consistent with that of the stars at $\sim1\, R_{\mathrm{e}}$, however the PNe are rotating more closely aligned to the photometric minor-axis of the galaxy with $\mathrm{PA}_{\mathrm{kin}}\sim150\degr$ at all radii, while the stars show rotation along the galaxy photometric major-axis out to $\sim1\, R_{\mathrm{e}}$. A similar misalignment between the $\mathrm{PA}_{\mathrm{kin}}$ of the PNe and the $\mathrm{PA}_{\mathrm{phot}}$ of the galaxy was previously also found in \citet{Coccato2009}. The GCs show decreasing rotation velocity profile out to $\sim4\, R_{\mathrm{e}}$, which is overall consistent with that of the PNe. 
    The 2D kinematic maps in Fig. A5 show that also the GCs are rotating along a $\mathrm{PA}_{\mathrm{kin}}$ which is misaligned from the $\mathrm{PA}_{\mathrm{phot}}$ of the galaxy, similarly to the PNe. The derived 1D kinematic profile of the $\mathrm{PA}_{\mathrm{kin}}$ of the GCs shows agreement with that of the stars (and, thus, with the $\mathrm{PA}_{\mathrm{phot}}$ of the galaxy), however if we consider the large $1$-sigma errors of the $\mathrm{PA}_{\mathrm{kin}}$ profile of the GCs, then it is also consistent to some extent with the $\mathrm{PA}_{\mathrm{kin}}$ profile of the PNe. \citet{Pota2013} found that the rotation along the photometric minor-axis of the galaxy is associated with the blue GC sub-population. Some GCs also show evidence of counter-rotation at $\sim3\, R_{\mathrm{e}}$ towards South-West (see Fig. A5). 
    
    The stellar velocity dispersion profile shows evidence of decreasing within $\sim0.5\, R_{\mathrm{e}}$ (ATLAS$^{\mathrm{3D}}$) and, then, it is roughly constant at $\sim140\, \mathrm{kms^{-1}}$ at $\sim1\, R_{\mathrm{e}}$ (SLUGGS). The PNe show a decreasing velocity dispersion profile out to $\sim1\, R_{\mathrm{e}}$, consistent with both ATLAS$^{\mathrm{3D}}$ and SLUGGS. Beyond $\sim1\, R_{\mathrm{e}}$, the velocity dispersion profile of the PNe keeps decreasing out to $\sim4\, R_{\mathrm{e}}$. The GCs show a similar decreasing velocity dispersion profile, as the PNe, out to $\sim4\, R_{\mathrm{e}}$, as also shown by \citet{Pota2013}. 
    
    The $V_{\mathrm{rot}}/\sigma$ profiles of the stars, PNe and GCs are very similar in shape to the corresponding rotation velocity profiles, with the stars showing decreasing $V_{\mathrm{rot}}/\sigma$ profiles out to $\sim1\, R_{\mathrm{e}}$ and the PNe and GCs out to $\sim4\, R_{\mathrm{e}}$.
    
    The $\mathrm{q}_{\mathrm{kin}}$ profile of the stars is consistent with the $\mathrm{q}_{\mathrm{phot}}$ of the galaxy at $\sim1\, R_{\mathrm{e}}$, while it shows lower values within $\sim0.5\, R_{\mathrm{e}}$. The $\mathrm{q}_{\mathrm{kin}}$ profile of the GCs agrees with that of the stars at $\sim1\, R_{\mathrm{e}}$, and, then, it smoothly decreases out to $\sim4\, R_{\mathrm{e}}$.

    \subsubsection{NGC 4649}
    \label{sec:N4649_galaxy}
    NGC 4649 (M60) is an elliptical galaxy of E2/S0 morphological type \citep{deVaucouleurs1976}, belonging to the Virgo galaxy cluster. NGC 4649 is in a pair with the nearby smaller spiral galaxy, NGC 4647. However there are no clear signs of interactions between the two galaxies, whose optical light morphology does not look disturbed (e.g. \citealt{Pota2015}).
    
    This is the galaxy with the largest GC system in our sample containing $431$ objects. \citet{Pota2015} found evidence of a clear bimodal colour distribution of the GC population of NGC 4649 for $(g-z)=1.2$, which we also adopt in this work, thus yielding $179$ red, metal-rich and $252$ blue, metal-poor GCs.
    
    The PNe system contains $298$ objects, which do not have complete coverage for position angles $280\degr < \mathrm{PA} < 310\degr$ (with PA measured North towards East). The kinematic properties of the PNe were studied in \citet{Coccato2013}, who found consistent rotation aligned with the stars along the galaxy photometric major-axis in the inner regions that becomes misaligned beyond $\sim200\arcsec$.
    
    Given the incomplete coverage for $280\degr < \mathrm{PA} < 310\degr$, we test the symmetric properties of the PNe system with respect to the galaxy photometric axes and produce a denser distribution of tracers by \textit{folding} the kinematic catalogue around its symmetry axis, similarly to NGC 3115 in \citet{Dolfi2020} and to other galaxies in previous literature works (e.g. \citealt{Peng2004,Cappellari2015,Pulsoni2018}). 
    
    Before testing the point-symmetry of the PNe system of NGC 4649, we rotate the galaxy coordinates such that the photometric major-axis of the galaxy is aligned to the $x$-axis of an orthogonal reference system centred on NGC 4649. The new coordinates of this rotated system are calculated using equation 2 in \citet{Dolfi2020}.
     
\begin{figure}
    \centering
    \includegraphics[width=0.45\textwidth]{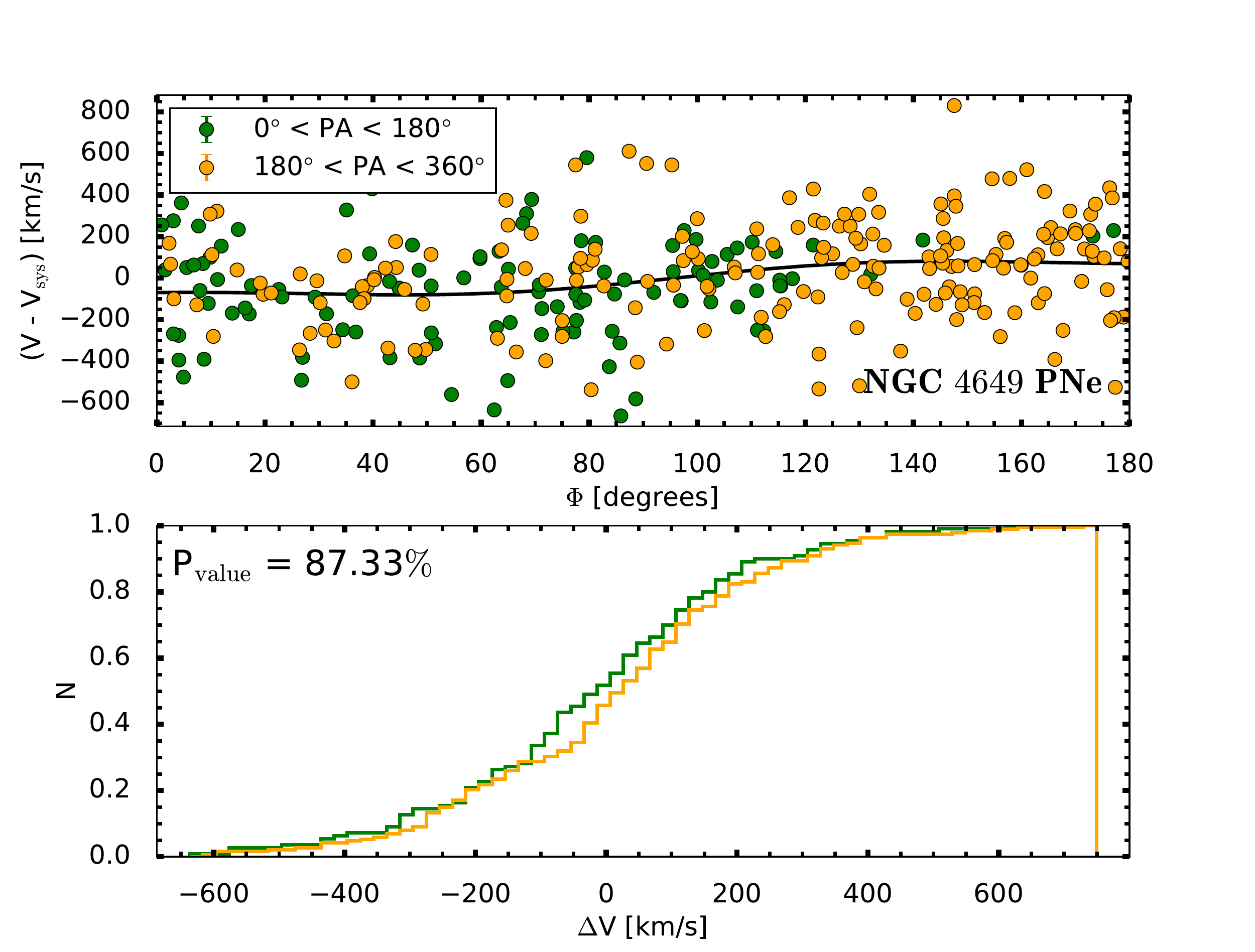}
    \caption{Point-symmetry for the PNe of NGC 4649. \textit{Top panel:} velocities of the PNe as a function of their position angle, $\Phi$. The PNe velocities on one side of the galaxy photometric major-axis (i.e. between $0\degr < \mathrm{PA} < 180\degr$; green points) are compared with the velocities (changed in sign) on the opposite side (i.e. between $180\degr < \mathrm{PA} < 360\degr$; orange points). The systemic velocity, $V_{\mathrm{sys}}$, of the galaxy has been subtracted from the PNe velocities. The black solid line represents the best-fit model rotation curve from Eq. \ref{eq:rotation_model} fitted to the PNe velocities. \textit{Bottom panel:} normalized cumulative distribution functions of the differences of the PNe velocities from the best-fit model rotation for the PNe with $0\degr < \mathrm{PA} < 180\degr$ (green line) and with $180\degr < \mathrm{PA} < 360\degr$ (yellow line). The $\mathrm{P}_{\mathrm{value}}$ indicates the two-sample KS-test probability value for the null hypothesis that the two functions are tracing similar underlying distributions.}
    \label{fig:NGC4649_PNe_point_symmetry}
    \end{figure}

    In the top panel of Fig. \ref{fig:NGC4649_PNe_point_symmetry}, we show the PNe velocities, $V(R, \Phi)$, with $0\degr < \Phi < 180\degr$ (green points) and those, $-V(R, \Phi-180\degr)$, with $180\degr < \Phi < 360\degr$ (orange points) at all radii as a function of the position angle, $\Phi$, of the tracers after subtracting the $V_{\mathrm{sys}}$ of NGC 4649. A significant overlap between the two velocity fields (i.e. $V(R, \Phi)$ and $-V(R, \Phi-180\degr)$) suggests that the system is symmetric with respect to the origin of the orthogonal reference system centred on NGC 4649.
    
    Following \citet{Pulsoni2018}, we fit our data with a rotation model function given by
    
    \begin{equation}
        V_{\mathrm{obs}} = V_{\mathrm{rot}}\cos(\phi - \mathrm{PA_{kin}}) + s_{3}\sin(3\phi - 3\mathrm{PA_{kin}}) + c_{3}\cos(3\phi - 3\mathrm{PA_{kin}}), 
        \label{eq:rotation_model}
    \end{equation}
    
    where $V_{\mathrm{rot}}$, $\mathrm{PA_{kin}}$, $s_{3}$ and $c_{3}$ are the free parameters in the fit and $\phi$ is the independent $x$ variable ranging between $[0,180]\degr$. In Eq. \ref{eq:rotation_model}, the parameters $s_{3}$ and $c_{3}$ are the coefficients associated with higher order harmonics to account for deviations from the simple cosine law rotation model. As already done in \citet{Dolfi2020}, and following \citet{Coccato2009}, we assume the kinematic axial-ratio, $\mathrm{q_{kin}}=1$, in Eq. \ref{eq:rotation_model} for the purpose of testing the point-symmetry. The best-fit rotation model curve is represented by the black solid line in Fig. \ref{fig:NGC4649_PNe_point_symmetry}. 
    
    To quantify the degree of point-symmetry of the PNe system of NGC 4649, we calculate the difference of the PNe velocities from the best-fit model rotation function and we show the results in the form of normalized cumulative distribution functions for the PNe with $0\degr < \Phi < 180\degr$ (green line) and $180\degr < \Phi < 360\degr$ (orange line) as shown in the bottom panel of Fig \ref{fig:NGC4649_PNe_point_symmetry}. We find that the two functions are tracing the same underlying distribution with a probability $\mathrm{P_{value}}=87.3\%$ from the non-parametric two-sample Kolmogorow-Smirnow (KS) test.
    Therefore, the PNe system of NGC 4649 is point-symmetric (as previously found also in \citealt{Coccato2013}) and we can create the point-symmetric kinematic catalogue by taking the original catalogue and applying the following transformation ($X$, $Y$, $V$) $\rightarrow$ ($-X$, $-Y$, $-V$). 
    The final PNe catalogue is twice as large as the original catalogue, being composed of the original catalogue and its point-symmetric counterpart.
    
    \myparagraph{Kinematic profiles}
    NGC 4649 has a smoothly rising stellar rotation velocity profile out to $\sim2$-$3\, R_{\mathrm{e}}$. This is consistent with ATLAS$^{\mathrm{3D}}$ that shows rotation for the stars within $\sim0.5\, R_{\mathrm{e}}$. The rotation velocity profile of the PNe is consistent with that of the stars out to $\sim1.5\, R_{\mathrm{e}}$, where the $\mathrm{PA}_{\mathrm{kin}}$ profiles of the stars and PNe show overall good kinematic alignment along the photometric major-axis of the galaxy. At $\sim2\, R_{\mathrm{e}}$, the rotation velocity of the PNe shows a dip in rotation. Beyond this radius, the rotation velocity of the PNe rises again and it is characterized by a new peak in rotation occurring beyond $\sim2.5\, R_{\mathrm{e}}$, where the PNe show rotation more closely aligned to the photometric minor-axis of the galaxy, as seen from the $\mathrm{PA}_{\mathrm{kin}}$ profile. The red GCs show a similar double-peaked rotation velocity profile as the the PNe, with strong rotation ($\sim350\, \mathrm{kms^{-1}}$) within $\sim1\, R_{\mathrm{e}}$, where the $\mathrm{PA}_{\mathrm{kin}}$ profile shows good kinematic alignment with both the stars and PNe, followed by a sharp drop in rotation out to $\sim3\, R_{\mathrm{e}}$. At larger radii, we see that the rotation velocity of the red GCs increases again and reaches a new peak in rotation between $\sim5$-$6\, R_{\mathrm{e}}$. 
    The $\mathrm{PA}_{\mathrm{kin}}$ profile of the red GCs does not show clear signs of kinematic twists as a function of galactocentric radius, suggesting that the rotation occurs along a similar kinematic axis at all radii. \citet{Pota2015} have also shown that the red GCs have overall constant $\mathrm{PA}_{\mathrm{kin}}$ at all radii, consistent with the $\mathrm{PA}_{\mathrm{phot}}$ of the galaxy. The blue GCs show a rotation velocity profile consistent with that of the stars between $\sim1$-$2\, R_{\mathrm{e}}$, beyond which it decreases out to large radii, i.e. $\sim7\, R_{\mathrm{e}}$. However, the blue GCs show rotation more closely aligned to the photometric minor-axis of the galaxy between $\sim2$-$4\, R_{\mathrm{e}}$, similarly to the PNe beyond $\sim2.5\, R_{\mathrm{e}}$, as was also previously found by \citet{Pota2015}.      
    
\begin{figure*}
    \centering
        \includegraphics[width=0.8\textwidth]{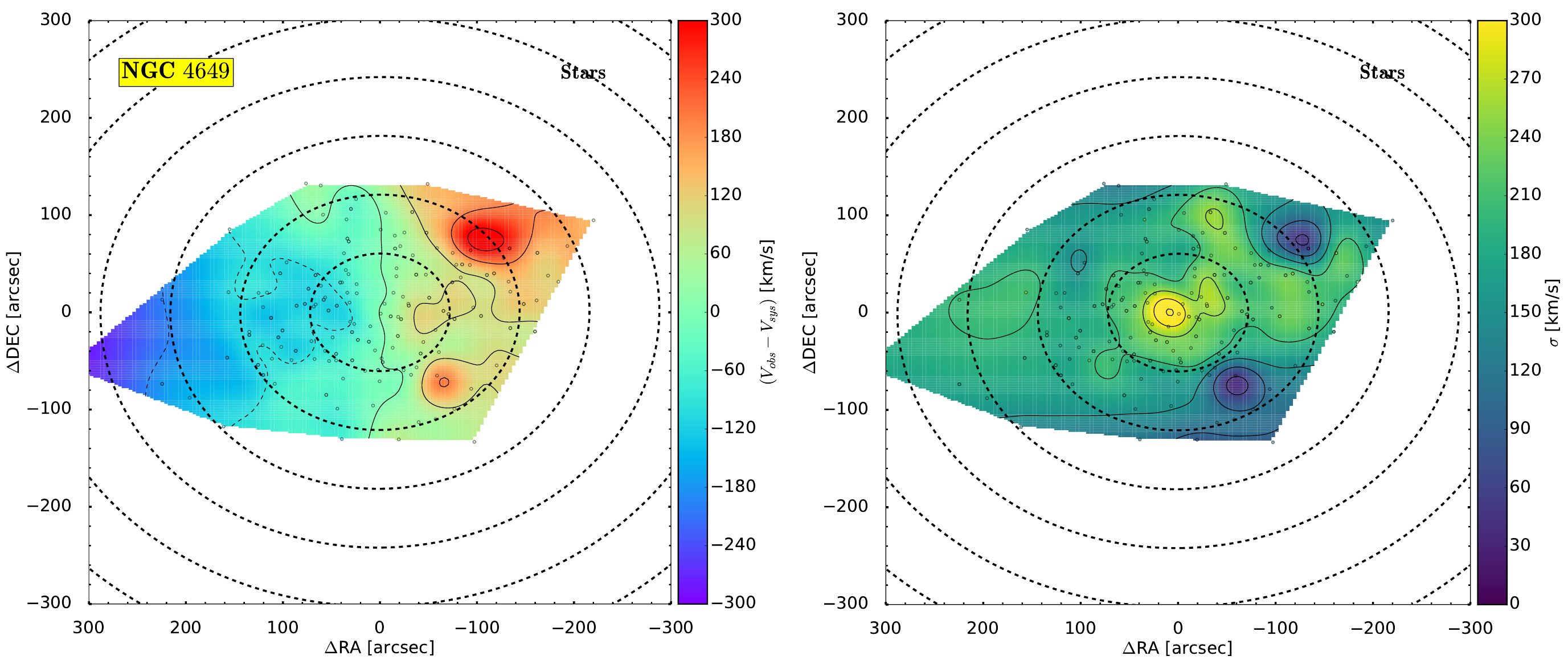}
        \includegraphics[width=0.8\textwidth]{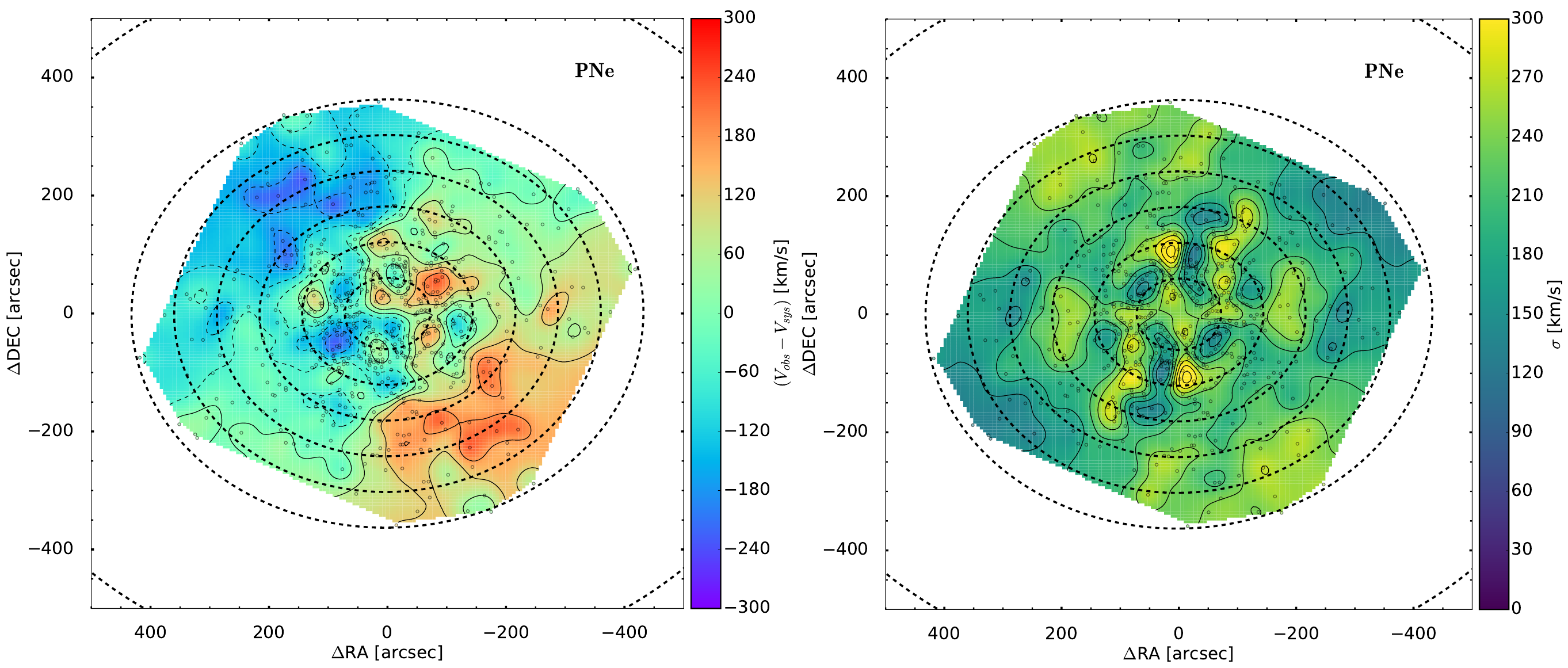}
        \includegraphics[width=0.8\textwidth]{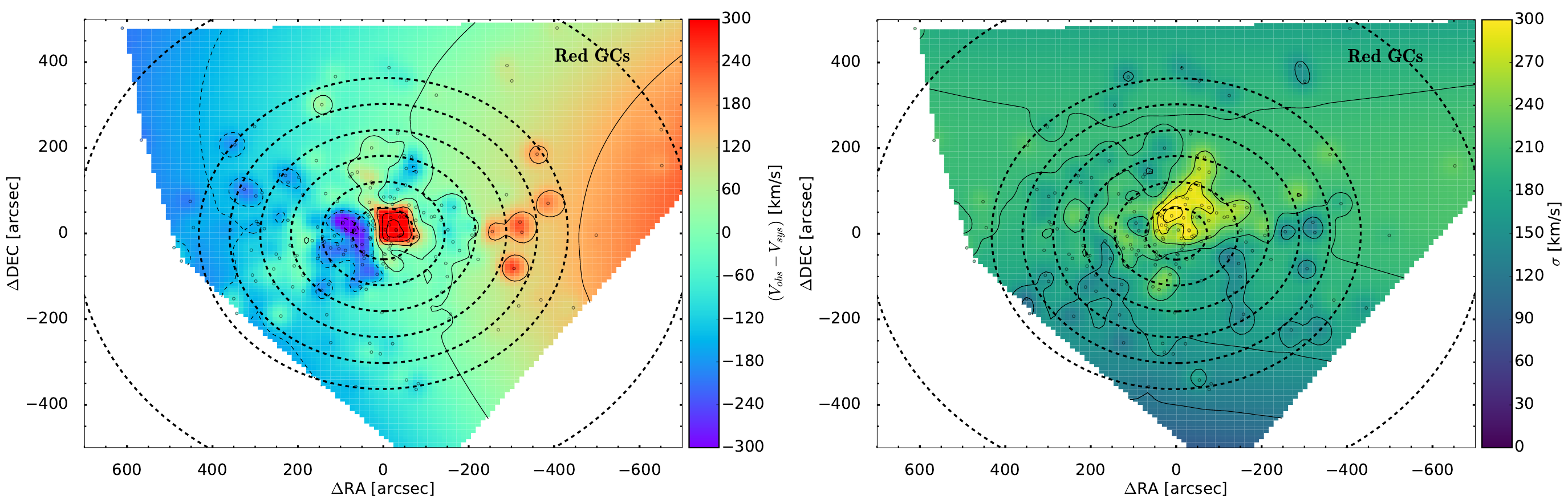}
        \includegraphics[width=0.8\textwidth]{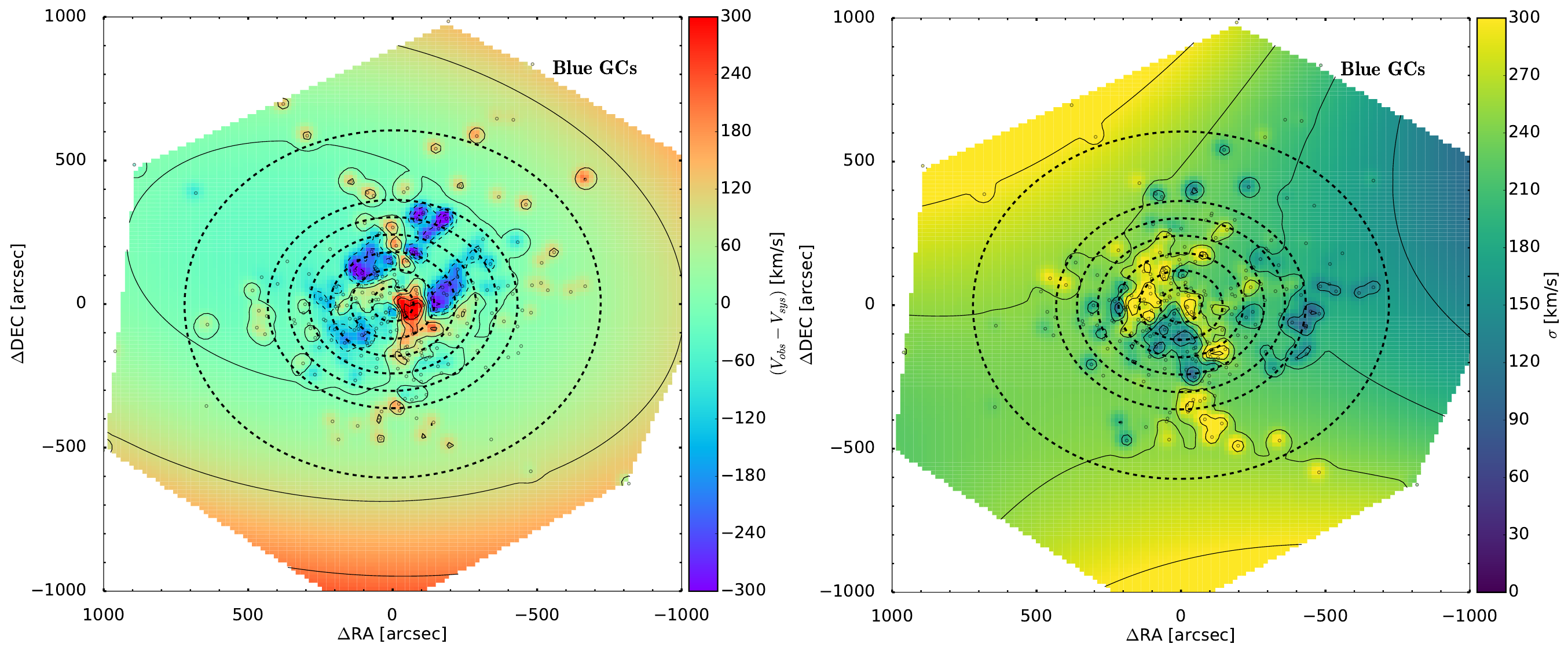}
        \caption{From top to bottom, 2D velocity (left-hand side) and velocity dispersion (right-hand side) maps of the SLUGGS stars, PNe, red and blue GCs for NGC 4649. The description is as in Fig. \ref{fig:NGC1023_2d_kinematic_maps}. We fold the PNe catalogue with respect to the galaxy centre (as described in Sec. \ref{sec:N4649_galaxy}) prior to producing the 2D kinematic maps, in order to account for the incomplete $\mathrm{PA}$ coverage of the PNe towards North-West (see Sec. \ref{sec:N4649_galaxy}). The description is as in Fig. 2.}
    \label{fig:NGC4649_2d_kinematic_maps}
    \end{figure*}

    The velocity dispersion profile of the stars smoothly decreases out to $\sim2$-$3\, R_{\mathrm{e}}$, after the prominent central peak at $\sim300\, \mathrm{kms^{-1}}$, shown by ATLAS$^{\mathrm{3D}}$. The velocity dispersion of the red GCs also shows a high central peak at $\sim300\, \mathrm{kms^{-1}}$ within $\sim1$-$2\, R_{\mathrm{e}}$, followed by a subsequent decrease out to $\sim7\, R_{\mathrm{e}}$, similarly to the stars (see Fig. \ref{fig:NGC4649_2d_kinematic_maps}) but with an offset of $\sim50\, \mathrm{kms^{-1}}$ to higher values. The PNe and blue GCs show similar velocity dispersion profiles out to $\sim5$-$7\, R_{\mathrm{e}}$, which are slightly decreasing along the photometric major-axis of the galaxy, while they remain roughly constant along the photometric minor-axis of the galaxy (see Fig. \ref{fig:NGC4649_2d_kinematic_maps}). 
    The velocity dispersion of the PNe and blue GCs is also slightly higher than that of the red GCs beyond $\sim2\, R_{\mathrm{e}}$ (see Fig. 3).
    
    The stellar $V_{\mathrm{rot}}/\sigma$ profile steeply rises out to $\sim2\, R_{\mathrm{e}}$, where it reaches its maximum value (i.e. $V_{\mathrm{rot}}/\sigma\sim1$). The blue GCs show a $V_{\mathrm{rot}}/\sigma$ profile consistent with the stars within $\sim2\, R_{\mathrm{e}}$, beyond which it decreases out to large radii (i.e. $\sim7\, R_{\mathrm{e}}$). The PNe and red GCs are characterized by a double-peaked $V_{\mathrm{rot}}/\sigma$ profile, with the first peak occurring at $\sim1$-$1.5\, R_{\mathrm{e}}$, where it is overall consistent with the stars. The second peak of the $V_{\mathrm{rot}}/\sigma$ profile of the PNe and red GCs occurs at larger radii, i.e. $\sim3\, R_{\mathrm{e}}$ and $\sim5$-$6\, R_{\mathrm{e}}$, respectively, where the rotation is happening along the photometric minor-axis of the galaxy for the PNe. 
    The dip in rotation velocity for the PNe and GCs was also previously found by \citet{Pota2015}, who argued that it could be either not a real feature and due to the presence of the neighbouring spiral galaxy, NGC 4647, located between $\sim2$-$3\, R_{\mathrm{e}}$ towards North-West that prevents us from observing the GCs and PNe within that spatial region or it could be a result of the past interaction with the ultra-compact dwarf galaxy, UCD 1 \citep{Strader2013}.

    \begin{figure*}
    \centering
        \includegraphics[width=0.82\textwidth]{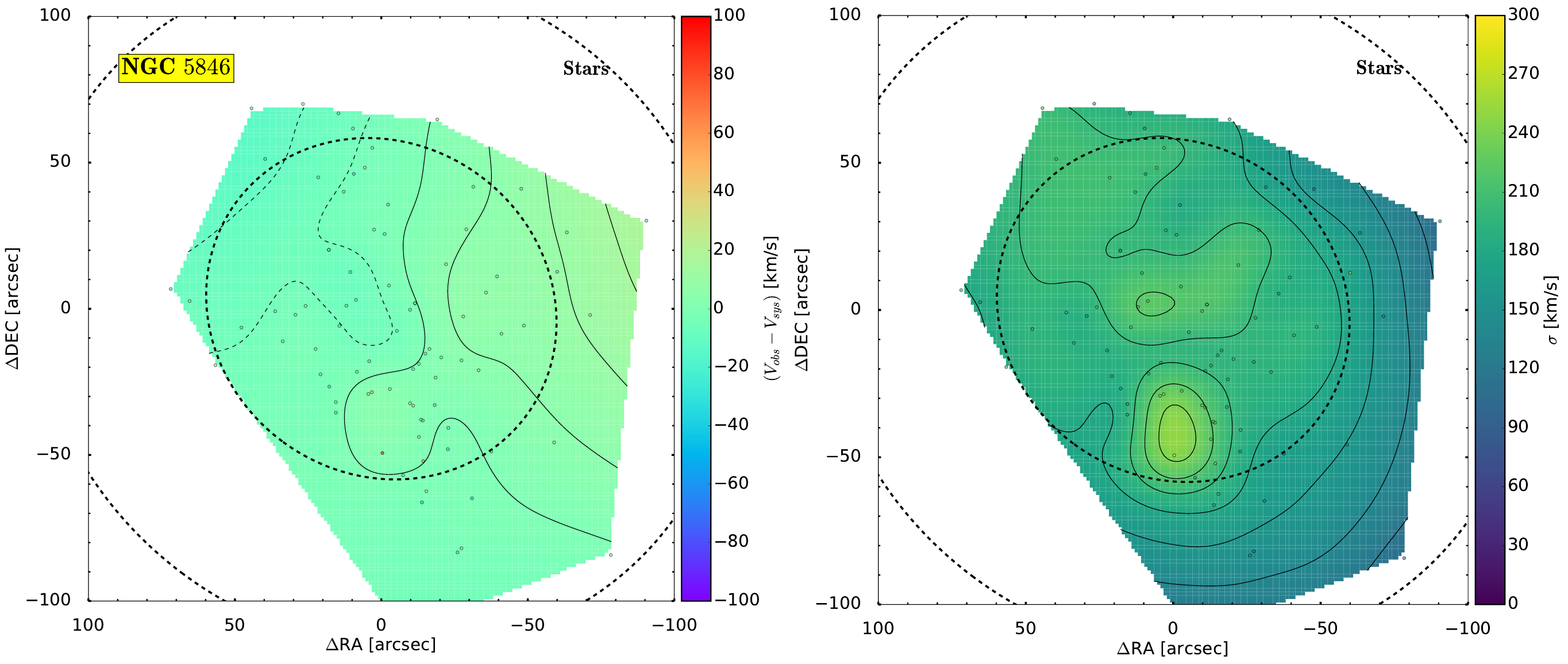}
        \includegraphics[width=0.82\textwidth]{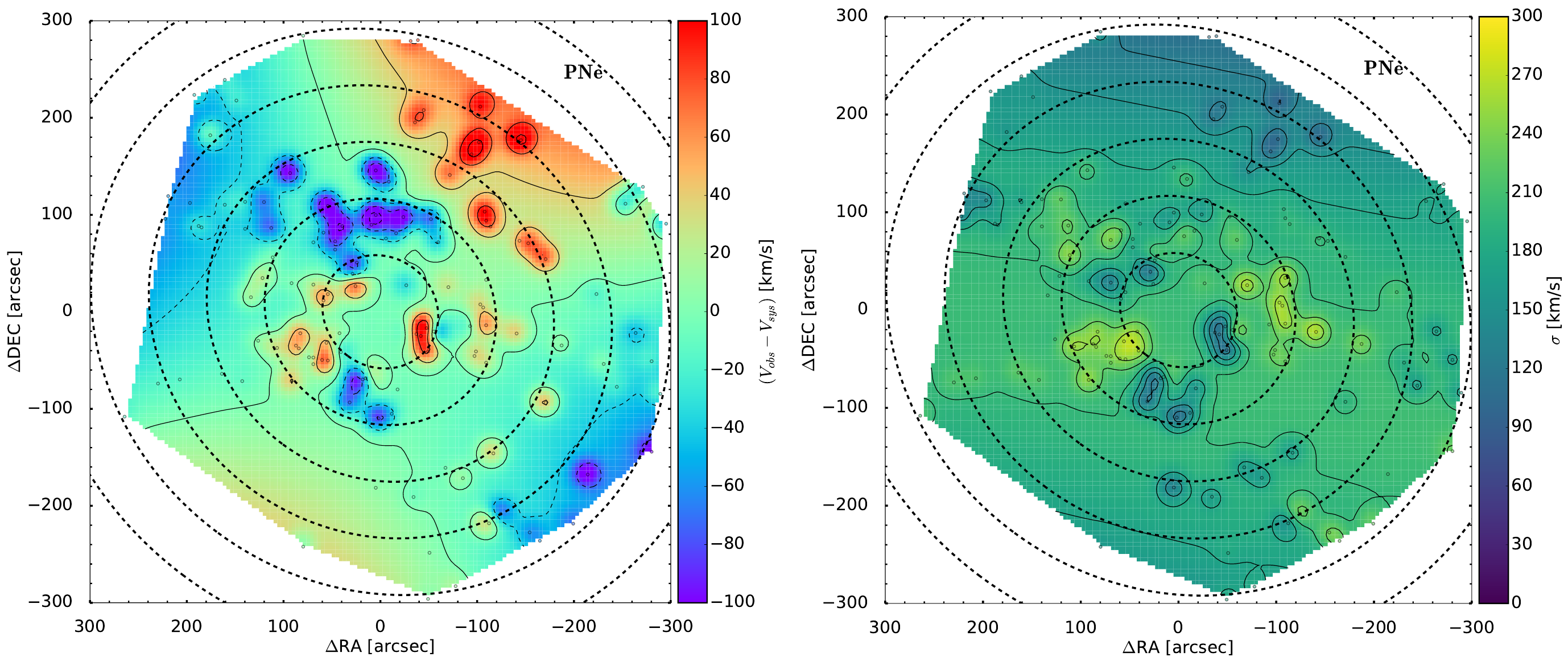}
        \includegraphics[width=0.82\textwidth]{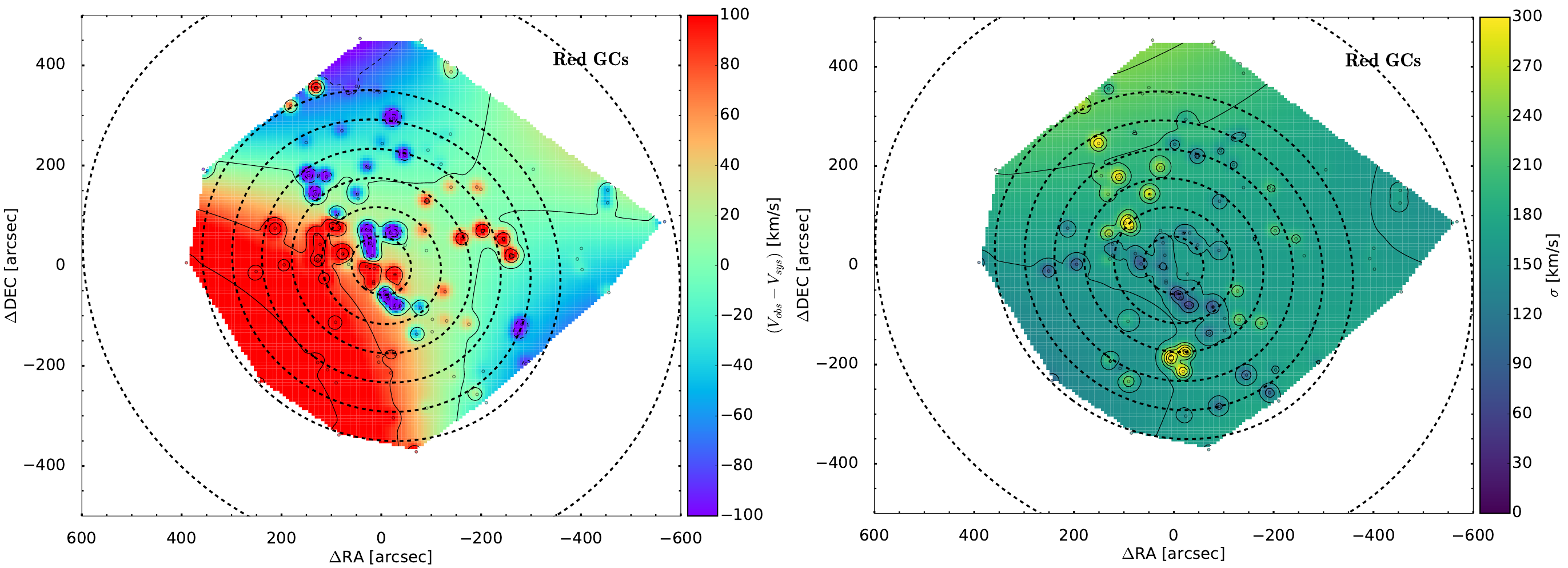}
        \includegraphics[width=0.82\textwidth]{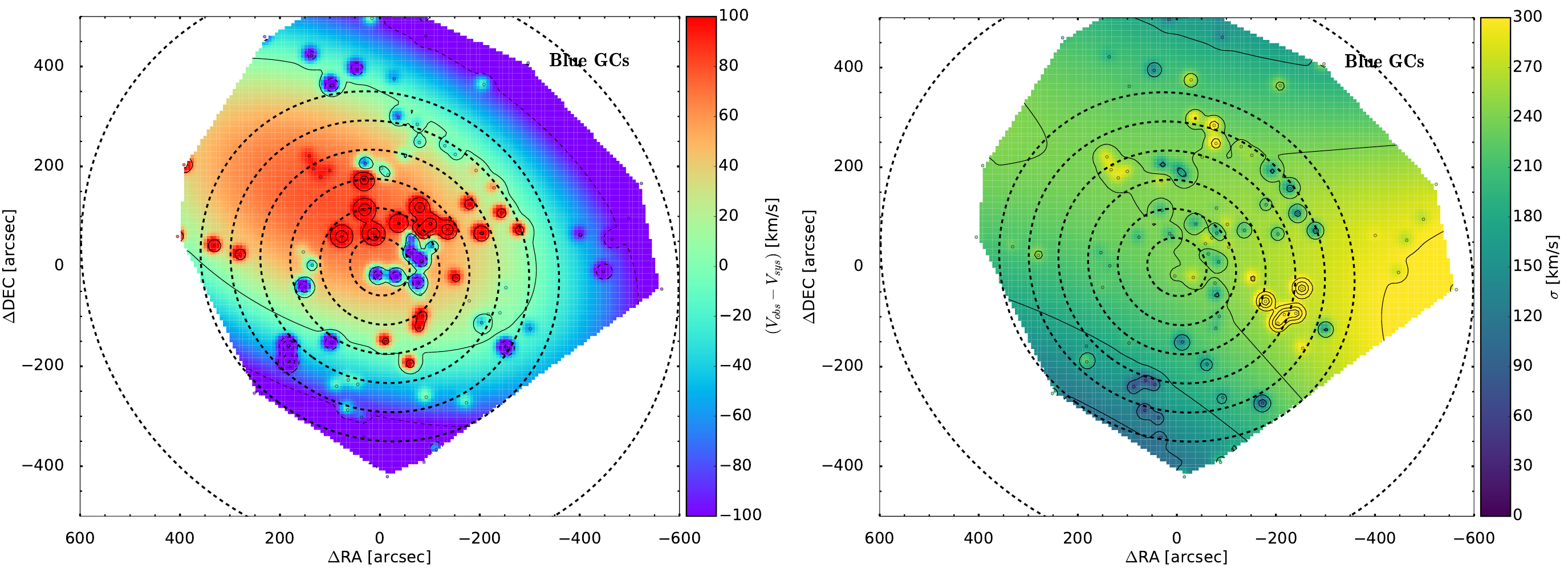}
        \caption{From top to bottom, 2D velocity (left-hand side) and velocity dispersion (right-hand side) maps of the SLUGGS stars, PNe, red and blue GCs of NGC 5846. The description is as in Fig. 2.}
    \label{fig:NGC5846_2d_kinematic_maps}
    \end{figure*}

    \subsubsection{NGC 5846}
    \label{sec:N5846_galaxy}
    NGC 5846 is a massive, bright early-type galaxy belonging to a galaxy group, which is located in the constellation Virgo. This galaxy has conflicting morphology, being classified as an E0-1 in RC2 and S0 in RSA \citep{Michard1994}. 
    Early works have identified very little to no rotation from stellar \citep{Emsellem2004,Foster2016} and GC kinematic \citep{Pota2013} studies of this galaxy.
    
    The GC system of NGC 5846 contains $215$ objects. \citet{Pota2013} obtained photometry in the $g$, $r$ and $i$ bands for $195$ GCs from their new Subaru observations and they identified a colour bi-modality distribution in the GC population for $(g-i) = 0.95\, \mathrm{mag}$. In this work, we adopt the same colour-split obtaining as final samples $103$ red, metal-rich and $92$ blue, metal-poor GCs.
    
    The PNe system of NGC 5846 contains $124$ objects and its kinematic properties were initially studied by \citet{Coccato2009}, who did not find significant rotation along the major- and minor-axes.
    
    \myparagraph{Kinematic profiles}
    NGC 5846 does not show any stellar rotation within $\sim1$-$2\, R_{\mathrm{e}}$, as previously shown by ATLAS$^{\mathrm{3D}}$ within $\sim0.5\, R_{\mathrm{e}}$. No net rotation is also seen for the PNe, whose rotation velocity profile is constant at $\sim50\, \mathrm{kms^{-1}}$ out to $\sim5\, R_{\mathrm{e}}$. The red GCs also show roughly constant rotation velocity profile out to $\sim6\, R_{\mathrm{e}}$, which is overall consistent with that of the PNe with a small offset in rotation. 
    On the other hand, the blue GCs show rotation velocity, peaking at $\sim200\, \mathrm{kms^{-1}}$, between $\sim1.5$-$2.5\, R_{\mathrm{e}}$, which decreases out to large radii, i.e. $\sim4\, R_{\mathrm{e}}$, beyond which it is consistent with both that of the PNe and red GCs. However, from the 2D kinematic maps in Fig. \ref{fig:NGC5846_2d_kinematic_maps}, we do not see any clear evidence of rotation around the galaxy for the blue GCs. \citet{Pota2013} also found rotation for the blue GCs from the 1D kinematic profiles, but they argue that it may likely be due to few outliers GCs that belong to the nearby companion galaxy, NGC 5846A. 
    
    The stellar velocity dispersion is roughly constant at $\sim200\, \mathrm{kms^{-1}}$ within $\sim1\, R_{\mathrm{e}}$ and shows good agreement with the velocity dispersion profile of the PNe, which remains overall constant out to $\sim5\, R_{\mathrm{e}}$. The red and blue GCs also show velocity dispersion profiles, which are roughly constant out to $\sim5\, R_{\mathrm{e}}$ and overall consistent with the velocity dispersion profile of the PNe within the $1$-sigma errors. However, the blue GCs show somewhat higher velocity dispersion than the red GCs and PNe (see Fig. \ref{fig:NGC5846_2d_kinematic_maps}).
    
    The $V_{\mathrm{rot}}/\sigma$ profiles of the stars and PNe show that NGC 5846 has spheroid-like kinematics out to $\sim5\, R_{\mathrm{e}}$, with flat $V_{\mathrm{rot}}/\sigma<0.5$ at all radii. Similar behaviour is also observed from the red and blue GCs $V_{\mathrm{rot}}/\sigma$ profiles, if the observed rotation between $\sim1$-$2\, R_{\mathrm{e}}$ (red GCs) and $\sim1.5$-$2.5\, R_{\mathrm{e}}$ (blue GCs) is, indeed, spurious and a consequence of outliers GCs.  
    For this galaxy, we do not fit for the $\mathrm{PA}_{\mathrm{kin}}$ and $\mathrm{q}_{\mathrm{kin}}$ profiles of any kinematic dataset, as they are not well constrained from the \texttt{kinemetry} fit. However, we did calculate the 1D kinematic profiles for NGC 5846, leaving the $\mathrm{PA}_{\mathrm{kin}}$ parameter free to vary for all our kinematic datasets, finding that both the rotation velocity and velocity dispersion profiles did not change significantly.




\bibliographystyle{mnras}
\bibliography{biblio} 



\bsp	
\label{lastpage}
\end{document}